\documentclass[10pt]{article}

\usepackage{amsmath}
\usepackage{amssymb}

\usepackage{graphicx,multirow}
\usepackage[hang,tight]{subfigure}
\usepackage{threeparttable}
\usepackage{lscape}
\usepackage{cite,url}

\usepackage{color}

\usepackage[labelfont=bf,labelsep=period,justification=raggedright]{caption}

\makeatletter
\renewcommand{\@biblabel}[1]{\quad#1.}
\makeatother

\date{}

\begin{document}
\vspace*{0.35in}

\begin{flushleft}
{\Large
\textbf{Profitability of contrarian strategies in the Chinese stock market}
}
\newline
\\

Huai-Long Shi\textsuperscript{1,2},
Zhi-Qiang Jiang\textsuperscript{1,2},
Wei-Xing Zhou\textsuperscript{1,2,3,*}

\bigskip

\bf{1} Department of Finance, School of Business, East China University of Science and Technology, Shanghai 200237, China
\\
\bf{2} Research Center for Econophysics, East China University of Science and Technology, Shanghai 300237, China
\\
\bf{3} Department of Mathematics, School of Science, East China University of Science and Technology, Shanghai 300237, China
\\
\bigskip

%
%




* wxzhou@ecust.edu.cn (WXZ)

\end{flushleft}

\section*{Abstract}

This paper reexamines the profitability of loser, winner and contrarian portfolios in the Chinese stock market using monthly data of all stocks traded on the Shanghai Stock Exchange and Shenzhen Stock Exchange covering the period from January 1997 to December 2012. We find evidence of short-term and long-term contrarian profitability in the whole sample period when the estimation and holding horizons are 1 month or longer than 12 months and the annualized returns of contrarian portfolios increases with the estimation and holding horizons. We perform subperiod analysis and find that the long-term contrarian effect is significant in both bullish and bearish states while the short-term contrarian effect disappears in bullish states. We compare the performance of contrarian portfolios based on different grouping manners in the estimation period and unveil that decile grouping outperforms quintile grouping and tertile grouping, which is more evident and robust in the long run. Generally, loser portfolios and winner portfolios have positive returns and loser portfolios perform much better than winner portfolios. Both loser and winner portfolios in bullish states perform better than those in the whole sample period. In contrast, loser and winner portfolios have smaller returns in bearish states in which loser portfolio returns are significant only in the long term and winner portfolio returns become insignificant. These results are robust to the one-month skipping between the estimation and holding periods and for the two stock exchanges. Our findings show that the Chinese stock market is not efficient in the weak form. These findings also have obvious practical implications for financial practitioners.

\section*{Introduction}

The Efficient Markets Hypothesis is a cornerstone of modern finance \cite{Fama-1970-JF,Fama-1991-JF}. However, there is accumulating evidence for the presence of market anomalies, such as the momentum effect and the contrarian effect. The momentum effect describes the empirically observed tendency for rising asset prices to rise further and falling prices to keep falling, while the contrarian effect describes the price reversal phenomenon stating that stocks that perform the best (worst) in the past tend to reverse to perform well (poorly) over the subsequent periods. The momentum and contrarian effects have attracted wide attention in the academic community and in the financial industry as well in the past two decades.

Jegadeesh and Iitman conduct the first research on the momentum effect \cite{Jegadeesh-Titman-1993-JF}. By setting $16$ combinations of different estimation and holding horizons, investors would get abnormal returns in the holding period through purchasing the best performing stock (winner) portfolio and selling the worst performing stock (loser) portfolio in the estimation period. They find that 15 out of the 16 arbitrage portfolios yield statistically significant returns in the next 3 to 12 months, which also confirms the existence of the intermediate-term momentum effect. The research on the contrarian effect was initially conducted in Ref.~\cite{DeBondt-Thaler-1985-JF}. They use monthly data of hundreds of individual stocks listed on the New York Stock Exchange from 1926 to 1982, and construct the winner portfolio of 35 best performed stocks in the past 3 years and the loser portfolio of 35 worst performed stocks in the past 3 years. They find empirical evidence that in the next three years, the loser portfolio performs better than the winner portfolio with a $25\%$ higher average cumulative return, which indicates the existence of a long-term contrarian effect.

Early investigations of momentum and contrarian effects focused on the US market. These two anomalies have also been found in  other markets later. Rouwenhorst investigates 12 European stock markets from 1980 to 1995 and reports that the winner portfolio results in an average monthly return $1\%$ higher than the loser portfolio \cite{Rouwenhorst-1998-JF}. Chan et al. \cite{Chan-Jegadeesh-Lakonishok-1996-JF} and Chou et al. \cite{Chou-Wei-Chung-2007-JEF} find the short-horizon contrarian effect in the Japanese market. Baytas and Cakici report the presence of a long-term contrarian effect in seven non-US markets \cite{Baytas-Cakici-1999-JBF}. Hameed and Ting find the price reversal in the Malaysian market \cite{Hameed-Ting-2000-PBFJ}. Kang et al. reveal that there was a short-term contrarian effect and an intermediate-term momentum effect in the Chinese market \cite{Kang-Liu-Ni-2002-PBFJ}. Naughton et al. also find the momentum effect in China \cite{Naughton-Truong-Veeraraghavan-2008-PBFJ}. Additionally, the momentum and contrarian effects have been discovered in the markets of different financial products. Grinblatt et al. utilize the data of 155 mutual fund companies from 1975 to 1984 and find that about $77\%$ mutual funds that applied momentum strategy gained statistically significant higher returns than the rest \cite{Grinblatt-Titman-Wermers-1995-AER}. Asness et al. investigate the correlation between the value and momentum effects in international markets, and discover a universal momentum effect in different regions and different asset classes \cite{Asness-Moskowitz-Pedersen-2013-JF}. The research in Ref.~\cite{NovyMarx-2012-JFE} draws similar conclusions.

The original momentum and contrarian effects were based on cross-sectional prices or returns of assets, also called ``price momentum'' \cite{Jegadeesh-Titman-1993-JF,Kang-Liu-Ni-2002-PBFJ,Chou-Wei-Chung-2007-JEF}. More and more kinds of momentum or contrarian effects were explored in a body of further studies about market anomalies, which in turn partially explained the presence of momentum and contrarian effects. On the basis of price momentum, various factors containing firm-specific information were taken into consideration. Investors can construct zero-cost arbitrage portfolios in terms of more information and get higher profits. These factors include firm capitalization \cite{Jegadeesh-Titman-1993-JF,Rouwenhorst-1998-JF}, stock price \cite{Lesmond-Schill-Zhou-2004-JFE,George-Hwang-2004-JF}, book-to-market ratio \cite{Daniel-Titman-2000}, trading volume \cite{Lee-Swaminathan-2000-JF,Naughton-Truong-Veeraraghavan-2008-PBFJ}, and so on. Moreover, the assets could be divided into portfolios with different styles according to these firm-specific information factors and the momentum or contrarian effect about style portfolio (style investing) also attracted wide attention as well \cite{Chen-DeBondt-2004-JEF,Levis-Liodakis-1999-JPM}. Studies on specific industrial sectors unveil that the momentum and contrarian effects can obtain much more profits in industrial sectors \cite{Moskowitz-Grinblatt-1999-JF,Lewellen-2002-RFS}. Note that the results about the momentum and contrarian effects vary with changing market states \cite{Cooper-Gutierrez-Hameed-2004-JF} and seasonality \cite{Yao-2012-JBF}. There are also studies on individual stocks and stock market indexes \cite{He-Li-2015-JBF}, which is beyond the scope of this work.

With the increasing importance of China in the world economy, more and more related researches have been carried out on the Chinese stock market. It has been shown that the Chinese stock market and the US stock market are uncorrelated \cite{Zhou-Sornette-2004a-PA} and even negatively correlated in some periods \cite{Kang-Liu-Ni-2002-PBFJ}. In early 1990s, two stock exchanges, the Shanghai Stock Exchange (SHSE) and the Shenzhen Stock Exchange (SZSE), were established in China. Compared with other mature financial markets, the data size of the Chinese market is relatively smaller, which may lead to imprecise results. The trading mechanisms of the Chinese market differ from other markets and keep self-improving. In the Chinese market, the majority of investors are retail investors, causing larger irrational and speculative behaviors. These situations may contribute differently to some anomaly phenomena. It is not surprising that studies about the momentum and contrarian effects in the Chinese stock market report mixed results. The conclusions of momentum and contrarian effects are often associated with the length of the estimation and holding periods and the sample periods under investigation. In general, the horizons can be divided into short term (3 months or less), intermediate term (3 to 12 months) and long term (more than 12 months).

Most studies report that there is a long-term contrarian effect in the Chinese stock market. Using monthly data of 53 individual stocks listed on the SHSE and the SZSE from January 1993 to December 2000, Wang and Zhao discover a statistically significant contrarian effect with the estimate period ranging from 1 to 3 years and the holding period from 1 to 5 years \cite{Wang-Zhao-2001-cnSMH}. Li and Li investigate A-shares on the SHSE and the SZSE from January 1996 to December 2002 and reveal that the market exhibits a contrarian effect in horizons more than 1 year \cite{Li-Li-2003-cnMR}. Using monthly data of A-shares traded on the SHSE and the SZSE from January 1995 to December 2002, Luo and Wang also draw the similar conclusion \cite{Luo-Wang-2004-cnSETMA}. Similar results can be found in later studies \cite{Wang-Chin-2004-PBFJ,Yang-Chen-2004-cnJTUST,Ma-Zhang-2005-cnJIEEM,Shao-Su-Yu-2005-cnCEM,Lu-Zou-2007-cnERJ,Liu-Pi-2007-cnJFR,Wu-2008-cnSCF,Pan-Xu-2011-cnJFR}.

In the short term and intermediate term, the conclusions are mixed. For example, Kang et al. use the data of individual stocks from 1993 to 2000 and find the existence of a short-term (1, 2, 4, 8, and 12 weeks) contrarian effect and a statistically significant momentum effect in the intermediate term (12, 16, 20, and 26 weeks) \cite{Kang-Liu-Ni-2002-PBFJ}. Based on A-share data in the SHSE and the SZSE from January 1995 to December 2001, Zhu et al. verify the presence of a significant momentum effect with both estimation and holding periods less than 4 weeks \cite{Zhu-Wu-Wang-2003-cnJWE}. Zhu et al. find a statistically significant contrarian effect with the estimation and holding periods less than 5 days \cite{Zhu-Wu-Xiao-2005-cnJSJU}. Liu and Qin use monthly data of constituent stocks of the SHSE 180 Index from July 2002 to September 2005 and report the presence of a momentum effect with the horizons less than 12 months \cite{Liu-Qin-2007-cnJSM}. Pan et al. find the existence of a momentum effect in weekly returns and a contrarian effect in monthly returns\cite{Pan-Xu-2011-cnJFR,Pan-Tang-Xu-2013-PBFJ}. There are also studies finding no significant momentum or contrarian effects in the short term or in the intermediate term \cite{Wang-Zhao-2001-cnSMH,Yang-Chen-2004-cnJTUST,Liu-Pi-2007-cnJFR}.

The different conclusions in the above-mentioned studies can be attributed to the following factors.
(1) Different data samples. Some studies use part of the individual stocks listed on the SHSE and the SZSE \cite{Wang-Zhao-2001-cnSMH}, while others use data of all A-shares \cite{Lu-Zou-2007-cnERJ,Liu-Pi-2007-cnJFR,Zhu-Wu-Wang-2003-cnJWE}.
(2) Different sample periods. For example, Wang and Zhao \cite{Wang-Zhao-2001-cnSMH} and Kang et al. \cite{Kang-Liu-Ni-2002-PBFJ} use the sample period from 1993 to 2000, Zhu et al. \cite{Zhu-Wu-Xiao-2005-cnJSJU} study the period from 1996 to 2001, Lu and Zou \cite{Lu-Zou-2007-cnERJ} investigate the period from 1998 to 2005, and Naughton et al. \cite{Naughton-Truong-Veeraraghavan-2008-PBFJ} consider the period from 1995 to 2005.
(3) Different sampling frequencies. Some researchers use monthly data \cite{Jegadeesh-Titman-1993-JF,Naughton-Truong-Veeraraghavan-2008-PBFJ}, while some others adopt daily and weekly data \cite{Kang-Liu-Ni-2002-PBFJ,Zhu-Wu-Wang-2003-cnJWE,Zhu-Wu-Xiao-2005-cnJSJU,Pan-Tang-Xu-2013-PBFJ}.
(4) Other factors. It is found that the bid-ask spread, non-synchronous trading as well as the lack of liquidity would enlarge the momentum and contrarian effects \cite{Lehmann-1990-QJE,Conrad-Gultekin-Kaul-1997-JBES,Ball-Kothari-Wasley-2004-JF}. To avoid these, the common approach is to skip certain time intervals between the estimation period and the holding period \cite{Jegadeesh-Titman-1993-JF,Zhu-Wu-Wang-2003-cnJWE}. Studies that do not adopt this interval-skipping approach may lead to different results \cite{Lu-Zou-2007-cnERJ,Wang-Zhao-2001-cnSMH}.

With more data available, it is worth to re-examine the contrarian and momentum effects in the Chinese stock market. We use monthly return data of all A-shares listed on the SHSE and the SZSE from January 1997 to December 2012 to construct the winner and loser portfolios, which forms zero-cost arbitrage portfolios. In the estimation period for stock ranking, one needs to determine the number of stock groups. Different studies on the momentum and contrarian effects have adopted different grouping ways, including decile grouping, quintile grouping and tertile grouping \cite{Jegadeesh-Titman-1993-JF,Wang-Zhao-2001-cnSMH,Asness-Moskowitz-Pedersen-2013-JF,Kang-Liu-Ni-2002-PBFJ,Pan-Tang-Xu-2013-PBFJ}. For various markets, different grouping ways may lead to significantly different results. This paper will take into account these three grouping ways for comparison. Most of the previous studies about the Chinese market take the A-share market as a whole. Since the features of A-shares listed on these two stock exchanges are not similar, we investigate the momentum and contrarian effects in the SHSE and the SZSE independently. For instance, the A-share stocks listed on the SHSE generally have higher market capitalization compared with those in the SZSE. However, empirical analysis in this study fails to verify any significant differences between the results of the two exchanges.

\section*{Materials and Methods}
\label{S1:Data:Method}

\subsection*{Data}

There are two stock exchanges -- Shanghai Stock Exchange (SHSE) and Shenzhen Stock Exchange (SZSE) -- in mainland China, and the Chinese stock market contains an A-share market and a B-share market. Most stocks are traded only in the A-share market, while a small proportion of stocks are traded in both markets. At the end of 2012, there are 944 A-share stocks and 54 B-share stocks in the SHSE and 1528 A-share stocks and 53 B-share stocks in the SZSE (Table \ref{Tab:StockInf}). Different from A-share stocks, B-shares were not accessible to domestic investors until February 2001, and the B-share stock market has lower liquidity and market value. As described in Table \ref{Tab:StockInf}, by the end of 2012, the B-share stocks accounted for $5.41\%$ and $3.35\%$ in the SHSE and the SZSE, the market value of B-share stocks only accounted for $0.5\%$ and $1.11\%$ in the SHSE and the SZSE, the trading value of B-share stocks took up $0.25\%$ in the SHSE and $0.30\%$ in the SZSE, and the A-share market had larger number of investors and higher turnover rate. Therefore, our analysis is carried out upon the Chinese A-share stock market, which can be representative of the Chinese domestic investment environment. Because the average market capitalizations of SHSE stocks (16.73 billion CNY per stock) and SZSE stocks (4.64 billion CNY per stock) are significantly different, we shall investigate separately the A-share stocks in the two exchanges for comparison.

\setlength\tabcolsep{2.5pt}
\begin{table*}[htb]
\centering
  \caption{{\textbf{Basic information about Chinese stock market by the end of 2012.}} The data are retrieved from the annual reports released by the SHSE and the SZSE.}
  \medskip
  \label{Tab:StockInf}%
  \begin{tabular}{ccccccccccccccccc}
    \hline
          && \multicolumn{2}{c}{No. of listed stocks} && \multicolumn{2}{c}{Market cap (B)} && \multicolumn{2}{c}{Trading value (B)} && \multicolumn{2}{c}{No. of investors (M)} && \multicolumn{2}{c}{Turnover rate} \\
          \cline{3-4} \cline{6-7} \cline{9-10} \cline{12-13} \cline{15-16}
          && A-share & B-share && A-share & B-share && A-share & B-share && A-share & B-share && A-share & B-share \\
    \hline
    SHSE  && 944   & 54    && 15791.27  & 78.58  && 16404.74 & 41.35 && 88.42 & 1.55 && 101.9\% & 57.5\% \\
    SZSE  && 1528  & 53    && ~7086.27  & 79.65  && 14966.78 & 45.47 && /    & /   && /   & / \\
    \hline
  \end{tabular}%
\end{table*}%

\begin{figure}[htb]
  \centering
  \includegraphics[width=8cm]{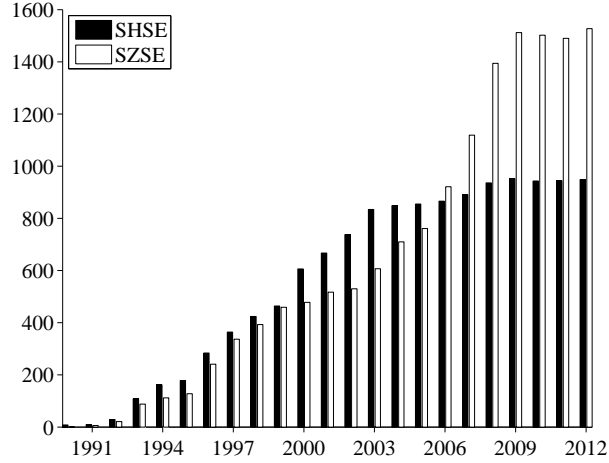}
  \caption{\label{Fig:StockNum} {\textbf{The evolution of stock amounts on the SHSE and the SZSE.}}}
\end{figure}

We use monthly data of all A-share stocks listed on the SHSE and the SZSE retrieved from the RESSET database (http://www.resset.cn). The data mainly contain the monthly adjusted returns of individual stocks, covering the period from January 1991 to December 2012. Very often, stock price jumps occur during the IPO month, which is attributed to the price determination of China Securities Regulatory Commission (CSRC), and it results in abnormal first-month returns for individual stocks. Therefore, the first-month return data of individual stocks are excluded from our analysis. During the early years after the establishment of the two exchanges, only a few stocks were available to investors, as shown in Fig.~\ref{Fig:StockNum}. There were less than 200 stocks listed on each exchange by the end of 1995 and the number of stocks increased steadily. The mature price limit trading rules became effective since December 1996, requiring that the maximum daily price fluctuation with respect to the last closing price is $\pm10\%$ for common stocks. Although there were also other price limit trading rules in the Chinese stock market for some periods before then, the implementation time periods were short. Combining these factors, we exclude the data before 1997 and consider the period from January 1997 to December 2012 in the empirical analysis.


\subsection*{Method}

\begin{figure}[h]
  \centering
  \includegraphics[width=0.6\linewidth]{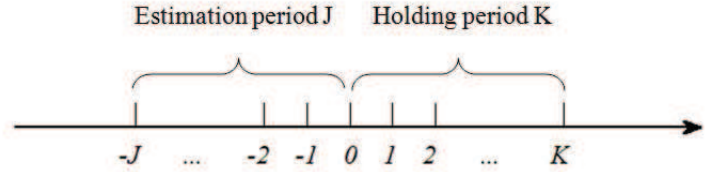}
  \caption{\label{Fig:Periods} {\textbf{The estimation and holding periods.}} The estimation period is $J$ months and the holding period is $K$ months.}
\end{figure}

Like most studies about the cross-sectional momentum or contrarian effect \cite{Naughton-Truong-Veeraraghavan-2008-PBFJ,Pan-Tang-Xu-2013-PBFJ}, we follow the procedure proposed by \cite{Jegadeesh-Titman-1993-JF} to construct $J-K$ portfolios. For a given ``current'' month $t=0$, all the stocks are sorted according to their returns in the past $J$ months from $t=-J$ to $t=0$ (Fig.~\ref{Fig:Periods}). We divide the stocks into several groups. For comparison, decile grouping, quintile grouping and tertile grouping are adopted. The group of stocks with the worst performance in the estimation is called loser portfolio LOS($J,K$) and the group with best performance is called winner portfolio WIN($J,K$). One then adopts the contrarian strategy by buying the loser portfolio and selling the winner portfolio. The contrarian portfolio CON($J,K$) is held for $K$ months. We examine the equal-weighted average returns per annum of the loser portfolio, the winner portfolio, and the contrarian portfolio during the holding period, denoted by $L_{J,K}$, $W_{J,K}$ and $C_{J,K}$ respectively. The contrarian effect is verified if the time series of returns for contrarian portfolios turn out to be statistically positive. Conversely, there would be the momentum effect.

\section*{Empirical results}

\subsection*{The case of identical estimation and holding horizons $(J=K)$}

We study the performance of the three portfolios in the whole period (1997-2012) with same estimation period and holding period for the three ranking groupings. The periods range from one month to four years: $J=K \in \{J,K|1,6,12,18,24,30,36,42,48\}$. Table \ref{TB:Empirics:J=K} reports the equal-weighted average annual returns for the loser, winner and contrarian portfolios by taking a long position of each portfolio in the holding period. Comparing the two panels, we find that the results are qualitatively similar with minor differences.

\setlength\tabcolsep{1.3pt}
\begin{landscape}
\begin{table}[!ht]
\caption{
{\bf The annualized returns of the loser, winner, and contrarian portfolios formed based on decile grouping with $J=K$ for the whole sample period 1997-2012.}}
   \begin{tabular}{cccccccccccccccccccccccccccccccccccc}
   \hline
    & \multicolumn{2}{c}{1} && \multicolumn{2}{c}{6} && \multicolumn{2}{c}{12} && \multicolumn{2}{c}{18} && \multicolumn{2}{c}{24} && \multicolumn{2}{c}{30} && \multicolumn{2}{c}{36} && \multicolumn{2}{c}{42} && \multicolumn{2}{c}{48} \\
   \cline{2-3} \cline{5-6} \cline{8-9} \cline{11-12} \cline{14-15} \cline{17-18} \cline{20-21} \cline{23-24} \cline{26-27}
       & Ret & $t$-stat && Ret & $t$-stat && Ret & $t$-stat && Ret & $t$-stat && Ret & $t$-stat && Ret & $t$-stat && Ret & $t$-stat && Ret & $t$-stat && Ret & $t$-stat \\
   \hline
   \multicolumn{19}{l}{Panel A: SHSE}\\
   \hline
   \vspace{-3mm}\\
   \multicolumn{27}{l}{{\textit{Panel A1: Decile grouping}}} \\
   LOS & 0.213 &  2.22$^{*~}$ && 0.187 &  1.90$^{~~}$ && 0.225 &  2.25$^{*~}$ && 0.272 &  2.64$^{**}$ && 0.302 &  3.17$^{**}$ && 0.328 &  3.83$^{**}$ && 0.328 &  4.81$^{**}$ && 0.350 &  5.25$^{**}$ && 0.361 &  4.86$^{**}$  \\
   WIN & 0.086 &  0.88$~~$ && 0.175 &  1.92$^{~~}$ && 0.185 &  2.03$^{*~}$ && 0.185 &  1.91$^{~~}$ && 0.190 &  1.96$^{~~}$ && 0.190 &  2.17$^{*~}$ && 0.172 &  2.45$^{*~}$ && 0.152 &  2.81$^{**}$ && 0.146 &  3.03$^{**}$  \\
   CON & 0.128 &  3.41$^{**}$ && 0.011 &  0.29$~~$ && 0.040 &  1.14$~~$ && 0.087 &  2.95$^{**}$ && 0.112 &  3.92$^{**}$ && 0.138 &  4.41$^{**}$ && 0.156 &  4.64$^{**}$ && 0.197 &  5.80$^{**}$ && 0.215 &  5.18$^{**}$  \\
    \vspace{-3mm}\\
   \multicolumn{27}{l}{{\textit{Panel A2: Quintile grouping}}} \\
   LOS & 0.228 &  2.33$^{*~}$ && 0.194 &  1.95$^{~~}$ && 0.241 &  2.40$^{*~}$ && 0.282 &  2.69$^{**}$ && 0.304 &  3.03$^{**}$ && 0.307 &  3.71$^{**}$ && 0.303 &  4.68$^{**}$ && 0.318 &  5.12$^{**}$ && 0.324 &  4.85$^{**}$  \\
   WIN & 0.107 &  1.07$~~$ && 0.186 &  1.99$^{*~}$ && 0.190 &  2.09$^{*~}$ && 0.200 &  2.08$^{*~}$ && 0.199 &  2.14$^{*~}$ && 0.199 &  2.39$^{*~}$ && 0.177 &  2.69$^{**}$ && 0.169 &  3.19$^{**}$ && 0.171 &  3.40$^{**}$  \\
   CON & 0.120 &  3.58$^{**}$ && 0.008 &  0.26$~~$ && 0.051 &  1.75$^{~~}$ && 0.082 &  3.24$^{**}$ && 0.106 &  4.43$^{**}$ && 0.108 &  4.26$^{**}$ && 0.126 &  4.49$^{**}$ && 0.148 &  5.18$^{**}$ && 0.153 &  4.87$^{**}$  \\
   \vspace{-3mm}\\
   \multicolumn{27}{l}{{\textit{Panel A3: Tertile grouping}}} \\
   LOS & 0.227 &  2.30$^{*~}$ && 0.203 &  2.04$^{*~}$ && 0.246 &  2.47$^{*~}$ && 0.281 &  2.73$^{**}$ && 0.300 &  3.00$^{**}$ && 0.301 &  3.65$^{**}$ && 0.290 &  4.56$^{**}$ && 0.297 &  5.05$^{**}$ && 0.298 &  4.88$^{**}$  \\
   WIN & 0.126 &  1.24$~~$ && 0.193 &  2.03$^{*~}$ && 0.204 &  2.20$^{*~}$ && 0.219 &  2.25$^{*~}$ && 0.222 &  2.34$^{*~}$ && 0.215 &  2.63$^{**}$ && 0.186 &  2.99$^{**}$ && 0.179 &  3.43$^{**}$ && 0.183 &  3.58$^{**}$  \\
   CON & 0.101 &  3.62$^{**}$ && 0.010 &  0.41$~~$ && 0.042 &  1.82$^{~~}$ && 0.061 &  3.02$^{**}$ && 0.078 &  4.01$^{**}$ && 0.086 &  4.28$^{**}$ && 0.104 &  4.82$^{**}$ && 0.118 &  5.54$^{**}$ && 0.114 &  5.35$^{**}$  \\
   \hline
   \multicolumn{19}{l}{Panel B: SZSE}\\
   \hline
   \vspace{-3mm}\\
   \multicolumn{27}{l}{{\textit{Panel B1: Decile grouping}}} \\
   LOS & 0.193 &  2.05$^{*~}$ && 0.175 &  1.81$^{~~}$ && 0.245 &  2.53$^{*~}$ && 0.286 &  2.92$^{**}$ && 0.347 &  3.05$^{**}$ && 0.357 &  3.64$^{**}$ && 0.338 &  4.51$^{**}$ && 0.342 &  5.06$^{**}$ && 0.344 &  4.72$^{**}$  \\
   WIN & 0.091 &  0.87$~~$ && 0.159 &  1.71$^{~~}$ && 0.187 &  1.86$^{~~}$ && 0.199 &  1.86$^{~~}$ && 0.217 &  1.89$^{~~}$ && 0.197 &  2.11$^{*~}$ && 0.177 &  2.56$^{*~}$ && 0.168 &  2.67$^{**}$ && 0.157 &  2.88$^{**}$  \\
   CON & 0.102 &  2.54$^{*~}$ && 0.017 &  0.40$~~$ && 0.058 &  1.36$~~$ && 0.087 &  2.43$^{*~}$ && 0.131 &  3.35$^{**}$ && 0.160 &  4.03$^{**}$ && 0.160 &  4.05$^{**}$ && 0.174 &  4.23$^{**}$ && 0.187 &  5.18$^{**}$  \\
   \vspace{-3mm}\\
   \multicolumn{27}{l}{{\textit{Panel B2: Quintile grouping}}} \\
   LOS & 0.203 &  2.13$^{*~}$ && 0.189 &  1.90$^{~~}$ && 0.250 &  2.52$^{*~}$ && 0.290 &  2.78$^{**}$ && 0.321 &  2.93$^{**}$ && 0.322 &  3.54$^{**}$ && 0.308 &  4.42$^{**}$ && 0.295 &  5.00$^{**}$ && 0.308 &  4.68$^{**}$  \\
   WIN & 0.094 &  0.92$~~$ && 0.170 &  1.79$^{~~}$ && 0.193 &  1.93$^{~~}$ && 0.207 &  1.96$^{~~}$ && 0.217 &  2.03$^{*~}$ && 0.208 &  2.33$^{*~}$ && 0.192 &  2.81$^{**}$ && 0.180 &  2.99$^{**}$ && 0.175 &  3.09$^{**}$  \\
   CON & 0.109 &  3.33$^{**}$ && 0.020 &  0.61$~~$ && 0.057 &  1.77$^{~~}$ && 0.083 &  2.97$^{**}$ && 0.104 &  3.56$^{**}$ && 0.114 &  3.80$^{**}$ && 0.116 &  3.66$^{**}$ && 0.115 &  3.44$^{**}$ && 0.133 &  4.60$^{**}$  \\
   \vspace{-3mm}\\
   \multicolumn{27}{l}{{\textit{Panel B3: Tertile grouping}}} \\
   LOS & 0.207 &  2.12$^{*~}$ && 0.198 &  1.96$^{~~}$ && 0.250 &  2.48$^{*~}$ && 0.281 &  2.67$^{**}$ && 0.307 &  2.90$^{**}$ && 0.302 &  3.51$^{**}$ && 0.286 &  4.34$^{**}$ && 0.275 &  4.89$^{**}$ && 0.280 &  4.54$^{**}$  \\
   WIN & 0.110 &  1.06$~~$ && 0.172 &  1.82$^{~~}$ && 0.202 &  2.05$^{*~}$ && 0.224 &  2.15$^{*~}$ && 0.230 &  2.23$^{*~}$ && 0.221 &  2.57$^{*~}$ && 0.200 &  3.02$^{**}$ && 0.189 &  3.29$^{**}$ && 0.186 &  3.34$^{**}$  \\
   CON & 0.097 &  3.71$^{**}$ && 0.026 &  1.09$~~$ && 0.049 &  1.94$^{~~}$ && 0.057 &  2.72$^{**}$ && 0.077 &  3.32$^{**}$ && 0.081 &  3.43$^{**}$ && 0.086 &  3.17$^{**}$ && 0.086 &  3.09$^{**}$ && 0.094 &  4.12$^{**}$  \\
   \hline
   \end{tabular}
\begin{flushleft} This table reports the average annualized returns and the corresponding t-statistics adjusted for heteroscedasticity and autocorrelation of the loser (LOS), winner (WIN) and contrarian (CON) portfolios, which are formed based on $J$-month lagged returns and held for $K$ months with $J=K$. In ranking the $J$-month lagged returns, decile grouping, quintile grouping and tertile grouping are adopted. The sample period is January 1997 to  December 2012. The superscripts * and ** denote the significance at 5\% and 1\% levels, respectively.
\end{flushleft}
  \label{TB:Empirics:J=K}
\end{table}
\end{landscape}

The returns of all the portfolios are positive. This finding is not surprising because the Chinese market has an overall rise. The returns of the loser portfolios are larger than those of the winner portfolios, indicating the presence of investor overreaction \cite{DeBondt-Thaler-1985-JF,DeBondt-Thaler-1987-JF,Thaler-1988a-JEP}. The performance of the contrarian strategy is worse than the simple strategy of buying loser portfolio. The positive returns of winner portfolios indicate that the contrarian strategy profits mainly come from the loser portfolio.


Nearly all the strategy portfolios could gain the statistically significant positive return when the holding horizons are beyond about 18 months or equal to 1 month. When the horizon is 6 or 12 months, there is no significant evidence for the presence of the contrarian effect or the momentum effect. The finding of long-term contrarian effect is consistent with most literature about the Chinese stock market. The short-term contrarian effect may be due to no time gap between the estimation and holding period, which could exaggerate the contrarian effect because of some measurement errors. We will discuss further in the rest of the paper.

A close scrutiny unveils that the return $C_{J,K}$ of contrarian portfolios decreases at first and then rises up with the horizon $J$ or $K$, which can be characterized as a U-shape relation. For example, on the basis of decile grouping for SHSE stocks, the CON$(1,1)$ portfolio produces an annual return of $12.8\%$, the return of the CON$(6,6)$ portfolio decreases to $1.1\%$, and the profit of the contrarian portfolios keep rising to $19.7\%$ for CON$(48,48)$ with increasing horizon $J$. It is worthy noting that \cite{Chou-Wei-Chung-2007-JEF} report a similar and more general U-shape relationship between the returns and the horizons when investigating the contrarian effect in the Japanese market.

Table \ref{TB:Empirics:J=K} also shows the impact of different grouping methods. The annualized return of loser portfolios on both exchanges decreases with the number of groups from tertile to decile for small $J$ and increases for long horizons $J$. For the winner portfolios on both exchanges, the annualized return is smaller when there are more groups. Combining these observations, the trend of returns of the contrarian portfolios on short horizons is mixed; however, on long horizons, the return increases with the number of groups.

\subsection*{The case of varying $J$ and $K$}

We now perform more comprehensive analysis with varying $J$ and $K$ for decile, quintile and tertile groupings. The results of decile grouping are presented in Table \ref{TB:Empirics:SHSE:Decile} for SHSE stocks and in Table \ref{TB:Empirics:SZSE:Decile} for SZSE stocks. The values of estimation horizons $J$ and holding horizons $K$ range from one month to four years. Panels A, B and C report the results for loser portfolios, winner portfolios, and contrarian portfolios, respectively. The results based on quintile grouping and tertile grouping for the two exchanges are similar to the case of decile grouping and shown in Table~\ref{TBS:Empirics:SHSE:Quintile} for quintile grouping of SHSE stocks, Table~\ref{TBS:Empirics:SZSE:Quintile} for quintile grouping of SZSE stocks, Table~\ref{TBS:Empirics:SHSE:tertile} for tertile grouping of SHSE stocks, and Table~\ref{TBS:Empirics:SZSE:tertile} for tertile grouping of SZSE stocks.

{\small{
According to Table \ref{TB:Empirics:SHSE:Decile} and Table \ref{TB:Empirics:SZSE:Decile}, the most intriguing feature is that all annualized returns are positive. Panel A illustrates that, for both exchanges, all the loser portfolios result in positive returns across all estimation and holding horizons, and all the returns are statistically significant at the 5\% level except for a few portfolios at small horizons. The profits of winner portfolios on both exchanges, as shown in panel B, are also positive for all combinations of $J$ and $K$, which suggests that the profits of contrarian portfolios are mainly contributed by the loser portfolios. The significance of results for winner portfolios differ by the holding horizons $K$. No returns of the winner portfolios are significant at the 5\% level when $K\leq6$, while all returns are significantly positive when $K\geq30$. The contrarian portfolios are formed by selling the winner portfolios and buying the loser portfolios. Panel C shows that all the contrarian portfolios yield positive returns, which implies the absence of the momentum effect on both exchanges during the whole sample period. All the returns are significantly positive when the holding horizon $K=1$, except for CON$(12,1)$ on the SHSE, which indicates the presence of the short-term contrarian effect, consistent with the most previous literatures such as \cite{Kang-Liu-Ni-2002-PBFJ} and \cite{Lu-Zou-2007-cnERJ} on shorter sample periods. When the estimation horizon $J$ is larger than one year, most returns are significantly positive despite of the holding periods, suggesting the presence of both short-term and long-term contrarian effects, which is again consistent with other works such as \cite{Wang-Zhao-2001-cnSMH}, \cite{Zou-Qian-2003-cnASM}, \cite{Li-Li-2003-cnMR}, and  \cite{Liu-Pi-2007-cnJFR} for shorter sample periods.}}

\begin{landscape}
\setlength\tabcolsep{1.3pt}
\begin{table}[!ht]
\caption{
{\bf The annualized returns of the loser, winner, and contrarian portfolios on the SHSE formed based on decile grouping with varying $J$ and $K$ for the whole sample period 1997-2012.}}
   \begin{tabular}{ccccccccccccccccccccccccccc}
   \hline
    & \multicolumn{2}{c}{$K=1$} && \multicolumn{2}{c}{6} && \multicolumn{2}{c}{12} && \multicolumn{2}{c}{18} && \multicolumn{2}{c}{24} && \multicolumn{2}{c}{30} && \multicolumn{2}{c}{36} && \multicolumn{2}{c}{42} && \multicolumn{2}{c}{48} \\
   \cline{2-3} \cline{5-6} \cline{8-9} \cline{11-12} \cline{14-15} \cline{17-18} \cline{20-21} \cline{23-24} \cline{26-27}
    $J$ & Ret & $t$-stat && Ret & $t$-stat && Ret & $t$-stat && Ret & $t$-stat && Ret & $t$-stat && Ret & $t$-stat && Ret & $t$-stat && Ret & $t$-stat && Ret & $t$-stat \\
   \hline
   \vspace{-3mm}\\
   \multicolumn{27}{l}{\textit{Panel A: Loser portfolio}} \\
   1& 0.213 &  2.22$^{*~}$ && 0.204 &  2.02$^{*~}$ && 0.225 &  2.29$^{*~}$ && 0.245 &  2.48$^{*~}$ && 0.258 &  2.62$^{*~}$ && 0.261 &  3.10$^{**}$ && 0.245 &  3.74$^{**}$ && 0.246 &  4.17$^{**}$ && 0.249 &  4.06$^{**}$  \\
   6& 0.237 &  2.18$^{*~}$ && 0.187 &  1.90$^{~~}$ && 0.220 &  2.20$^{*~}$ && 0.241 &  2.39$^{*~}$ && 0.257 &  2.64$^{**}$ && 0.257 &  3.16$^{**}$ && 0.244 &  3.83$^{**}$ && 0.248 &  4.26$^{**}$ && 0.252 &  4.16$^{**}$  \\
   12& 0.221 &  2.05$^{*~}$ && 0.194 &  1.91$^{~~}$ && 0.225 &  2.25$^{*~}$ && 0.253 &  2.46$^{*~}$ && 0.267 &  2.81$^{**}$ && 0.274 &  3.43$^{**}$ && 0.271 &  4.07$^{**}$ && 0.270 &  4.38$^{**}$ && 0.278 &  4.25$^{**}$  \\
   18& 0.248 &  2.27$^{*~}$ && 0.206 &  1.99$^{*~}$ && 0.247 &  2.39$^{*~}$ && 0.272 &  2.64$^{**}$ && 0.280 &  2.99$^{**}$ && 0.286 &  3.57$^{**}$ && 0.279 &  4.20$^{**}$ && 0.285 &  4.57$^{**}$ && 0.293 &  4.42$^{**}$  \\
   24& 0.249 &  2.26$^{*~}$ && 0.226 &  2.19$^{*~}$ && 0.268 &  2.63$^{*~}$ && 0.292 &  2.87$^{**}$ && 0.302 &  3.17$^{**}$ && 0.303 &  3.70$^{**}$ && 0.295 &  4.42$^{**}$ && 0.299 &  4.86$^{**}$ && 0.310 &  4.62$^{**}$  \\
   30& 0.278 &  2.54$^{*~}$ && 0.263 &  2.51$^{*~}$ && 0.291 &  2.86$^{**}$ && 0.314 &  3.05$^{**}$ && 0.323 &  3.24$^{**}$ && 0.328 &  3.83$^{**}$ && 0.315 &  4.64$^{**}$ && 0.318 &  5.07$^{**}$ && 0.331 &  4.75$^{**}$  \\
   36& 0.303 &  2.64$^{**}$ && 0.277 &  2.62$^{*~}$ && 0.304 &  2.94$^{**}$ && 0.326 &  3.05$^{**}$ && 0.340 &  3.35$^{**}$ && 0.339 &  3.99$^{**}$ && 0.328 &  4.81$^{**}$ && 0.335 &  5.18$^{**}$ && 0.345 &  4.83$^{**}$  \\
   42& 0.301 &  2.75$^{**}$ && 0.283 &  2.69$^{**}$ && 0.313 &  3.01$^{**}$ && 0.341 &  3.18$^{**}$ && 0.351 &  3.44$^{**}$ && 0.353 &  4.06$^{**}$ && 0.343 &  4.86$^{**}$ && 0.350 &  5.25$^{**}$ && 0.357 &  4.90$^{**}$  \\
   48& 0.304 &  2.62$^{*~}$ && 0.291 &  2.65$^{**}$ && 0.322 &  3.00$^{**}$ && 0.345 &  3.15$^{**}$ && 0.360 &  3.38$^{**}$ && 0.362 &  3.99$^{**}$ && 0.353 &  4.85$^{**}$ && 0.353 &  5.26$^{**}$ && 0.361 &  4.86$^{**}$  \\
    \vspace{-3mm}\\
   \multicolumn{27}{l}{\textit{Panel B: Winner portfolio}}  \\
   1& 0.086 &  0.88$^{~~}$ && 0.164 &  1.80$^{~~}$ && 0.210 &  2.21$^{*~}$ && 0.231 &  2.32$^{*~}$ && 0.237 &  2.47$^{*~}$ && 0.237 &  2.86$^{**}$ && 0.226 &  3.33$^{**}$ && 0.229 &  3.83$^{**}$ && 0.230 &  3.86$^{**}$  \\
   6& 0.113 &  1.20$^{~~}$ && 0.175 &  1.92$^{~~}$ && 0.192 &  2.13$^{*~}$ && 0.215 &  2.23$^{*~}$ && 0.219 &  2.32$^{*~}$ && 0.218 &  2.62$^{*~}$ && 0.214 &  3.00$^{**}$ && 0.215 &  3.33$^{**}$ && 0.212 &  3.53$^{**}$  \\
   12& 0.140 &  1.48$^{~~}$ && 0.165 &  1.87$^{~~}$ && 0.185 &  2.03$^{*~}$ && 0.201 &  2.07$^{*~}$ && 0.203 &  2.15$^{*~}$ && 0.209 &  2.43$^{*~}$ && 0.203 &  2.71$^{**}$ && 0.197 &  3.03$^{**}$ && 0.186 &  3.33$^{**}$  \\
   18& 0.118 &  1.23$^{~~}$ && 0.147 &  1.66$^{~~}$ && 0.169 &  1.87$^{~~}$ && 0.185 &  1.91$^{~~}$ && 0.193 &  2.02$^{*~}$ && 0.196 &  2.25$^{*~}$ && 0.187 &  2.49$^{*~}$ && 0.177 &  2.92$^{**}$ && 0.172 &  3.12$^{**}$  \\
   24& 0.115 &  1.17$^{~~}$ && 0.142 &  1.59$^{~~}$ && 0.164 &  1.82$^{~~}$ && 0.177 &  1.87$^{~~}$ && 0.190 &  1.96$^{~~}$ && 0.191 &  2.19$^{*~}$ && 0.177 &  2.49$^{*~}$ && 0.170 &  2.88$^{**}$ && 0.167 &  3.05$^{**}$  \\
   30& 0.092 &  0.97$^{~~}$ && 0.131 &  1.48$^{~~}$ && 0.156 &  1.74$^{~~}$ && 0.174 &  1.80$^{~~}$ && 0.188 &  1.92$^{~~}$ && 0.190 &  2.17$^{*~}$ && 0.175 &  2.45$^{*~}$ && 0.166 &  2.82$^{**}$ && 0.159 &  2.96$^{**}$  \\
   36& 0.103 &  1.06$^{~~}$ && 0.139 &  1.55$^{~~}$ && 0.163 &  1.78$^{~~}$ && 0.179 &  1.84$^{~~}$ && 0.191 &  1.91$^{~~}$ && 0.192 &  2.15$^{*~}$ && 0.172 &  2.45$^{*~}$ && 0.158 &  2.82$^{**}$ && 0.148 &  2.89$^{**}$  \\
   42& 0.110 &  1.13$^{~~}$ && 0.132 &  1.50$^{~~}$ && 0.158 &  1.72$^{~~}$ && 0.182 &  1.83$^{~~}$ && 0.194 &  1.90$^{~~}$ && 0.192 &  2.18$^{*~}$ && 0.167 &  2.47$^{*~}$ && 0.152 &  2.81$^{**}$ && 0.143 &  2.90$^{**}$  \\
   48& 0.095 &  0.96$^{~~}$ && 0.118 &  1.33$^{~~}$ && 0.158 &  1.70$^{~~}$ && 0.184 &  1.85$^{~~}$ && 0.190 &  1.93$^{~~}$ && 0.186 &  2.23$^{*~}$ && 0.170 &  2.58$^{*~}$ && 0.154 &  2.89$^{**}$ && 0.146 &  3.03$^{**}$  \\
   \vspace{-3mm}\\
   \multicolumn{27}{l}{\textit{Panel C: Contrarian portfolio}} \\
   1 & 0.128 &  3.41$^{**}$ && 0.040 &  1.97$^{~~}$ && 0.014 &  0.93$^{~~}$ && 0.013 &  1.13$^{~~}$ && 0.021 &  1.95$^{~~}$ && 0.024 &  1.95$^{~~}$ && 0.019 &  1.46$^{~~}$ && 0.017 &  1.35$^{~~}$ && 0.019 &  1.22$^{~~}$  \\
   6 & 0.123 &  2.45$^{*~}$ && 0.011 &  0.29$^{~~}$ && 0.028 &  0.94$^{~~}$ && 0.026 &  1.17$^{~~}$ && 0.038 &  1.73$^{~~}$ && 0.039 &  1.77$^{~~}$ && 0.030 &  1.13$^{~~}$ && 0.033 &  1.20$^{~~}$ && 0.040 &  1.43$^{~~}$  \\
   12 & 0.081 &  1.56$^{~~}$ && 0.029 &  0.67$^{~~}$ && 0.040 &  1.14$^{~~}$ && 0.052 &  1.88$^{~~}$ && 0.064 &  2.50$^{*~}$ && 0.065 &  2.26$^{*~}$ && 0.068 &  2.13$^{*~}$ && 0.073 &  2.20$^{*~}$ && 0.091 &  3.03$^{**}$  \\
   18 & 0.129 &  2.20$^{*~}$ && 0.058 &  1.21$^{~~}$ && 0.077 &  2.04$^{*~}$ && 0.087 &  2.95$^{**}$ && 0.087 &  3.18$^{**}$ && 0.090 &  2.85$^{**}$ && 0.092 &  2.68$^{**}$ && 0.108 &  3.36$^{**}$ && 0.121 &  3.85$^{**}$  \\
   24 & 0.134 &  2.38$^{*~}$ && 0.084 &  1.96$^{~~}$ && 0.104 &  2.90$^{**}$ && 0.115 &  3.90$^{**}$ && 0.112 &  3.92$^{**}$ && 0.112 &  3.76$^{**}$ && 0.117 &  3.43$^{**}$ && 0.129 &  3.69$^{**}$ && 0.143 &  4.24$^{**}$  \\
   30 & 0.186 &  3.37$^{**}$ && 0.132 &  2.98$^{**}$ && 0.135 &  3.72$^{**}$ && 0.140 &  4.87$^{**}$ && 0.135 &  4.65$^{**}$ && 0.138 &  4.41$^{**}$ && 0.140 &  3.88$^{**}$ && 0.152 &  4.30$^{**}$ && 0.172 &  4.92$^{**}$  \\
   36 & 0.200 &  3.28$^{**}$ && 0.139 &  3.14$^{**}$ && 0.141 &  4.12$^{**}$ && 0.147 &  5.23$^{**}$ && 0.150 &  5.00$^{**}$ && 0.147 &  4.56$^{**}$ && 0.156 &  4.64$^{**}$ && 0.177 &  5.19$^{**}$ && 0.198 &  5.48$^{**}$  \\
   42 & 0.191 &  3.26$^{**}$ && 0.151 &  3.36$^{**}$ && 0.155 &  4.28$^{**}$ && 0.160 &  5.45$^{**}$ && 0.157 &  5.27$^{**}$ && 0.161 &  5.33$^{**}$ && 0.176 &  5.48$^{**}$ && 0.197 &  5.80$^{**}$ && 0.214 &  5.69$^{**}$  \\
   48 & 0.209 &  3.29$^{**}$ && 0.173 &  3.54$^{**}$ && 0.164 &  4.43$^{**}$ && 0.161 &  5.24$^{**}$ && 0.170 &  5.76$^{**}$ && 0.175 &  5.59$^{**}$ && 0.183 &  5.29$^{**}$ && 0.199 &  5.44$^{**}$ && 0.215 &  5.18$^{**}$  \\
   \hline
   \end{tabular}
   \label{TB:Empirics:SHSE:Decile}
\end{table}
\end{landscape}

\setlength\tabcolsep{1.3pt}
\begin{landscape}
\begin{table}[!ht]
\caption{
{\bf The annualized returns of the loser, winner, and contrarian portfolios on the SZSE formed based on decile grouping with varying $J$ and $K$ for the whole sample period 1997-2012.}}
   \begin{tabular}{ccccccccccccccccccccccccccc}
   \hline
    & \multicolumn{2}{c}{$K=1$} && \multicolumn{2}{c}{6} && \multicolumn{2}{c}{12} && \multicolumn{2}{c}{18} && \multicolumn{2}{c}{24} && \multicolumn{2}{c}{30} && \multicolumn{2}{c}{36} && \multicolumn{2}{c}{42} && \multicolumn{2}{c}{48} \\
   \cline{2-3} \cline{5-6} \cline{8-9} \cline{11-12} \cline{14-15} \cline{17-18} \cline{20-21} \cline{23-24} \cline{26-27}
    $J$ & Ret & $t$-stat && Ret & $t$-stat && Ret & $t$-stat && Ret & $t$-stat && Ret & $t$-stat && Ret & $t$-stat && Ret & $t$-stat && Ret & $t$-stat && Ret & $t$-stat \\
   \hline
   \vspace{-3mm}\\
   \multicolumn{27}{l}{\textit{Panel A: Loser portfolio}} \\
   1& 0.193 &  2.05$^{*~}$ && 0.183 &  1.81$^{~~}$ && 0.202 &  2.08$^{*~}$ && 0.228 &  2.36$^{*~}$ && 0.240 &  2.56$^{*~}$ && 0.245 &  3.03$^{**}$ && 0.229 &  3.55$^{**}$ && 0.230 &  4.10$^{**}$ && 0.228 &  3.97$^{**}$  \\
   6& 0.204 &  1.94$^{~~}$ && 0.175 &  1.81$^{~~}$ && 0.217 &  2.24$^{*~}$ && 0.235 &  2.48$^{*~}$ && 0.253 &  2.63$^{**}$ && 0.267 &  3.05$^{**}$ && 0.249 &  3.54$^{**}$ && 0.247 &  4.10$^{**}$ && 0.245 &  3.97$^{**}$  \\
   12& 0.264 &  2.22$^{*~}$ && 0.211 &  2.09$^{*~}$ && 0.245 &  2.53$^{*~}$ && 0.260 &  2.78$^{**}$ && 0.294 &  2.87$^{**}$ && 0.306 &  3.20$^{**}$ && 0.286 &  3.78$^{**}$ && 0.278 &  4.33$^{**}$ && 0.281 &  4.17$^{**}$  \\
   18& 0.261 &  2.33$^{*~}$ && 0.232 &  2.24$^{*~}$ && 0.261 &  2.64$^{**}$ && 0.286 &  2.92$^{**}$ && 0.319 &  2.98$^{**}$ && 0.329 &  3.41$^{**}$ && 0.307 &  3.98$^{**}$ && 0.297 &  4.52$^{**}$ && 0.299 &  4.28$^{**}$  \\
   24& 0.284 &  2.49$^{*~}$ && 0.240 &  2.36$^{*~}$ && 0.267 &  2.76$^{**}$ && 0.304 &  2.98$^{**}$ && 0.347 &  3.05$^{**}$ && 0.347 &  3.48$^{**}$ && 0.320 &  4.15$^{**}$ && 0.313 &  4.66$^{**}$ && 0.313 &  4.29$^{**}$  \\
   30& 0.281 &  2.59$^{*~}$ && 0.265 &  2.56$^{*~}$ && 0.295 &  2.96$^{**}$ && 0.327 &  3.16$^{**}$ && 0.357 &  3.24$^{**}$ && 0.357 &  3.64$^{**}$ && 0.329 &  4.31$^{**}$ && 0.320 &  4.79$^{**}$ && 0.318 &  4.38$^{**}$  \\
   36& 0.300 &  2.71$^{**}$ && 0.289 &  2.67$^{**}$ && 0.317 &  2.96$^{**}$ && 0.343 &  3.20$^{**}$ && 0.365 &  3.30$^{**}$ && 0.365 &  3.77$^{**}$ && 0.338 &  4.51$^{**}$ && 0.331 &  4.84$^{**}$ && 0.335 &  4.54$^{**}$  \\
   42& 0.336 &  2.89$^{**}$ && 0.310 &  2.72$^{**}$ && 0.328 &  2.97$^{**}$ && 0.349 &  3.22$^{**}$ && 0.373 &  3.38$^{**}$ && 0.369 &  3.84$^{**}$ && 0.343 &  4.57$^{**}$ && 0.342 &  5.06$^{**}$ && 0.339 &  4.58$^{**}$  \\
   48& 0.325 &  2.83$^{**}$ && 0.306 &  2.72$^{**}$ && 0.319 &  2.95$^{**}$ && 0.342 &  3.27$^{**}$ && 0.369 &  3.41$^{**}$ && 0.365 &  3.94$^{**}$ && 0.352 &  4.82$^{**}$ && 0.344 &  5.16$^{**}$ && 0.344 &  4.72$^{**}$  \\
    \vspace{-3mm}\\
   \multicolumn{27}{l}{\textit{Panel B: Winner portfolio}}  \\
   1& 0.091 &  0.87$^{~~}$ && 0.157 &  1.68$^{~~}$ && 0.195 &  2.05$^{*~}$ && 0.224 &  2.14$^{*~}$ && 0.232 &  2.37$^{*~}$ && 0.230 &  2.81$^{**}$ && 0.213 &  3.35$^{**}$ && 0.211 &  3.78$^{**}$ && 0.215 &  3.70$^{**}$  \\
   6& 0.104 &  1.14$^{~~}$ && 0.159 &  1.71$^{~~}$ && 0.190 &  2.00$^{*~}$ && 0.221 &  2.06$^{*~}$ && 0.225 &  2.20$^{*~}$ && 0.214 &  2.53$^{*~}$ && 0.201 &  3.02$^{**}$ && 0.209 &  3.33$^{**}$ && 0.212 &  3.30$^{**}$  \\
   12& 0.141 &  1.44$^{~~}$ && 0.158 &  1.65$^{~~}$ && 0.187 &  1.86$^{~~}$ && 0.208 &  1.95$^{~~}$ && 0.210 &  2.09$^{*~}$ && 0.211 &  2.43$^{*~}$ && 0.202 &  2.84$^{**}$ && 0.201 &  3.06$^{**}$ && 0.201 &  3.15$^{**}$  \\
   18& 0.109 &  1.11$^{~~}$ && 0.146 &  1.54$^{~~}$ && 0.176 &  1.78$^{~~}$ && 0.199 &  1.86$^{~~}$ && 0.213 &  1.97$^{~~}$ && 0.211 &  2.29$^{*~}$ && 0.195 &  2.64$^{**}$ && 0.191 &  2.90$^{**}$ && 0.189 &  3.04$^{**}$  \\
   24& 0.120 &  1.20$^{~~}$ && 0.150 &  1.57$^{~~}$ && 0.173 &  1.74$^{~~}$ && 0.203 &  1.79$^{~~}$ && 0.217 &  1.89$^{~~}$ && 0.205 &  2.17$^{*~}$ && 0.183 &  2.48$^{*~}$ && 0.180 &  2.79$^{**}$ && 0.181 &  2.90$^{**}$  \\
   30& 0.116 &  1.11$^{~~}$ && 0.145 &  1.49$^{~~}$ && 0.185 &  1.79$^{~~}$ && 0.217 &  1.85$^{~~}$ && 0.220 &  1.86$^{~~}$ && 0.197 &  2.11$^{*~}$ && 0.174 &  2.47$^{*~}$ && 0.178 &  2.84$^{**}$ && 0.182 &  2.89$^{**}$  \\
   36& 0.150 &  1.42$^{~~}$ && 0.167 &  1.65$^{~~}$ && 0.201 &  1.86$^{~~}$ && 0.218 &  1.85$^{~~}$ && 0.215 &  1.89$^{~~}$ && 0.198 &  2.20$^{*~}$ && 0.177 &  2.56$^{*~}$ && 0.174 &  2.81$^{**}$ && 0.173 &  2.84$^{**}$  \\
   42& 0.130 &  1.24$^{~~}$ && 0.173 &  1.65$^{~~}$ && 0.201 &  1.79$^{~~}$ && 0.212 &  1.81$^{~~}$ && 0.208 &  1.88$^{~~}$ && 0.191 &  2.18$^{*~}$ && 0.170 &  2.44$^{*~}$ && 0.168 &  2.67$^{**}$ && 0.165 &  2.85$^{**}$  \\
   48& 0.115 &  1.08$^{~~}$ && 0.157 &  1.48$^{~~}$ && 0.184 &  1.62$^{~~}$ && 0.197 &  1.71$^{~~}$ && 0.196 &  1.83$^{~~}$ && 0.183 &  2.09$^{*~}$ && 0.163 &  2.31$^{*~}$ && 0.159 &  2.66$^{**}$ && 0.157 &  2.88$^{**}$  \\
   \vspace{-3mm}\\
   \multicolumn{27}{l}{\textit{Panel C: Contrarian portfolio}} \\
   1 & 0.102 &  2.54$^{*~}$ && 0.026 &  1.33$^{~~}$ && 0.007 &  0.41$^{~~}$ && 0.004 &  0.18$^{~~}$ && 0.007 &  0.43$^{~~}$ && 0.014 &  1.12$^{~~}$ && 0.016 &  1.48$^{~~}$ && 0.019 &  1.55$^{~~}$ && 0.013 &  0.89$^{~~}$  \\
   6 & 0.100 &  2.02$^{*~}$ && 0.017 &  0.40$^{~~}$ && 0.028 &  0.85$^{~~}$ && 0.014 &  0.41$^{~~}$ && 0.028 &  1.20$^{~~}$ && 0.053 &  2.54$^{*~}$ && 0.048 &  2.31$^{*~}$ && 0.038 &  1.38$^{~~}$ && 0.032 &  1.13$^{~~}$  \\
   12 & 0.123 &  2.02$^{*~}$ && 0.053 &  1.15$^{~~}$ && 0.058 &  1.36$^{~~}$ && 0.052 &  1.36$^{~~}$ && 0.084 &  3.10$^{**}$ && 0.095 &  3.52$^{**}$ && 0.085 &  2.60$^{*~}$ && 0.077 &  2.07$^{*~}$ && 0.080 &  2.11$^{*~}$  \\
   18 & 0.153 &  2.62$^{*~}$ && 0.086 &  1.74$^{~~}$ && 0.085 &  2.05$^{*~}$ && 0.087 &  2.43$^{*~}$ && 0.106 &  3.31$^{**}$ && 0.118 &  3.39$^{**}$ && 0.112 &  2.85$^{**}$ && 0.106 &  2.52$^{*~}$ && 0.110 &  2.98$^{**}$  \\
   24 & 0.165 &  2.39$^{*~}$ && 0.090 &  1.68$^{~~}$ && 0.094 &  2.26$^{*~}$ && 0.101 &  2.43$^{*~}$ && 0.131 &  3.35$^{**}$ && 0.142 &  3.73$^{**}$ && 0.137 &  3.27$^{**}$ && 0.134 &  3.34$^{**}$ && 0.131 &  3.57$^{**}$  \\
   30 & 0.165 &  2.23$^{*~}$ && 0.120 &  2.15$^{*~}$ && 0.110 &  2.15$^{*~}$ && 0.110 &  2.15$^{*~}$ && 0.137 &  3.01$^{**}$ && 0.160 &  4.03$^{**}$ && 0.155 &  3.78$^{**}$ && 0.142 &  3.58$^{**}$ && 0.136 &  3.40$^{**}$  \\
   36 & 0.151 &  2.21$^{*~}$ && 0.122 &  2.08$^{*~}$ && 0.117 &  2.00$^{*~}$ && 0.125 &  2.41$^{*~}$ && 0.150 &  3.39$^{**}$ && 0.167 &  4.39$^{**}$ && 0.160 &  4.05$^{**}$ && 0.156 &  3.79$^{**}$ && 0.162 &  3.80$^{**}$  \\
   42 & 0.206 &  2.72$^{**}$ && 0.137 &  2.13$^{*~}$ && 0.127 &  2.11$^{*~}$ && 0.137 &  2.69$^{**}$ && 0.164 &  3.69$^{**}$ && 0.178 &  4.61$^{**}$ && 0.173 &  4.35$^{**}$ && 0.174 &  4.23$^{**}$ && 0.174 &  4.53$^{**}$  \\
   48 & 0.211 &  2.60$^{*~}$ && 0.149 &  2.23$^{*~}$ && 0.135 &  2.25$^{*~}$ && 0.144 &  2.78$^{**}$ && 0.173 &  4.00$^{**}$ && 0.182 &  4.76$^{**}$ && 0.189 &  4.82$^{**}$ && 0.185 &  4.68$^{**}$ && 0.187 &  5.18$^{**}$  \\
   \hline
   \end{tabular}
   \label{TB:Empirics:SZSE:Decile}
\end{table}
\end{landscape}

\begin{figure}[ht]
  \centering
  \includegraphics[width=0.95\linewidth]{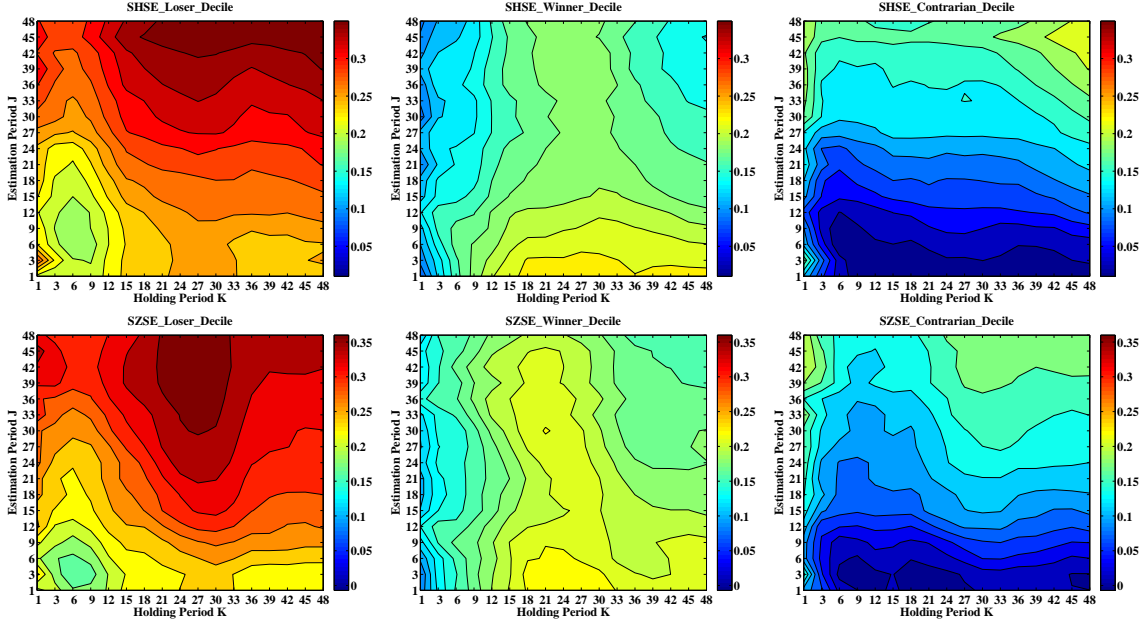}
  \caption{\label{Fig:Empirics:Contour:Decile} {\textbf{Contour plots of the average annualized returns based on decile grouping with varying estimation and holding horizons.}} The left panel is for loser portfolios, the middle panel is for winner portfolios, and the right panel is for contrarian portfolios. The top panel is for the SHSE stocks and the bottom is for the SZSE stocks.}
\end{figure}

Table \ref{TB:Empirics:SHSE:Decile} and Table \ref{TB:Empirics:SZSE:Decile} show that the average annualized returns depend on the values of $J$ and $K$. To better explore this dependence, we illustrate the contours of returns based on decile grouping in Fig.~\ref{Fig:Empirics:Contour:Decile}. We also show the results in Figure~\ref{FigS:Empirics:Contour:Quintile} for quintile grouping and in Figure~\ref{FigS:Empirics:Contour:Tertile} for tertile grouping. The values of $J$ and $K$ are 1 month and multiples of 3 months up to 4 years. Similar results are obtained for quintile grouping and tertile grouping. The top panel is for the SHSE stocks and the bottom panel is for the SZSE stocks. The corresponding contour patterns are roughly similar for the same type of portfolio. However, different types of portfolio exhibit very different contour patterns.

For the loser portfolios, the return $L_{J,K}$ increases roughly with $J$ and $K$. A closer scrutiny unveils more details. There is a valley around $J=9$ and $K=6$, which are the portfolios with the worst performance. The best performance is achieved if one ranks the stocks according to their profits in the past four years and holds the loser portfolio for two years (SZSE) or longer (SHSE). The highest annualized return is 36.2\% for the LOS$(48,36)$ portfolio on the SHSE and 36.9\% for the LOS$(48,24)$ and LOS$(42,30)$ portfolios on the SZSE.

When the holding horizon $K$ is short, say no longer than 1 year, the annualized returns of the winner portfolios are almost independent of the holding horizon $J$. When the holding horizon is longer than 1 year, the winner returns decrease with increasing estimation horizon. When the estimation horizon $J$ is fixed, the winner portfolio return increases first and decreases with the holding horizon $K$. When the estimation horizon is very short (say, $J=1$) and the holding horizon is around 2 years, the winner portfolio return is the highest.

The annualized returns of contrarian portfolios depend more on the estimation horizon $J$ than on the holding horizon $K$. A evident trend is that $C_{J,K}$ increases with $J$ for fixed $K$. For fixed $J$, the return $C_{J,K}$ decreases first and then increases, showing a V-shape pattern. Hence, the best performance is achieved for portfolios formed on long estimation horizons and held for very short (1-3 months) and especially for very long (3-4 years) horizons. Under these conditions, the profits are very significant. For instance, the CON$(48,1)$ portfolio provide an average annualized return of $20.9\%$ for the SHSE stocks and $21.1\%$ for the SZSE stocks.

\subsection*{The relative profits of contrarian portfolios with different ranking groups}

In the literature, three grouping methods are adopted in the formation of portfolios, including decile grouping \cite{Jegadeesh-Titman-1993-JF,Wang-Zhao-2001-cnSMH}, quintile grouping \cite{Kang-Liu-Ni-2002-PBFJ,Pan-Tang-Xu-2013-PBFJ}, and tertile grouping \cite{Asness-Moskowitz-Pedersen-2013-JF}. For the Chinese stock market, the results are qualitatively similar for different grouping methods. We now investigate the relative performances of these grouping methods. For each of the three grouping methods, we form winner, loser and contrarian portfolios based on different combinations of estimation and holding horizons with $J$ and $K$ belonging to $\{1,6,12,18,24,30,36,42,48\}$. For each portfolio, say WIN$(J,K)$, we obtain the return time series $W_{J,K,g}(t)$ for each grouping $g$ and determine the averages of return differences between different $g$'s, $\langle{W_{J,K,g_1}(t)-W_{J,K,g_2}(t)}\rangle_t$. The resulting average return differences and the associated t-statistics of the winner, loser and contrarian portfolios for the two exchanges are presented in Table~\ref{TBS:Empirics:Diff:Group:LOS:SHSE}-Table~\ref{TBS:Empirics:Diff:Group:CON:SZSE}. Table \ref{TB:Empirics:Diff:Group:J=K} shows the results for $J=K$.

Panel A of Table \ref{TB:Empirics:Diff:Group:J=K} reports the average return differences for the loser portfolios with $J=K$ in the two exchanges. Most of the return differences are negative for short horizons and positive for long horizons. For short horizons, some of return differences are significantly negative while other are insignificant. For long horizons, especially when $J=K\geq 30$ for SHSE stocks and $J=K\geq 24$ for SZSE stocks, all return differences are significantly positive. This panel suggests that, if one buys loser portfolios, she should adopt long horizons and use the decile grouping to form her portfolios to obtain high and robust profits. These findings also hold when the estimation horizon and the holding horizon are not fixed identical.

Panel B of Table \ref{TB:Empirics:Diff:Group:J=K} reports the average return differences for the winner portfolios with $J=K$ in the two exchanges. The most significant feature is that all return differences are negative. The negative return difference is more likely to be significant if the horizons are longer. When $J$ is not necessary equal to $K$, the results have only slight differences, as shown in Table~\ref{TBS:Empirics:Diff:Group:WIN:SHSE} and Table~\ref{TBS:Empirics:Diff:Group:WIN:SZSE}. For the comparison of quintile grouping and tertile grouping $(G_5-G_3)$, all 81 combinations of $J$ and $K$ have negative return differences for the SHSE stocks, and only 3 out of the 81 combinations (WIN$(12,1)$, WIN$(30,1)$ and WIN$(42,1)$) have positive return differences for the SZSE stocks which are however not significant. For the comparison of decile grouping and quintile grouping $(G_{10}-G_5)$, only 2 combinations (WIN$(1,42)$ and WIN$(1,48)$) have positive return differences for the SHSE stocks which are insignificant, and there are 14 combinations have positive return differences for the SZSE stocks with one (WIN$(42,6)$) being significant at the 5\% level. For the comparison of decile grouping and tertile grouping $(G_{10}-G_3)$, all 81 combinations have negative return differences for the SHSE stocks, and there are 3 combinations have positive return differences for the SZSE stocks which are not significant at the 5\% level. Therefore, if one adopts the strategy to buy winner portfolios, it is better to use tertile grouping and long estimation and holding horizons.

Panel C of Table \ref{TB:Empirics:Diff:Group:J=K} reports the average return differences for the contrarian portfolios with $J=K$ in the two exchanges. There are positive and negative return differences. All negative return differences are not significant at the 5\% level and the corresponding horizons are not longer than one year. For horizons longer than one year, all the return differences are positive and most are significant at the 1\% level. If we do not fix $J=K$, the results are similar. Therefore, if an investor wants to adopt the contrarian strategy, she should use decile grouping and long-term lagged returns to rank stocks to construct her portfolio and hold it for a long period.

\setlength\tabcolsep{0.5pt}
\begin{landscape}
\begin{table}[!ht]
\caption{
{\bf Relative performance of different grouping methods based on the same strategy of portfolios with $J=K$ for the whole sample period 1997-2012.}}
   {\small{
   \begin{tabular}{ccccccccccccccccccccccccccccccc}
   \hline
    && \multicolumn{2}{c}{1} && \multicolumn{2}{c}{6} && \multicolumn{2}{c}{12} && \multicolumn{2}{c}{18} && \multicolumn{2}{c}{24} && \multicolumn{2}{c}{30} && \multicolumn{2}{c}{36} && \multicolumn{2}{c}{42} && \multicolumn{2}{c}{48} \\
   \cline{3-4} \cline{6-7} \cline{9-10} \cline{12-13} \cline{15-16} \cline{18-19} \cline{21-22} \cline{24-25} \cline{27-28}
       && $\Delta{R}$ & $t$-stat && $\Delta{R}$ & $t$-stat && $\Delta{R}$ & $t$-stat && $\Delta{R}$ & $t$-stat && $\Delta{R}$ & $t$-stat && $\Delta{R}$ & $t$-stat && $\Delta{R}$ & $t$-stat && $\Delta{R}$ & $t$-stat && $\Delta{R}$ & $t$-stat \\
   \hline
   \multicolumn{27}{l}{Panel A: Loser portfolios}\\
   \hline
   \vspace{-3mm}\\
   \multicolumn{27}{l}{\textit{Panel A1: SHSE}} \\
 $G_{5}-G_{3}$&& 0.001 &  0.14$^{~~}$ && -0.009 & -3.14$^{**}$ && -0.005 & -1.96$^{~~}$ && 0.001 &  0.49$^{~~}$ && 0.004 &  1.59$^{~~}$ && 0.006 &  1.99$^{*~}$ && 0.013 &  4.71$^{**}$ && 0.021 &  6.38$^{**}$ && 0.026 &  6.99$^{**}$  \\
 $G_{10}-G_{5}$&& -0.014 & -1.60$^{~~}$ && -0.007 & -1.56$^{~~}$ && -0.016 & -4.29$^{**}$ && -0.010 & -2.01$^{*~}$ && -0.002 & -0.51$^{~~}$ && 0.021 &  5.43$^{**}$ && 0.024 &  5.96$^{**}$ && 0.032 &  5.41$^{**}$ && 0.037 &  5.83$^{**}$  \\
 $G_{10}-G_{3}$&& -0.014 & -1.03$^{~~}$ && -0.016 & -2.59$^{*~}$ && -0.021 & -4.06$^{**}$ && -0.009 & -1.52$^{~~}$ && 0.002 &  0.28$^{~~}$ && 0.027 &  4.97$^{**}$ && 0.037 &  6.42$^{**}$ && 0.053 &  7.66$^{**}$ && 0.063 &  7.03$^{**}$  \\
   \vspace{-3mm}\\
   \multicolumn{27}{l}{\textit{Panel A2: SZSE}} \\
 $G_{5}-G_{3}$&& -0.004 & -0.54$^{~~}$ && -0.009 & -2.08$^{*~}$ && -0.001 & -0.19$^{~~}$ && 0.010 &  2.70$^{**}$ && 0.014 &  2.92$^{**}$ && 0.020 &  4.89$^{**}$ && 0.021 &  5.97$^{**}$ && 0.020 &  6.06$^{**}$ && 0.028 &  7.62$^{**}$  \\
 $G_{10}-G_{5}$&& -0.009 & -0.84$^{~~}$ && -0.014 & -2.34$^{*~}$ && -0.005 & -1.12$^{~~}$ && -0.004 & -0.58$^{~~}$ && 0.026 &  3.79$^{**}$ && 0.035 &  5.00$^{**}$ && 0.030 &  4.28$^{**}$ && 0.047 &  7.10$^{**}$ && 0.035 &  4.81$^{**}$  \\
 $G_{10}-G_{3}$&& -0.013 & -0.85$^{~~}$ && -0.023 & -2.61$^{*~}$ && -0.006 & -0.85$^{~~}$ && 0.006 &  0.74$^{~~}$ && 0.040 &  4.08$^{**}$ && 0.055 &  5.55$^{**}$ && 0.051 &  5.84$^{**}$ && 0.067 &  7.44$^{**}$ && 0.063 &  6.53$^{**}$  \\
   \hline
   \multicolumn{27}{l}{Panel B: Winner portfolios}\\
   \hline
   \vspace{-3mm}\\
   \multicolumn{27}{l}{\textit{Panel B1: SHSE}} \\
 $G_{5}-G_{3}$&& -0.018 & -2.40$^{*~}$ && -0.007 & -2.07$^{*~}$ && -0.014 & -4.43$^{**}$ && -0.020 & -7.57$^{**}$ && -0.024 & -8.80$^{**}$ && -0.016 & -5.68$^{**}$ && -0.009 & -3.16$^{**}$ && -0.009 & -3.26$^{**}$ && -0.012 & -3.46$^{**}$  \\
 $G_{10}-G_{5}$&& -0.022 & -1.79$^{~~}$ && -0.010 & -2.07$^{*~}$ && -0.005 & -1.27$^{~~}$ && -0.015 & -3.98$^{**}$ && -0.009 & -2.31$^{*~}$ && -0.009 & -2.49$^{*~}$ && -0.005 & -1.63$^{~~}$ && -0.017 & -4.51$^{**}$ && -0.025 & -4.75$^{**}$  \\
 $G_{10}-G_{3}$&& -0.040 & -2.45$^{*~}$ && -0.018 & -2.50$^{*~}$ && -0.019 & -3.15$^{**}$ && -0.035 & -6.74$^{**}$ && -0.032 & -6.87$^{**}$ && -0.025 & -4.71$^{**}$ && -0.014 & -2.65$^{**}$ && -0.026 & -6.00$^{**}$ && -0.037 & -8.01$^{**}$  \\
   \vspace{-3mm}\\
   \multicolumn{27}{l}{\textit{Panel B2: SZSE}} \\
 $G_{5}-G_{3}$&& -0.016 & -1.76$^{~~}$ && -0.003 & -0.65$^{~~}$ && -0.009 & -2.66$^{**}$ && -0.017 & -4.74$^{**}$ && -0.013 & -2.72$^{**}$ && -0.013 & -3.25$^{**}$ && -0.008 & -3.08$^{**}$ && -0.009 & -2.86$^{**}$ && -0.011 & -4.41$^{**}$  \\
 $G_{10}-G_{5}$&& -0.003 & -0.25$^{~~}$ && -0.011 & -1.76$^{~~}$ && -0.005 & -1.09$^{~~}$ && -0.008 & -1.80$^{~~}$ && -0.000 & -0.03$^{~~}$ && -0.011 & -2.29$^{*~}$ && -0.015 & -3.74$^{**}$ && -0.013 & -2.77$^{**}$ && -0.018 & -3.73$^{**}$  \\
 $G_{10}-G_{3}$&& -0.019 & -1.00$^{~~}$ && -0.014 & -1.56$^{~~}$ && -0.015 & -2.06$^{*~}$ && -0.025 & -3.86$^{**}$ && -0.013 & -1.47$^{~~}$ && -0.024 & -3.16$^{**}$ && -0.023 & -4.44$^{**}$ && -0.021 & -3.60$^{**}$ && -0.030 & -5.36$^{**}$  \\
   \hline
   \multicolumn{27}{l}{Panel C: Contrarian portfolios}\\
   \hline
   \vspace{-3mm}\\
   \multicolumn{27}{l}{\textit{Panel C1: SHSE}} \\
 $G_{5}-G_{3}$&& 0.019 &  1.95$^{~~}$ && -0.002 & -0.38$^{~~}$ && 0.009 &  2.05$^{*~}$ && 0.021 &  5.41$^{**}$ && 0.028 &  7.36$^{**}$ && 0.022 &  5.19$^{**}$ && 0.022 &  5.32$^{**}$ && 0.030 &  6.39$^{**}$ && 0.038 &  6.51$^{**}$  \\
 $G_{10}-G_{5}$&& 0.007 &  0.51$^{~~}$ && 0.003 &  0.44$^{~~}$ && -0.011 & -1.93$^{~~}$ && 0.005 &  0.67$^{~~}$ && 0.006 &  0.86$^{~~}$ && 0.030 &  5.83$^{**}$ && 0.030 &  5.95$^{**}$ && 0.049 &  7.66$^{**}$ && 0.062 &  9.34$^{**}$  \\
 $G_{10}-G_{3}$&& 0.027 &  1.31$^{~~}$ && 0.002 &  0.15$^{~~}$ && -0.002 & -0.21$^{~~}$ && 0.026 &  3.10$^{**}$ && 0.034 &  4.27$^{**}$ && 0.052 &  7.02$^{**}$ && 0.052 &  6.75$^{**}$ && 0.079 &  9.46$^{**}$ && 0.100 &  9.22$^{**}$  \\
   \vspace{-3mm}\\
   \multicolumn{27}{l}{\textit{Panel C2: SZSE}} \\
 $G_{5}-G_{3}$&& 0.011 &  1.04$^{~~}$ && -0.006 & -0.98$^{~~}$ && 0.009 &  1.56$^{~~}$ && 0.026 &  5.01$^{**}$ && 0.027 &  4.38$^{**}$ && 0.033 &  6.27$^{**}$ && 0.029 &  7.18$^{**}$ && 0.028 &  6.84$^{**}$ && 0.039 &  8.73$^{**}$  \\
 $G_{10}-G_{5}$&& -0.006 & -0.40$^{~~}$ && -0.003 & -0.35$^{~~}$ && 0.000 &  0.04$^{~~}$ && 0.004 &  0.53$^{~~}$ && 0.026 &  2.92$^{**}$ && 0.046 &  6.12$^{**}$ && 0.044 &  5.87$^{**}$ && 0.059 &  8.66$^{**}$ && 0.054 &  7.85$^{**}$  \\
 $G_{10}-G_{3}$&& 0.005 &  0.24$^{~~}$ && -0.009 & -0.72$^{~~}$ && 0.009 &  0.80$^{~~}$ && 0.031 &  2.76$^{**}$ && 0.053 &  4.42$^{**}$ && 0.079 &  7.18$^{**}$ && 0.074 &  7.90$^{**}$ && 0.088 &  9.81$^{**}$ && 0.093 & 10.87$^{**}$  \\
   \hline
   \end{tabular}
   }}
\begin{flushleft} This table reports the differences of the average annualized returns and the corresponding $t$-statistics of two strategies that are different only in the grouping methods. The three panels are for the loser, winner and contrarian portfolios, respectively. In the first row, $G_3$, $G_5$ and $G_{10}$ stand for tertile, quintile and decile groupings. The sample period is January 1997 to  December 2012. The superscripts * and ** denote the significance at 5\% and 1\% levels, respectively.
\end{flushleft}
   \label{TB:Empirics:Diff:Group:J=K}
\end{table}
\end{landscape}

\subsection*{The profits of contrarian portfolios in different exchanges}

As mentioned in {\textit{Materials and Methods}}, the SHSE and the SZSE have different features, such as market value per stock, which may lead to different results in the two exchanges. However, the results resented so far are qualitatively similar for both exchanges. Now we intend to investigate quantitatively the contrarian return differences between the SHSE and the SZSE. We first obtain the time series of returns difference for each contrarian portfolio with the stocks listed on different exchanges, and then calculate their averages. The results are reported in the Table \ref{TB:Empirics:Diff:Market}.

Panel A shows the results for contrarian portfolios based on decile grouping. There are three contrarian portfolios (CON$(12,30)$, CON$(18,30)$ and CON$(24,30)$) having negative return differences that are significant at the 5\% level and four portfolios (CON$(30,48)$, CON$(36,48)$, CON$(42,48)$, CON$(48,48)$) with significant positive return differences (three at the 1\% level and one at the 5\% level). The remaining 74 portfolios do not exhibit significant differences between the two exchanges.

Panel B presents the results for contrarian portfolios based on quintile grouping. There are 17 contrarian portfolios showing significant differences between the two exchanges, in which 4 are negative and 13 are positive. The remaining 64 portfolios do not exhibit significant differences between the two exchanges.

Panel C depicts the results for contrarian portfolios based on tertile grouping. There are 26 contrarian portfolios that have significant differences between the two exchanges in which 3 are negative and 23 are positive, while the other 55 portfolios do not exhibit significant differences. The portfolios with positive return differences between the two exchanges concentrate mainly at the right-bottom corner of the panel, showing that contrarian portfolios with long-term estimation and holding horizons have significantly different average annualized returns on the two exchanges and they are more profitable on the SHSE that have large-cap stocks on average.

\setlength\tabcolsep{0.7pt}
\begin{landscape}
\begin{table}[!ht]
\caption{
{\bf Comparison of the performance of contrarian portfolios in the two exchanges for the whole sample period 1997-2012.}}
   \begin{tabular}{ccccccccccccccccccccccccccc}
   \hline
    & \multicolumn{2}{c}{$K=1$} && \multicolumn{2}{c}{6} && \multicolumn{2}{c}{12} && \multicolumn{2}{c}{18} && \multicolumn{2}{c}{24} && \multicolumn{2}{c}{30} && \multicolumn{2}{c}{36} && \multicolumn{2}{c}{42} && \multicolumn{2}{c}{48} \\
   \cline{2-3} \cline{5-6} \cline{8-9} \cline{11-12} \cline{14-15} \cline{17-18} \cline{20-21} \cline{23-24} \cline{26-27}
    $J$ & Ret & $t$-stat && Ret & $t$-stat && Ret & $t$-stat && Ret & $t$-stat && Ret & $t$-stat && Ret & $t$-stat && Ret & $t$-stat && Ret & $t$-stat && Ret & $t$-stat \\
   \hline
   \vspace{-3mm}\\
   \multicolumn{27}{l}{\textit{Panel A: Decile grouping}} \\
 1& 0.025 &  0.88$^{~~}$ && 0.014 &  0.99$^{~~}$ && 0.008 &  0.69$^{~~}$ && 0.009 &  0.61$^{~~}$ && 0.014 &  0.99$^{~~}$ && 0.009 &  0.87$^{~~}$ && 0.003 &  0.33$^{~~}$ && -0.002 & -0.15$^{~~}$ && 0.006 &  0.57$^{~~}$  \\
 6& 0.023 &  0.67$^{~~}$ && -0.006 & -0.33$^{~~}$ && 0.000 &  0.02$^{~~}$ && 0.012 &  0.73$^{~~}$ && 0.011 &  0.76$^{~~}$ && -0.014 & -1.02$^{~~}$ && -0.018 & -1.51$^{~~}$ && -0.005 & -0.45$^{~~}$ && 0.007 &  0.74$^{~~}$  \\
 12& -0.042 & -1.14$^{~~}$ && -0.024 & -1.68$^{~~}$ && -0.017 & -1.26$^{~~}$ && 0.001 &  0.04$^{~~}$ && -0.020 & -1.56$^{~~}$ && -0.030 & -2.42$^{*~}$ && -0.017 & -1.49$^{~~}$ && -0.004 & -0.43$^{~~}$ && 0.012 &  1.03$^{~~}$  \\
 18& -0.023 & -0.59$^{~~}$ && -0.028 & -1.74$^{~~}$ && -0.007 & -0.52$^{~~}$ && -0.000 & -0.02$^{~~}$ && -0.019 & -1.62$^{~~}$ && -0.029 & -2.34$^{*~}$ && -0.021 & -1.78$^{~~}$ && 0.002 &  0.20$^{~~}$ && 0.011 &  1.14$^{~~}$  \\
 24& -0.030 & -0.67$^{~~}$ && -0.006 & -0.35$^{~~}$ && 0.010 &  0.63$^{~~}$ && 0.014 &  0.77$^{~~}$ && -0.019 & -1.40$^{~~}$ && -0.030 & -2.61$^{*~}$ && -0.019 & -1.90$^{~~}$ && -0.005 & -0.61$^{~~}$ && 0.012 &  1.15$^{~~}$  \\
 30& 0.021 &  0.45$^{~~}$ && 0.013 &  0.68$^{~~}$ && 0.025 &  1.34$^{~~}$ && 0.030 &  1.44$^{~~}$ && -0.002 & -0.14$^{~~}$ && -0.022 & -1.91$^{~~}$ && -0.015 & -1.43$^{~~}$ && 0.009 &  0.97$^{~~}$ && 0.036 &  2.88$^{**}$  \\
 36& 0.049 &  1.01$^{~~}$ && 0.016 &  0.71$^{~~}$ && 0.024 &  1.05$^{~~}$ && 0.021 &  0.97$^{~~}$ && -0.001 & -0.06$^{~~}$ && -0.020 & -1.62$^{~~}$ && -0.005 & -0.44$^{~~}$ && 0.021 &  1.85$^{~~}$ && 0.036 &  2.67$^{**}$  \\
 42& -0.015 & -0.25$^{~~}$ && 0.014 &  0.56$^{~~}$ && 0.027 &  1.11$^{~~}$ && 0.023 &  1.09$^{~~}$ && -0.007 & -0.46$^{~~}$ && -0.017 & -1.51$^{~~}$ && 0.003 &  0.28$^{~~}$ && 0.023 &  1.71$^{~~}$ && 0.040 &  3.19$^{**}$  \\
 48& -0.002 & -0.03$^{~~}$ && 0.025 &  0.88$^{~~}$ && 0.029 &  1.23$^{~~}$ && 0.016 &  0.78$^{~~}$ && -0.003 & -0.23$^{~~}$ && -0.007 & -0.62$^{~~}$ && -0.006 & -0.47$^{~~}$ && 0.014 &  0.95$^{~~}$ && 0.028 &  2.09$^{*~}$  \\
    \vspace{-3mm}\\
   \multicolumn{27}{l}{\textit{Panel B: Quintile grouping}}  \\
 1& 0.012 &  0.53$^{~~}$ && 0.012 &  1.26$^{~~}$ && 0.000 &  0.03$^{~~}$ && 0.002 &  0.25$^{~~}$ && 0.003 &  0.36$^{~~}$ && 0.003 &  0.37$^{~~}$ && 0.001 &  0.06$^{~~}$ && -0.000 & -0.02$^{~~}$ && 0.006 &  0.88$^{~~}$  \\
 6& 0.022 &  0.90$^{~~}$ && -0.012 & -1.12$^{~~}$ && 0.000 &  0.02$^{~~}$ && 0.005 &  0.48$^{~~}$ && 0.013 &  1.29$^{~~}$ && -0.000 & -0.02$^{~~}$ && -0.002 & -0.27$^{~~}$ && 0.005 &  0.56$^{~~}$ && 0.021 &  2.78$^{**}$  \\
 12& -0.018 & -0.75$^{~~}$ && -0.022 & -2.50$^{*~}$ && -0.006 & -0.67$^{~~}$ && 0.003 &  0.26$^{~~}$ && -0.011 & -1.14$^{~~}$ && -0.015 & -2.02$^{*~}$ && -0.014 & -1.97$^{~~}$ && -0.004 & -0.58$^{~~}$ && 0.008 &  1.10$^{~~}$  \\
 18& -0.032 & -1.11$^{~~}$ && -0.028 & -2.88$^{**}$ && -0.009 & -1.11$^{~~}$ && -0.001 & -0.11$^{~~}$ && -0.007 & -0.91$^{~~}$ && -0.017 & -2.33$^{*~}$ && -0.011 & -1.76$^{~~}$ && 0.005 &  0.79$^{~~}$ && 0.015 &  2.18$^{*~}$  \\
 24& -0.025 & -0.89$^{~~}$ && -0.008 & -0.80$^{~~}$ && 0.006 &  0.58$^{~~}$ && 0.019 &  1.56$^{~~}$ && 0.001 &  0.13$^{~~}$ && -0.009 & -1.37$^{~~}$ && 0.000 &  0.02$^{~~}$ && 0.016 &  2.15$^{*~}$ && 0.026 &  3.42$^{**}$  \\
 30& 0.008 &  0.26$^{~~}$ && 0.011 &  1.00$^{~~}$ && 0.025 &  2.03$^{*~}$ && 0.022 &  1.69$^{~~}$ && 0.002 &  0.23$^{~~}$ && -0.006 & -0.80$^{~~}$ && 0.006 &  1.01$^{~~}$ && 0.024 &  3.26$^{**}$ && 0.044 &  5.00$^{**}$  \\
 36& 0.012 &  0.33$^{~~}$ && 0.023 &  1.60$^{~~}$ && 0.028 &  1.93$^{~~}$ && 0.029 &  2.18$^{*~}$ && 0.010 &  0.99$^{~~}$ && 0.001 &  0.10$^{~~}$ && 0.010 &  1.44$^{~~}$ && 0.034 &  3.54$^{**}$ && 0.049 &  4.75$^{**}$  \\
 42& 0.002 &  0.04$^{~~}$ && 0.001 &  0.09$^{~~}$ && 0.022 &  1.35$^{~~}$ && 0.025 &  1.79$^{~~}$ && 0.006 &  0.58$^{~~}$ && 0.005 &  0.73$^{~~}$ && 0.025 &  3.22$^{**}$ && 0.034 &  3.50$^{**}$ && 0.044 &  4.47$^{**}$  \\
 48& -0.011 & -0.27$^{~~}$ && 0.011 &  0.64$^{~~}$ && 0.025 &  1.73$^{~~}$ && 0.015 &  1.17$^{~~}$ && -0.005 & -0.45$^{~~}$ && -0.003 & -0.31$^{~~}$ && 0.008 &  0.86$^{~~}$ && 0.016 &  1.51$^{~~}$ && 0.020 &  1.90$^{~~}$  \\
   \vspace{-3mm}\\
   \multicolumn{27}{l}{\textit{Panel C: Tertile grouping}} \\
 1 & 0.004 &  0.21$^{~~}$ && 0.008 &  1.02$^{~~}$ && 0.001 &  0.15$^{~~}$ && -0.001 & -0.09$^{~~}$ && 0.001 &  0.12$^{~~}$ && 0.001 &  0.15$^{~~}$ && 0.001 &  0.09$^{~~}$ && 0.003 &  0.43$^{~~}$ && 0.008 &  1.45$^{~~}$  \\
 6 & 0.010 &  0.55$^{~~}$ && -0.016 & -2.09$^{*~}$ && -0.006 & -0.94$^{~~}$ && -0.003 & -0.35$^{~~}$ && -0.001 & -0.10$^{~~}$ && -0.001 & -0.22$^{~~}$ && -0.002 & -0.51$^{~~}$ && -0.002 & -0.35$^{~~}$ && 0.015 &  2.95$^{**}$  \\
 12 & -0.021 & -1.03$^{~~}$ && -0.024 & -3.76$^{**}$ && -0.007 & -1.07$^{~~}$ && 0.003 &  0.43$^{~~}$ && 0.001 &  0.28$^{~~}$ && -0.001 & -0.14$^{~~}$ && -0.004 & -0.82$^{~~}$ && 0.004 &  0.80$^{~~}$ && 0.017 &  2.80$^{**}$  \\
 18 & -0.020 & -0.94$^{~~}$ && -0.020 & -2.69$^{**}$ && -0.002 & -0.26$^{~~}$ && 0.004 &  0.75$^{~~}$ && 0.001 &  0.15$^{~~}$ && -0.006 & -1.24$^{~~}$ && -0.002 & -0.54$^{~~}$ && 0.011 &  2.43$^{*~}$ && 0.022 &  3.85$^{**}$  \\
 24 & -0.016 & -0.66$^{~~}$ && -0.007 & -0.93$^{~~}$ && 0.006 &  0.94$^{~~}$ && 0.016 &  2.17$^{*~}$ && 0.001 &  0.12$^{~~}$ && -0.006 & -0.99$^{~~}$ && 0.006 &  1.26$^{~~}$ && 0.018 &  3.22$^{**}$ && 0.028 &  4.77$^{**}$  \\
 30 & -0.008 & -0.31$^{~~}$ && 0.006 &  0.86$^{~~}$ && 0.025 &  3.11$^{**}$ && 0.020 &  2.86$^{**}$ && 0.004 &  0.66$^{~~}$ && 0.005 &  1.00$^{~~}$ && 0.012 &  3.05$^{**}$ && 0.024 &  4.63$^{**}$ && 0.040 &  6.65$^{**}$  \\
 36 & 0.023 &  0.89$^{~~}$ && 0.025 &  2.77$^{**}$ && 0.024 &  2.75$^{**}$ && 0.012 &  1.52$^{~~}$ && 0.006 &  0.78$^{~~}$ && 0.006 &  1.00$^{~~}$ && 0.018 &  2.96$^{**}$ && 0.036 &  4.55$^{**}$ && 0.046 &  6.01$^{**}$  \\
 42 & -0.009 & -0.32$^{~~}$ && -0.002 & -0.21$^{~~}$ && 0.009 &  0.95$^{~~}$ && 0.006 &  0.79$^{~~}$ && -0.002 & -0.26$^{~~}$ && 0.004 &  0.70$^{~~}$ && 0.025 &  3.53$^{**}$ && 0.032 &  4.30$^{**}$ && 0.036 &  5.29$^{**}$  \\
 48 & -0.004 & -0.13$^{~~}$ && 0.005 &  0.44$^{~~}$ && 0.015 &  1.36$^{~~}$ && 0.002 &  0.24$^{~~}$ && -0.001 & -0.06$^{~~}$ && 0.010 &  1.16$^{~~}$ && 0.019 &  2.27$^{*~}$ && 0.018 &  2.32$^{*~}$ && 0.020 &  2.91$^{**}$  \\
   \hline
   \end{tabular}
\begin{flushleft} This table reports the differences of average annualized returns of contrarian portfolios in the SHSE and SZSE and the corresponding $t$-statistics. The return differences are the SHSE returns minus the SZSE returns. The contrarian portfolios are formed based on $J$-month lagged returns and held for $K$ months. 
\end{flushleft}
   \label{TB:Empirics:Diff:Market}
\end{table}
\end{landscape}

\section*{Robustness check}
\label{S1:Robust}

In this section, two approaches are adopted to check the robustness of the findings. First, we perform subperiod analysis through dividing the whole sample period into two subperiods delimited by the onset of the largest and most infamous crash in the history of China's stock market. One subperiod is from January 1997 to September 2007 and the other is from October 2007 to December 2012. Second, the case of skipping one month between the estimation and holding periods is considered to avoid possible measurement errors.

\subsection*{Subperiod analysis}

\begin{figure}[htb]
  \centering
  \includegraphics[width=10cm]{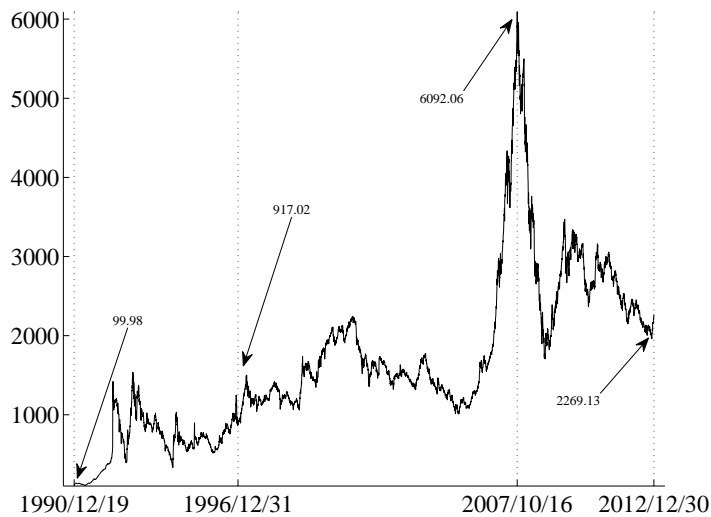}
  \caption{\label{Fig:Figure4} Evolution of the Shanghai Stock Exchange Composite Index from December 1990 to December 2012.}
\end{figure}

Figure~\ref{Fig:Figure4} illustrates the evolution of daily closing prices of the Shanghai Stock Exchange Composite Index. The SHSE composite index was merely $99.98$ on 19 December 1990, which is the first day when SHSE took into operation. By the end of 1996, the index reached 917.02, about ten times of the initial index. The market then experienced bulls and bears and kept climbing to the peak about 6092.06 on 16 October 2007. The historical intraday high was 6124.04 on the same day. Afterwards, the index declined sharply, and entered the long-term adjustment stage. In 2009, the market experienced a bubble \cite{Jiang-Zhou-Sornette-Woodard-Bastiaensen-Cauwels-2010-JEBO}. However, the index is far lower than the historical high. At the end of 2012, the index was $2269.13$.

It is thus necessary to check the robustness of our findings during subperiods. Subperiod analysis is frequently-used in literatures \cite{Jegadeesh-Titman-1993-JF,Chou-Wei-Chung-2007-JEF,Pan-Tang-Xu-2013-PBFJ}. We divide the whole sample period (1997-2012) into two subperiods, January 1997 to September 2007 and October 2007 to December 2012. The former period represents a long-term rising stage of the market, while the latter one represents an adjustment stage. The results are presented in Table \ref{TB:Subperiods:SHSE:J=K} for the SHSE stocks  and Table \ref{TB:Subperiods:SZSE:J=K} for SZSE stocks, in which all portfolios are formed with identical estimation and holding horizons, that is, $J=K$.

For the loser portfolios in the first subperiod (Panel A of both tables), almost all annualized returns are significantly positive at the 5\% level for the SHSE stocks except for two LOS$(6,6)$ portfolios based on decile grouping and quintile grouping, while almost all annualized returns are significantly positive for the SZSE stocks. For the loser portfolios in the second subperiod (Panel B of both tables), the annualized returns are positive but not significant when $J\leq 24$ and significantly positive when $J\geq 30$ for the SHSE stocks, while the annualized returns for the SZSE stocks are significantly positive only when $J\geq 24$. A loser portfolio results in significantly higher profits in the first subperiod that is bullish on average. However, despite the market status, strategies of buying loser portfolios are unlikely to incur losses and are very likely to earn money. We also observe that loser portfolios have better performances if the horizons are longer. For the winner portfolios in the first subperiod, all the annualized returns are positive for both exchanges. Moreover, winner portfolios with $K\geq30$ give significant positive returns at the 5\% level and other portfolios with shorter horizons may have significant and insignificant positive returns. In contrast, although most annualized returns in the second subperiod are positive, no return is significantly positive nor negative. These observations have a simple intuitive explanation. When the market is bullish, loser portfolios will rebound to rise and winner portfolios will continue to rise. When the market is bearish, loser portfolios will reverse with a higher probability, while winner portfolios will bear pressure to continue rising.

\setlength\tabcolsep{1.3pt}
\begin{landscape}
\begin{table}[!ht]
\caption{
{\bf The annualized returns of the loser, winner and contrarian portfolios formed based on $J$-month lagged returns and held for $K$ months with $J=K$ for SHSE stocks in two subperiods.}}
   {\small{
   \begin{tabular}{ccccccccccccccccccccccccccc}
   \hline
    & \multicolumn{2}{c}{1} && \multicolumn{2}{c}{6} && \multicolumn{2}{c}{12} && \multicolumn{2}{c}{18} && \multicolumn{2}{c}{24} && \multicolumn{2}{c}{30} && \multicolumn{2}{c}{36} && \multicolumn{2}{c}{42} && \multicolumn{2}{c}{48} \\
   \cline{2-3} \cline{5-6} \cline{8-9} \cline{11-12} \cline{14-15} \cline{17-18} \cline{20-21} \cline{23-24} \cline{26-27}
    & Ret & $t$-stat && Ret & $t$-stat && Ret & $t$-stat && Ret & $t$-stat && Ret & $t$-stat && Ret & $t$-stat && Ret & $t$-stat && Ret & $t$-stat && Ret & $t$-stat \\
   \hline
   \multicolumn{27}{l}{Panel A: 1997/01-2007/09}\\
   \hline
   \vspace{-3mm}\\
   \multicolumn{27}{l}{{\textit{Panel A1: Decile grouping}}} \\
 LOS& 0.258 &  2.29$^{*~}$ && 0.216 &  1.84$^{~~}$ && 0.267 &  2.09$^{*~}$ && 0.320 &  2.38$^{*~}$ && 0.335 &  2.78$^{**}$ && 0.352 &  3.35$^{**}$ && 0.347 &  4.23$^{**}$ && 0.371 &  4.71$^{**}$ && 0.383 &  4.52$^{**}$  \\
 WIN& 0.191 &  1.67$^{~~}$ && 0.271 &  2.37$^{*~}$ && 0.260 &  2.21$^{*~}$ && 0.233 &  1.82$^{~~}$ && 0.232 &  1.84$^{~~}$ && 0.227 &  2.06$^{*~}$ && 0.203 &  2.37$^{*~}$ && 0.177 &  2.79$^{**}$ && 0.168 &  3.11$^{**}$  \\
 CON& 0.066 &  1.55$^{~~}$ && -0.055 & -1.42$^{~~}$ && 0.007 &  0.17$^{~~}$ && 0.087 &  2.45$^{*~}$ && 0.103 &  3.27$^{**}$ && 0.125 &  3.57$^{**}$ && 0.144 &  3.67$^{**}$ && 0.194 &  4.85$^{**}$ && 0.215 &  4.50$^{**}$  \\
   \vspace{-3mm}\\
   \multicolumn{27}{l}{{\textit{Panel A2: Quintile grouping}}} \\
 LOS& 0.272 &  2.35$^{*~}$ && 0.231 &  1.93$^{~~}$ && 0.283 &  2.21$^{*~}$ && 0.331 &  2.41$^{*~}$ && 0.342 &  2.67$^{**}$ && 0.332 &  3.26$^{**}$ && 0.325 &  4.17$^{**}$ && 0.341 &  4.66$^{**}$ && 0.345 &  4.54$^{**}$  \\
 WIN& 0.213 &  1.79$^{~~}$ && 0.275 &  2.34$^{*~}$ && 0.260 &  2.22$^{*~}$ && 0.249 &  1.96$^{~~}$ && 0.239 &  1.99$^{*~}$ && 0.235 &  2.25$^{*~}$ && 0.208 &  2.58$^{*~}$ && 0.196 &  3.15$^{**}$ && 0.195 &  3.45$^{**}$  \\
 CON& 0.059 &  1.54$^{~~}$ && -0.044 & -1.49$^{~~}$ && 0.023 &  0.70$^{~~}$ && 0.083 &  2.70$^{**}$ && 0.103 &  3.83$^{**}$ && 0.097 &  3.39$^{**}$ && 0.118 &  3.56$^{**}$ && 0.145 &  4.31$^{**}$ && 0.150 &  4.16$^{**}$  \\
   \vspace{-3mm}\\
   \multicolumn{27}{l}{{\textit{Panel A3: Tertile grouping}}} \\
 LOS& 0.275 &  2.37$^{*~}$ && 0.245 &  2.03$^{*~}$ && 0.290 &  2.29$^{*~}$ && 0.328 &  2.43$^{*~}$ && 0.338 &  2.64$^{**}$ && 0.328 &  3.24$^{**}$ && 0.315 &  4.11$^{**}$ && 0.321 &  4.63$^{**}$ && 0.318 &  4.59$^{**}$  \\
 WIN& 0.223 &  1.85$^{~~}$ && 0.276 &  2.31$^{*~}$ && 0.271 &  2.27$^{*~}$ && 0.270 &  2.10$^{*~}$ && 0.264 &  2.15$^{*~}$ && 0.250 &  2.45$^{*~}$ && 0.214 &  2.82$^{**}$ && 0.203 &  3.33$^{**}$ && 0.206 &  3.57$^{**}$  \\
 CON& 0.052 &  1.55$^{~~}$ && -0.031 & -1.38$^{~~}$ && 0.019 &  0.74$^{~~}$ && 0.058 &  2.40$^{*~}$ && 0.074 &  3.34$^{**}$ && 0.078 &  3.43$^{**}$ && 0.101 &  3.95$^{**}$ && 0.117 &  4.67$^{**}$ && 0.113 &  4.57$^{**}$  \\
   \hline
   \multicolumn{27}{l}{Panel B: 2007/10-2012/12}\\
   \hline
   \vspace{-3mm}\\
   \multicolumn{27}{l}{{\textit{Panel B1: Decile grouping}}} \\
 LOS& 0.122 &  0.71$^{~~}$ && 0.126 &  0.75$^{~~}$ && 0.130 &  0.96$^{~~}$ && 0.148 &  1.41$^{~~}$ && 0.205 &  1.84$^{~~}$ && 0.248 &  2.28$^{*~}$ && 0.247 &  3.25$^{**}$ && 0.244 &  5.20$^{**}$ && 0.218 &  5.63$^{**}$  \\
 WIN& -0.131 & -0.78$^{~~}$ && -0.023 & -0.18$^{~~}$ && 0.012 &  0.12$^{~~}$ && 0.059 &  0.72$^{~~}$ && 0.067 &  0.89$^{~~}$ && 0.063 &  1.03$^{~~}$ && 0.044 &  1.13$^{~~}$ && 0.029 &  0.85$^{~~}$ && 0.004 &  0.14$^{~~}$  \\
 CON& 0.253 &  3.92$^{**}$ && 0.149 &  2.23$^{*~}$ && 0.118 &  2.06$^{*~}$ && 0.088 &  1.81$^{~~}$ && 0.138 &  2.44$^{*~}$ && 0.184 &  3.10$^{**}$ && 0.203 &  4.33$^{**}$ && 0.215 &  6.47$^{**}$ && 0.214 & 11.43$^{**}$  \\
   \vspace{-3mm}\\
   \multicolumn{27}{l}{{\textit{Panel B2: Quintile grouping}}} \\
 LOS& 0.137 &  0.78$^{~~}$ && 0.116 &  0.70$^{~~}$ && 0.144 &  1.06$^{~~}$ && 0.155 &  1.50$^{~~}$ && 0.193 &  1.84$^{~~}$ && 0.222 &  2.16$^{*~}$ && 0.215 &  2.94$^{**}$ && 0.203 &  4.51$^{**}$ && 0.188 &  4.50$^{**}$  \\
 WIN& -0.108 & -0.63$^{~~}$ && -0.000 & -0.00$^{~~}$ && 0.028 &  0.27$^{~~}$ && 0.074 &  0.88$^{~~}$ && 0.080 &  1.03$^{~~}$ && 0.075 &  1.15$^{~~}$ && 0.054 &  1.32$^{~~}$ && 0.040 &  1.23$^{~~}$ && 0.018 &  0.53$^{~~}$  \\
 CON& 0.245 &  4.53$^{**}$ && 0.116 &  2.15$^{*~}$ && 0.116 &  2.30$^{*~}$ && 0.082 &  1.94$^{~~}$ && 0.113 &  2.47$^{*~}$ && 0.147 &  3.12$^{**}$ && 0.160 &  4.08$^{**}$ && 0.163 &  6.34$^{**}$ && 0.169 & 12.40$^{**}$  \\
   \vspace{-3mm}\\
   \multicolumn{27}{l}{{\textit{Panel B3: Tertile grouping}}} \\
 LOS& 0.129 &  0.73$^{~~}$ && 0.115 &  0.70$^{~~}$ && 0.144 &  1.08$^{~~}$ && 0.159 &  1.56$^{~~}$ && 0.190 &  1.84$^{~~}$ && 0.208 &  2.08$^{*~}$ && 0.191 &  2.76$^{**}$ && 0.180 &  4.27$^{**}$ && 0.165 &  4.02$^{**}$  \\
 WIN& -0.073 & -0.42$^{~~}$ && 0.021 &  0.15$^{~~}$ && 0.050 &  0.46$^{~~}$ && 0.089 &  1.04$^{~~}$ && 0.099 &  1.23$^{~~}$ && 0.093 &  1.35$^{~~}$ && 0.076 &  1.66$^{~~}$ && 0.058 &  1.64$^{~~}$ && 0.038 &  1.09$^{~~}$  \\
 CON& 0.202 &  4.97$^{**}$ && 0.094 &  2.13$^{*~}$ && 0.094 &  2.43$^{*~}$ && 0.069 &  2.05$^{*~}$ && 0.090 &  2.46$^{*~}$ && 0.115 &  3.08$^{**}$ && 0.115 &  3.84$^{**}$ && 0.122 &  6.31$^{**}$ && 0.126 & 11.71$^{**}$  \\
   \hline
   \end{tabular}
   }}
\begin{flushleft} This table reports the results of strategy portfolios in the SHSE during two subperiods. The values of $J$ and $K$ for different strategies are indicated in the first row. The top panel is for first subperiod, from January 1997 to September 2007, and the bottom is for the second subperiod from October 2007 to December 2012. The superscripts * and ** denote the significance at 5\% and 1\% levels, respectively.
\end{flushleft}
   \label{TB:Subperiods:SHSE:J=K}
\end{table}
\end{landscape}

We now turn to the contrarian portfolios. In the first subperiod, the returns with $J\geq 18$ are significantly positive and that with $J\leq12$ are insignificant for SHSE stocks, and the returns with $J\geq 24$ are significantly positive and those with $J\leq18$ are insignificant for SZSE stocks. In addition, a given portfolio performs better in the SHSE than in the SZSE. We also find that contrarian portfolios based on more groups have higher annualized returns than portfolios based on less groups. For instance, the annualized return of CON$(48,48)$ based on decile grouping is greater than that on tertile grouping by 0.102 on the SHSE and by 0.083 on the SZSE. In the second subperiod, almost all annualized returns are significantly positive, except for the two CON$(18,18)$ portfolios based on decile grouping and quintile grouping on the SHSE and for CON$(6,6)$ based on decile, quintile and tertile groupings on the SZSE. A given portfolio performs better in the SZSE than in the SHSE. Again, portfolios based on more groups result in higher returns. For instance, the annualized return of CON$(48,48)$ based on decile grouping is greater than that on tertile grouping by 0.087 on the SHSE and by 0.157 on the SZSE. Hence, when an investor adopts contrarian strategies, it is better to invest in SHSE stocks during bullish states and in SZSE stocks during bearish periods.

\setlength\tabcolsep{1.3pt}
\begin{landscape}
\begin{table}[!ht]
\caption{
{\bf The annualized returns of the loser, winner and contrarian portfolios formed based on $J$-month lagged returns and held for $K$ months with $J=K$  for SZSE stocks in two subperiods.}}
   {\small{
   \begin{tabular}{ccccccccccccccccccccccccccc}
   \hline
    & \multicolumn{2}{c}{1} && \multicolumn{2}{c}{6} && \multicolumn{2}{c}{12} && \multicolumn{2}{c}{18} && \multicolumn{2}{c}{24} && \multicolumn{2}{c}{30} && \multicolumn{2}{c}{36} && \multicolumn{2}{c}{42} && \multicolumn{2}{c}{48} \\
   \cline{2-3} \cline{5-6} \cline{8-9} \cline{11-12} \cline{14-15} \cline{17-18} \cline{20-21} \cline{23-24} \cline{26-27}
    & Ret & $t$-stat && Ret & $t$-stat && Ret & $t$-stat && Ret & $t$-stat && Ret & $t$-stat && Ret & $t$-stat && Ret & $t$-stat && Ret & $t$-stat && Ret & $t$-stat \\
   \hline
   \multicolumn{27}{l}{Panel A: 1997/01-2007/09}\\
   \hline
   \vspace{-3mm}\\
   \multicolumn{27}{l}{{\textit{Panel A1: Decile grouping}}} \\
 LOS& 0.245 &  2.27$^{*~}$ && 0.201 &  1.78$^{~~}$ && 0.274 &  2.29$^{*~}$ && 0.311 &  2.43$^{*~}$ && 0.356 &  2.45$^{*~}$ && 0.366 &  3.02$^{**}$ && 0.340 &  3.78$^{**}$ && 0.337 &  4.18$^{**}$ && 0.345 &  4.11$^{**}$  \\
 WIN& 0.209 &  1.66$^{~~}$ && 0.246 &  2.10$^{*~}$ && 0.263 &  1.98$^{*~}$ && 0.257 &  1.79$^{~~}$ && 0.271 &  1.80$^{~~}$ && 0.240 &  2.03$^{*~}$ && 0.216 &  2.57$^{*~}$ && 0.201 &  2.73$^{**}$ && 0.179 &  2.93$^{**}$  \\
 CON& 0.035 &  0.70$^{~~}$ && -0.046 & -1.14$^{~~}$ && 0.011 &  0.22$^{~~}$ && 0.054 &  1.28$^{~~}$ && 0.086 &  2.43$^{*~}$ && 0.126 &  3.11$^{**}$ && 0.125 &  2.95$^{**}$ && 0.136 &  3.03$^{**}$ && 0.165 &  4.21$^{**}$  \\
   \vspace{-3mm}\\
   \multicolumn{27}{l}{{\textit{Panel A2: Quintile grouping}}} \\
 LOS& 0.246 &  2.24$^{*~}$ && 0.220 &  1.87$^{~~}$ && 0.280 &  2.27$^{*~}$ && 0.318 &  2.33$^{*~}$ && 0.338 &  2.41$^{*~}$ && 0.339 &  3.00$^{**}$ && 0.321 &  3.81$^{**}$ && 0.300 &  4.26$^{**}$ && 0.317 &  4.20$^{**}$  \\
 WIN& 0.193 &  1.57$^{~~}$ && 0.245 &  2.06$^{*~}$ && 0.258 &  1.98$^{~~}$ && 0.260 &  1.85$^{~~}$ && 0.267 &  1.91$^{~~}$ && 0.249 &  2.21$^{*~}$ && 0.228 &  2.73$^{**}$ && 0.211 &  2.99$^{**}$ && 0.200 &  3.15$^{**}$  \\
 CON& 0.054 &  1.32$^{~~}$ && -0.025 & -0.82$^{~~}$ && 0.022 &  0.60$^{~~}$ && 0.058 &  1.83$^{~~}$ && 0.071 &  2.68$^{**}$ && 0.090 &  2.80$^{**}$ && 0.093 &  2.63$^{**}$ && 0.089 &  2.36$^{*~}$ && 0.117 &  3.66$^{**}$  \\
   \vspace{-3mm}\\
   \multicolumn{27}{l}{{\textit{Panel A3: Tertile grouping}}} \\
 LOS& 0.251 &  2.22$^{*~}$ && 0.231 &  1.90$^{~~}$ && 0.281 &  2.21$^{*~}$ && 0.308 &  2.25$^{*~}$ && 0.328 &  2.42$^{*~}$ && 0.322 &  3.02$^{**}$ && 0.304 &  3.81$^{**}$ && 0.286 &  4.28$^{**}$ && 0.292 &  4.13$^{**}$  \\
 WIN& 0.191 &  1.54$^{~~}$ && 0.237 &  2.00$^{*~}$ && 0.261 &  2.04$^{*~}$ && 0.275 &  2.00$^{*~}$ && 0.275 &  2.05$^{*~}$ && 0.258 &  2.38$^{*~}$ && 0.233 &  2.88$^{**}$ && 0.217 &  3.23$^{**}$ && 0.210 &  3.36$^{**}$  \\
 CON& 0.059 &  1.80$^{~~}$ && -0.006 & -0.27$^{~~}$ && 0.020 &  0.73$^{~~}$ && 0.033 &  1.46$^{~~}$ && 0.053 &  2.31$^{*~}$ && 0.064 &  2.46$^{*~}$ && 0.071 &  2.25$^{*~}$ && 0.070 &  2.16$^{*~}$ && 0.082 &  3.25$^{**}$  \\
   \hline
   \multicolumn{27}{l}{Panel B: 2007/10-2012/12}\\
   \hline
   \vspace{-3mm}\\
   \multicolumn{27}{l}{{\textit{Panel B1: Decile grouping}}} \\
 LOS& 0.089 &  0.51$^{~~}$ && 0.123 &  0.71$^{~~}$ && 0.177 &  1.19$^{~~}$ && 0.224 &  2.05$^{*~}$ && 0.322 &  2.52$^{*~}$ && 0.327 &  2.84$^{**}$ && 0.326 &  3.54$^{**}$ && 0.367 &  8.67$^{**}$ && 0.337 &  9.16$^{**}$  \\
 WIN& -0.151 & -0.90$^{~~}$ && -0.024 & -0.19$^{~~}$ && 0.012 &  0.12$^{~~}$ && 0.050 &  0.64$^{~~}$ && 0.059 &  0.88$^{~~}$ && 0.049 &  0.91$^{~~}$ && 0.022 &  0.60$^{~~}$ && 0.005 &  0.19$^{~~}$ && 0.008 &  0.25$^{~~}$  \\
 CON& 0.240 &  4.68$^{**}$ && 0.146 &  1.80$^{~~}$ && 0.166 &  2.40$^{*~}$ && 0.174 &  3.19$^{**}$ && 0.262 &  2.96$^{**}$ && 0.278 &  3.38$^{**}$ && 0.304 &  4.40$^{**}$ && 0.361 & 11.52$^{**}$ && 0.328 & 11.57$^{**}$  \\
   \vspace{-3mm}\\
   \multicolumn{27}{l}{{\textit{Panel B2: Quintile grouping}}} \\
 LOS& 0.114 &  0.65$^{~~}$ && 0.126 &  0.72$^{~~}$ && 0.180 &  1.22$^{~~}$ && 0.220 &  1.92$^{~~}$ && 0.273 &  2.25$^{*~}$ && 0.265 &  2.52$^{*~}$ && 0.254 &  3.19$^{**}$ && 0.271 &  6.47$^{**}$ && 0.250 &  5.90$^{**}$  \\
 WIN& -0.107 & -0.62$^{~~}$ && 0.012 &  0.09$^{~~}$ && 0.041 &  0.38$^{~~}$ && 0.071 &  0.83$^{~~}$ && 0.072 &  1.00$^{~~}$ && 0.069 &  1.14$^{~~}$ && 0.049 &  1.28$^{~~}$ && 0.027 &  0.92$^{~~}$ && 0.013 &  0.40$^{~~}$  \\
 CON& 0.221 &  5.36$^{**}$ && 0.113 &  1.76$^{~~}$ && 0.139 &  2.56$^{*~}$ && 0.149 &  3.09$^{**}$ && 0.201 &  2.97$^{**}$ && 0.196 &  3.41$^{**}$ && 0.206 &  4.16$^{**}$ && 0.244 & 11.01$^{**}$ && 0.237 & 14.21$^{**}$  \\
   \vspace{-3mm}\\
   \multicolumn{27}{l}{{\textit{Panel B3: Tertile grouping}}} \\
 LOS& 0.117 &  0.65$^{~~}$ && 0.130 &  0.76$^{~~}$ && 0.180 &  1.24$^{~~}$ && 0.210 &  1.82$^{~~}$ && 0.247 &  2.13$^{*~}$ && 0.236 &  2.32$^{*~}$ && 0.218 &  2.98$^{**}$ && 0.219 &  5.35$^{**}$ && 0.203 &  4.73$^{**}$  \\
 WIN& -0.057 & -0.32$^{~~}$ && 0.037 &  0.27$^{~~}$ && 0.066 &  0.58$^{~~}$ && 0.092 &  1.02$^{~~}$ && 0.098 &  1.24$^{~~}$ && 0.094 &  1.44$^{~~}$ && 0.070 &  1.65$^{~~}$ && 0.050 &  1.48$^{~~}$ && 0.032 &  0.87$^{~~}$  \\
 CON& 0.174 &  5.14$^{**}$ && 0.093 &  1.90$^{~~}$ && 0.114 &  2.65$^{*~}$ && 0.118 &  3.16$^{**}$ && 0.149 &  2.99$^{**}$ && 0.141 &  3.20$^{**}$ && 0.148 &  4.00$^{**}$ && 0.169 & 10.51$^{**}$ && 0.171 & 13.69$^{**}$  \\
   \hline
   \end{tabular}
   }}
\begin{flushleft} This table reports the results of strategy portfolios in the SZSE during two subperiods. The values of $J$ and $K$ for different strategies are indicated in the first row. The top panel is for first subperiod, from January 1997 to September 2007, and the bottom is for the second subperiod from October 2007 to December 2012. The superscripts * and ** denote the significance at 5\% and 1\% levels, respectively.
\end{flushleft}
   \label{TB:Subperiods:SZSE:J=K}
\end{table}
\end{landscape}

We also study the relationship between the returns of contrarian portfolios based on varying $J$ and $K$ in the two subperiods. The results are illustrated in Figure~\ref{FigS:Subperiod1:Contour} for the first subperiod and in Figure~\ref{FigS:Subperiod2:Contour} for the second subperiod. The main findings are qualitatively the same as in Tables \ref{TB:Subperiods:SHSE:J=K} and \ref{TB:Subperiods:SZSE:J=K}. The contour plots provide more information. It is evident that the returns of the cases based on three different grouping ways co-move positively with $J$. The dependence of the returns on the holding horizon $K$ differ in different subperiods. Specifically, during the first subperiod, the relation between contrarian returns and $K$ can be depicted as a U-shape, that is, the returns reduce at first and then increase with $K$. However, during the second subperiod, the relationship can be depicted as an L-shape, which suggests that the returns would decrease at first and then become steady with increasing $K$.

\subsection*{Robustness to measurement}

We have observed significant short-term contrarian effect when $J=K=1$ in the full period 1997-2012 and in the bearish subperiod 2007-2012, but not in the bullish subperiod 1997-2007. The short-term contrarian effect might be attributed to some factors such as bid-ask bounce and lagged reaction \cite{Jegadeesh-1990-JF,Lehmann-1990-QJE}. In order to avoid measurement errors caused by these factors, a common approach is to skip some time interval between the estimation period and the holding period \cite{Kang-Liu-Ni-2002-PBFJ,Chou-Wei-Chung-2007-JEF}. We use one month as the skipping interval and perform the analysis on the whole sample period for loser, winner and contrarian portfolios with $J=K$. The results are reported in Table \ref{TB:Empirics:Skip:J=K}, which are compared with Table \ref{TB:Empirics:J=K}.

For the loser portfolios on the SHSE (Panel A of Table \ref{TB:Empirics:Skip:J=K}), most of the returns reduce after the one-month skipping. The most remarkable reduction is happened for the LOS$(6,6)$ portfolios (0.019 for decile grouping, 0.014 for quintile grouping and 0.012 for the tertile grouping). The reduction of returns for LOS$(1,1)$ portfolios based on the quintile and tertile groupings is slightly lower. For other loser portfolios, the return reduction is minor. For the loser portfolios on the SZSE (Panel B of Table \ref{TB:Empirics:Skip:J=K}), we observe slightly different behaviors. After skipping one month, the returns of LOS$(1,1)$ portfolios increase slightly and the returns of LOS$(6,6)$ portfolios decrease mildly. For other loser portfolios, both increase and decrease in the annualized returns are observed; However, the degree of change is minor. Overall, skipping one month does not change the significance of the annualized returns of loser portfolios. For the winner portfolios, most of the annualized returns increase after skipping one month. When $K\geq12$, the differences are ignorable. When $J=1$ and $J=6$, we observe a large increase. However, the returns are still insignificant.

For the contrarian portfolios, the returns with $J\geq12$ after skipping one month change slightly of order 0.001 and are thus not significant. However, the returns of CON$(1,1)$ and CON$(6,6)$ reduce remarkably. For instance, in the case of decile grouping for SHSE stocks, the average annual returns of CON$(1,1)$ and CON$(6,6)$ in Table \ref{TB:Empirics:J=K} are $0.128$ and $0.011$, while in the case of skipping one month, the average annual returns of CON$(1,1)$ and CON$(6,6)$ decrease to $0.108$ and $-0.022$, respectively. Meanwhile, CON$(1,1)$ still has statistically significant return, which indicates that there is no measurement errors. The reduction of contrarian returns is mainly due to the higher profits of winner portfolios after skipping one month. It is clear that the short-term contrarian effect still exists though the average returns of contrarian portfolios reduce. It is trivial that skipping one month will not impact the profitability of trading strategies on the long run so that the long-term contrarian effect also exist.

It is also interesting to investigate the return differences of contrarian portfolios based on the three different grouping ways. In the short run, there is evidence showing that quintile grouping performs better than tertile grouping at the 5\% significance level. However, no significant difference is observed between decile grouping and other two groupings. In the long run, decile grouping outperforms quintile grouping and quintile grouping outperforms tertile grouping. We also study the relationship between returns of the contrarian portfolios with varying $J$ and $K$ in the case of one-month skipping (Figure~\ref{FigS:Skip:Contour}). We find that, the returns of contrarian portfolios increase with the estimation horizon $J$. The patterns with respect to the holding horizon $K$ are more complicated and similar to those for the original strategies without one-month skipping.

\setlength\tabcolsep{1.3pt}
\begin{landscape}
\begin{table}[!ht]
\caption{
{\bf The results for skipping one month between $J$ and $K$ for the whole sample period 1997-2012.}}
   \begin{tabular}{ccccccccccccccccccccccccccc}
   \hline
    & \multicolumn{2}{c}{1} && \multicolumn{2}{c}{6} && \multicolumn{2}{c}{12} && \multicolumn{2}{c}{18} && \multicolumn{2}{c}{24} && \multicolumn{2}{c}{30} && \multicolumn{2}{c}{36} && \multicolumn{2}{c}{42} && \multicolumn{2}{c}{48} \\
   \cline{2-3} \cline{5-6} \cline{8-9} \cline{11-12} \cline{14-15} \cline{17-18} \cline{20-21} \cline{23-24} \cline{26-27}
    & Ret & $t$-stat && Ret & $t$-stat && Ret & $t$-stat && Ret & $t$-stat && Ret & $t$-stat && Ret & $t$-stat && Ret & $t$-stat && Ret & $t$-stat && Ret & $t$-stat \\
   \hline
   \multicolumn{27}{l}{Panel A: SHSE}\\
   \hline
   \vspace{-3mm}\\
   \multicolumn{27}{l}{{\textit{Panel A1: Decile grouping}}} \\
 LOS& 0.214 &  2.15$^{*~}$ && 0.168 &  1.75$^{~~}$ && 0.225 &  2.24$^{*~}$ && 0.269 &  2.64$^{**}$ && 0.303 &  3.18$^{**}$ && 0.325 &  3.83$^{**}$ && 0.328 &  4.82$^{**}$ && 0.347 &  5.22$^{**}$ && 0.357 &  4.86$^{**}$  \\
 WIN& 0.106 &  1.04$^{~~}$ && 0.189 &  2.01$^{*~}$ && 0.184 &  2.02$^{*~}$ && 0.183 &  1.91$^{~~}$ && 0.190 &  1.94$^{~~}$ && 0.190 &  2.17$^{*~}$ && 0.170 &  2.48$^{*~}$ && 0.153 &  2.87$^{**}$ && 0.148 &  3.10$^{**}$  \\
 CON& 0.108 &  3.55$^{**}$ && -0.022 & -0.56$^{~~}$ && 0.041 &  1.22$^{~~}$ && 0.086 &  3.07$^{**}$ && 0.113 &  4.01$^{**}$ && 0.135 &  4.37$^{**}$ && 0.158 &  5.00$^{**}$ && 0.194 &  5.76$^{**}$ && 0.209 &  5.16$^{**}$  \\
   \vspace{-3mm}\\
   \multicolumn{27}{l}{{\textit{Panel A2: Quintile grouping}}} \\
 LOS& 0.222 &  2.24$^{*~}$ && 0.180 &  1.85$^{~~}$ && 0.240 &  2.38$^{*~}$ && 0.280 &  2.69$^{**}$ && 0.304 &  3.03$^{**}$ && 0.305 &  3.72$^{**}$ && 0.301 &  4.67$^{**}$ && 0.316 &  5.11$^{**}$ && 0.321 &  4.85$^{**}$  \\
 WIN& 0.126 &  1.23$^{~~}$ && 0.196 &  2.05$^{*~}$ && 0.190 &  2.09$^{*~}$ && 0.200 &  2.10$^{*~}$ && 0.201 &  2.15$^{*~}$ && 0.199 &  2.40$^{*~}$ && 0.177 &  2.71$^{**}$ && 0.170 &  3.23$^{**}$ && 0.175 &  3.45$^{**}$  \\
 CON& 0.097 &  3.87$^{**}$ && -0.016 & -0.51$^{~~}$ && 0.050 &  1.83$^{~~}$ && 0.080 &  3.22$^{**}$ && 0.104 &  4.41$^{**}$ && 0.106 &  4.24$^{**}$ && 0.125 &  4.67$^{**}$ && 0.146 &  5.28$^{**}$ && 0.146 &  4.65$^{**}$  \\
   \vspace{-3mm}\\
   \multicolumn{27}{l}{{\textit{Panel A3: Tertile grouping}}} \\
 LOS& 0.215 &  2.16$^{*~}$ && 0.191 &  1.96$^{~~}$ && 0.245 &  2.46$^{*~}$ && 0.280 &  2.72$^{**}$ && 0.299 &  3.00$^{**}$ && 0.299 &  3.65$^{**}$ && 0.288 &  4.53$^{**}$ && 0.295 &  5.05$^{**}$ && 0.296 &  4.88$^{**}$  \\
 WIN& 0.145 &  1.44$^{~~}$ && 0.203 &  2.10$^{*~}$ && 0.206 &  2.22$^{*~}$ && 0.220 &  2.26$^{*~}$ && 0.225 &  2.35$^{*~}$ && 0.216 &  2.65$^{**}$ && 0.187 &  3.02$^{**}$ && 0.180 &  3.47$^{**}$ && 0.186 &  3.62$^{**}$  \\
 CON& 0.070 &  3.52$^{**}$ && -0.011 & -0.48$^{~~}$ && 0.039 &  1.79$^{~~}$ && 0.060 &  3.01$^{**}$ && 0.074 &  3.86$^{**}$ && 0.084 &  4.27$^{**}$ && 0.101 &  4.93$^{**}$ && 0.115 &  5.66$^{**}$ && 0.111 &  5.36$^{**}$  \\
   \hline
   \multicolumn{27}{l}{Panel B: SZSE}\\
   \hline
   \vspace{-3mm}\\
   \multicolumn{27}{l}{{\textit{Panel B1: Decile grouping}}} \\
 LOS& 0.221 &  2.09$^{*~}$ && 0.172 &  1.76$^{~~}$ && 0.245 &  2.53$^{*~}$ && 0.287 &  2.86$^{**}$ && 0.346 &  3.03$^{**}$ && 0.351 &  3.62$^{**}$ && 0.335 &  4.51$^{**}$ && 0.338 &  5.03$^{**}$ && 0.340 &  4.69$^{**}$  \\
 WIN& 0.095 &  0.99$^{~~}$ && 0.173 &  1.78$^{~~}$ && 0.186 &  1.83$^{~~}$ && 0.199 &  1.85$^{~~}$ && 0.215 &  1.88$^{~~}$ && 0.194 &  2.12$^{*~}$ && 0.174 &  2.53$^{*~}$ && 0.164 &  2.66$^{**}$ && 0.154 &  2.89$^{**}$  \\
 CON& 0.126 &  3.41$^{**}$ && -0.002 & -0.05$^{~~}$ && 0.059 &  1.45$^{~~}$ && 0.088 &  2.60$^{*~}$ && 0.132 &  3.43$^{**}$ && 0.157 &  4.07$^{**}$ && 0.160 &  4.20$^{**}$ && 0.175 &  4.43$^{**}$ && 0.186 &  5.34$^{**}$  \\
   \vspace{-3mm}\\
   \multicolumn{27}{l}{{\textit{Panel B2: Quintile grouping}}} \\
 LOS& 0.208 &  2.03$^{*~}$ && 0.183 &  1.85$^{~~}$ && 0.248 &  2.51$^{*~}$ && 0.288 &  2.74$^{**}$ && 0.320 &  2.92$^{**}$ && 0.317 &  3.51$^{**}$ && 0.303 &  4.42$^{**}$ && 0.292 &  4.98$^{**}$ && 0.306 &  4.68$^{**}$  \\
 WIN& 0.102 &  1.07$^{~~}$ && 0.179 &  1.85$^{~~}$ && 0.192 &  1.91$^{~~}$ && 0.211 &  1.99$^{*~}$ && 0.217 &  2.04$^{*~}$ && 0.208 &  2.35$^{*~}$ && 0.192 &  2.80$^{**}$ && 0.179 &  2.98$^{**}$ && 0.174 &  3.11$^{**}$  \\
 CON& 0.106 &  3.54$^{**}$ && 0.004 &  0.13$^{~~}$ && 0.056 &  1.80$^{~~}$ && 0.076 &  2.83$^{**}$ && 0.102 &  3.59$^{**}$ && 0.110 &  3.82$^{**}$ && 0.112 &  3.59$^{**}$ && 0.114 &  3.49$^{**}$ && 0.132 &  4.83$^{**}$  \\
   \vspace{-3mm}\\
   \multicolumn{27}{l}{{\textit{Panel B3: Tertile grouping}}} \\
 LOS& 0.209 &  2.04$^{*~}$ && 0.193 &  1.92$^{~~}$ && 0.248 &  2.46$^{*~}$ && 0.277 &  2.64$^{**}$ && 0.304 &  2.90$^{**}$ && 0.299 &  3.49$^{**}$ && 0.284 &  4.33$^{**}$ && 0.273 &  4.87$^{**}$ && 0.278 &  4.51$^{**}$  \\
 WIN& 0.128 &  1.30$^{~~}$ && 0.180 &  1.88$^{~~}$ && 0.203 &  2.06$^{*~}$ && 0.226 &  2.17$^{*~}$ && 0.230 &  2.23$^{*~}$ && 0.222 &  2.58$^{*~}$ && 0.201 &  3.02$^{**}$ && 0.189 &  3.29$^{**}$ && 0.186 &  3.36$^{**}$  \\
 CON& 0.081 &  3.42$^{**}$ && 0.013 &  0.55$^{~~}$ && 0.045 &  1.88$^{~~}$ && 0.051 &  2.53$^{*~}$ && 0.074 &  3.22$^{**}$ && 0.077 &  3.40$^{**}$ && 0.083 &  3.11$^{**}$ && 0.084 &  3.09$^{**}$ && 0.092 &  4.26$^{**}$  \\
   \hline
   \end{tabular}
\begin{flushleft} This table reports the average annualized returns and the corresponding $t$-statistics adjusted for heteroscedasticity and autocorrelation of the loser, winner and contrarian portfolios, which are formed based on $J$-month lagged returns and held for $K$ months with $J=K$. There is a one-month skip between the estimation and holding horizons.
\end{flushleft}
   \label{TB:Empirics:Skip:J=K}
\end{table}
\end{landscape}

\section*{Conclusion}

Following the seminal work of Jegadeesh and Titman \cite{Jegadeesh-Titman-1993-JF}, we investigate the performance of loser portfolios, winner portfolios and contrarian portfolios in the Chinese stock market. The analysis is performed on the monthly returns of all A-share stocks listed on the Shanghai Stock Exchange and the Shenzhen Stock Exchanges. The two samples of SHSE stocks and SZSE stocks cover the period from January 1997 to December 2012.

We find the presence of the contrarian effect in both exchanges across short, intermediate and long horizons. The profits of portfolios depend on the estimation and holding horizons. Especially, longer estimation and holding horizons lead to more profitable contrarian portfolios. When adding one-month time interval between the estimation and holding periods, the results still suggest the existence of significant short-term and long-term contrarian effects even though the profits are decreasing substantially when $J=K=1$. We also conduct subperiod analysis. The long-term contrarian effect is very robust to the subperiods analysis, while the results for short estimation and holding horizons vary with different market states. The short-term contrarian effect is more explicit when the market is in a bearish stage.

Additionally, we study the impact of grouping ways on the performance of portfolios. Specifically, decile, quintile and tertile groupings are adopted. We find that the contrarian portfolios based on decile grouping are more profitable than those based on quintile and tertile groupings. This conclusion is more explicit when the estimation and holding horizons are longer than 12 months. These findings remain valid to the robustness checks based on subperiod analysis and one-month skipping.

\section*{Acknowledgments}

This work was partially supported by the National Natural Science Foundation of China (11075054 and 71131007), Shanghai ``Chen Guang'' Project (2012CG34), Program for Changjiang Scholars and Innovative Research Team in University (IRT1028), and the Fundamental Research Funds for the Central Universities.

\newpage
\setcounter{page}{1}
\setcounter{figure}{0}
\setcounter{table}{0}
\renewcommand\thefigure{S\arabic{figure}}
\renewcommand\thetable{S\arabic{table}}

\begin{figure}[htb]
  \centering
  \includegraphics[width=0.9\linewidth]{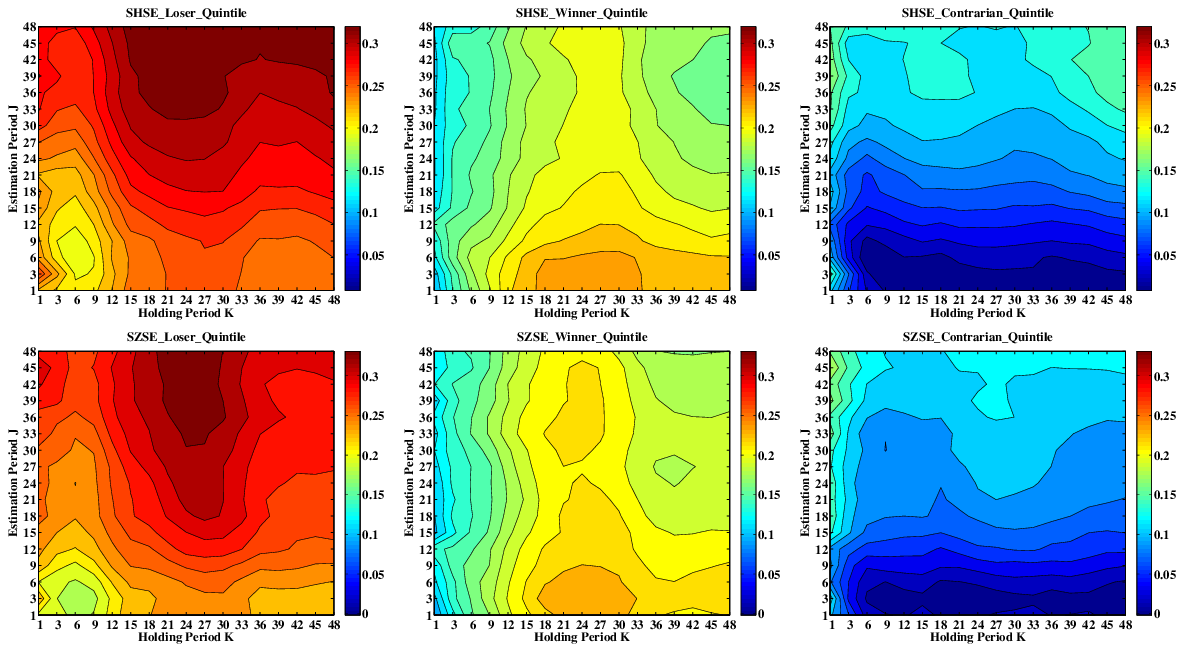}
  \caption{\label{FigS:Empirics:Contour:Quintile} The contour figures, including the top panel for the case of SHSE and the bottom panel for the SZSE, show the results of strategy portfolios based on the quintile grouping. The figures from left to right corresponds to the case of $LOS^{K}_{J}$,$WIN^{K}_{J}$ and $CON^{K}_{J}$, $J,K \in \{J,K|J,K=1,3*N,N=1,2,3...,16\}$. The sample period is from January $1997$ to December $2012$. }
\end{figure}

\begin{figure}[htb]
  \centering
  \includegraphics[width=0.9\linewidth]{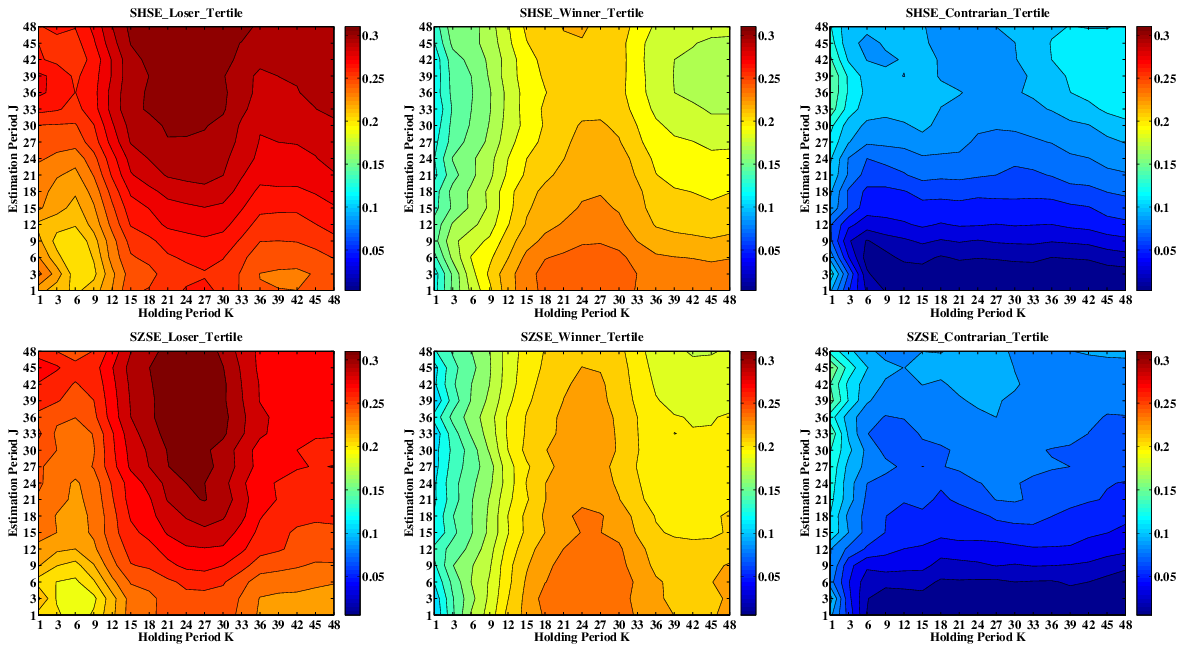}
  \caption{\label{FigS:Empirics:Contour:Tertile} The contour figures, including the top panel for the case of SHSE and the bottom panel for the SZSE, show the results of strategy portfolios based on the tertile grouping. The figures from left to right corresponds to the case of $LOS^{K}_{J}$,$WIN^{K}_{J}$ and $CON^{K}_{J}$, $J,K \in \{J,K|J,K=1,3*N,N=1,2,3...,16\}$. The sample period is from January $1997$ to December $2012$. }
\end{figure}


\begin{figure}[htb]
  \centering
  \includegraphics[width=0.9\linewidth]{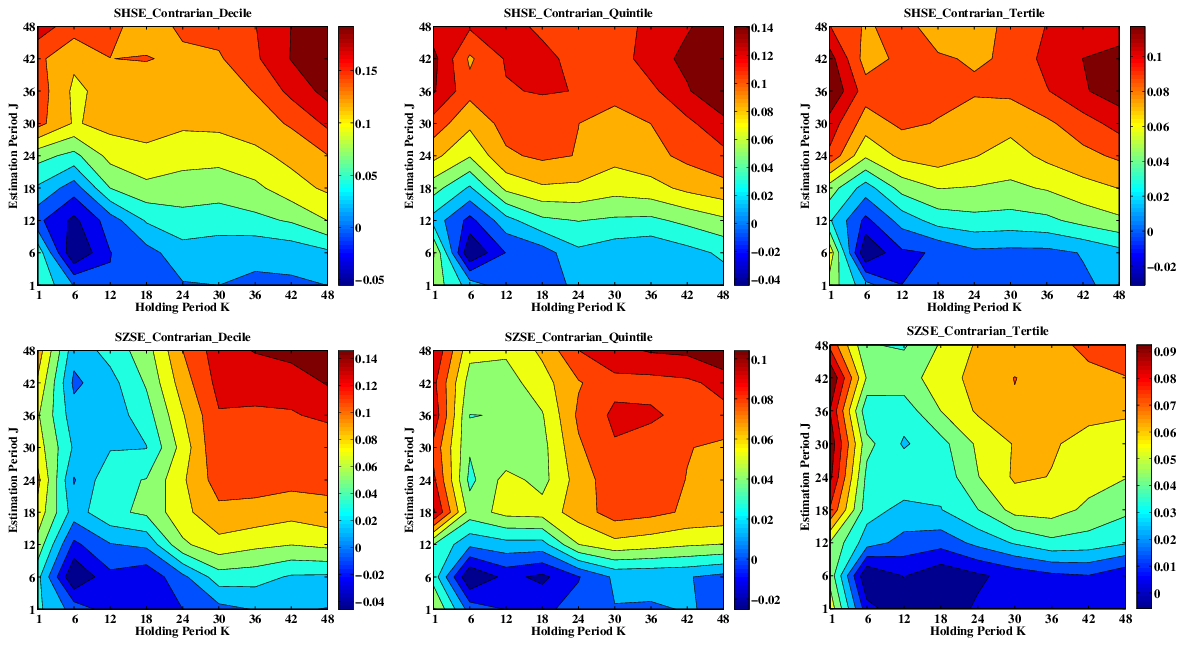}
  \caption{\label{FigS:Subperiod1:Contour} The contour figures, including the top panel for the case of SHSE and the bottom panel for the SZSE, show the results of $CON$s based on the three different grouping ways during the first subperiod. The sample period is from January $1997$ to September $2007$. The figures from left to right corresponds to the cases based on decile grouping, quintile grouping and tertile grouping. $J,K \in \{J,K|J,K=1,3*N,N=1,2,3...,16\}$}
\end{figure}

\begin{figure}[htb]
  \centering
  \includegraphics[width=0.9\linewidth]{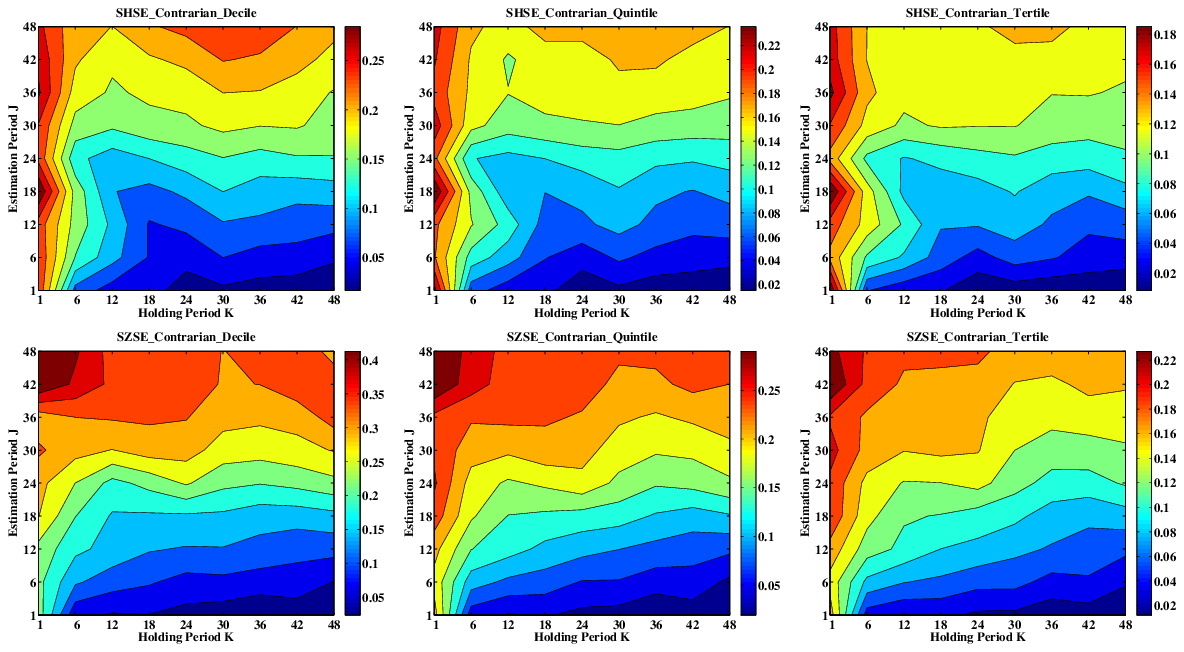}
  \caption{\label{FigS:Subperiod2:Contour} The contour figures, including the top panel for the case of SHSE and the bottom panel for the SZSE, show the results of $CON$s based on the three different grouping ways during the second subperiod. The sample period is from October $2007$ to December $2012$. The figures from left to right corresponds to the cases based on decile grouping, quintile grouping and tertile grouping. $J,K \in \{J,K|J,K=1,3*N,N=1,2,3...,16\}$.}
\end{figure}


\begin{figure}[htb]
\centering
\includegraphics[width=0.9\linewidth]{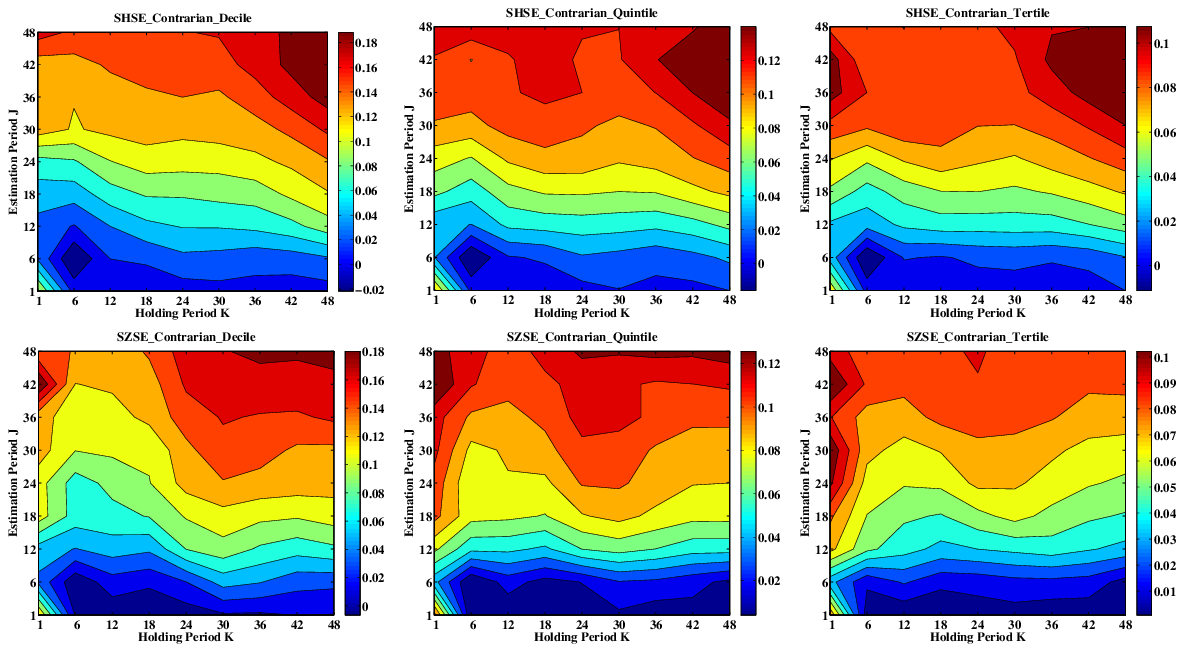}

\caption{\label{FigS:Skip:Contour} The contour figures, including the top panel for the case of SHSE and the bottom panel for the SZSE, show the results of $CON$s based on the three different grouping ways when skipping one month between estimation and holding period. The sample period is from January $1997$ to December $2012$. The figures from left to right corresponds to the cases based on decile grouping, quintile grouping and tertile grouping. $J,K \in \{J,K|J,K=1,3*N,N=1,2,3...,16\}$.}
\end{figure}




\setlength\tabcolsep{2.0pt}
\begin{landscape}
\begin{table}[htb]
\centering
\begin{threeparttable}[b]
   \small
   \caption{The annualized returns of the loser, winner, and contrarian portfolios on the SHSE formed based on $J$-month lagged returns and held for $K$ months by adopting the quintile grouping for the whole sample period 1997-2012.}
   \label{TBS:Empirics:SHSE:Quintile}
   \begin{tabular}{ccccccccccccccccccccccccccc}
   \hline
   \cline{2-27}
    & \multicolumn{2}{c}{$K=1$} && \multicolumn{2}{c}{6} && \multicolumn{2}{c}{12} && \multicolumn{2}{c}{18} && \multicolumn{2}{c}{24} && \multicolumn{2}{c}{30} && \multicolumn{2}{c}{36} && \multicolumn{2}{c}{42} && \multicolumn{2}{c}{48} \\
   \cline{2-3} \cline{5-6} \cline{8-9} \cline{11-12} \cline{14-15} \cline{17-18} \cline{20-21} \cline{23-24} \cline{26-27}
    $J$ & Return & t-stat && Return & t-stat && Return & t-stat && Return & t-stat && Return & t-stat && Return & t-stat && Return & t-stat && Return & t-stat && Return & t-stat \\
   \hline
   \multicolumn{27}{l}{\textit{Panel A: Loser portfolio}} \\
   1& 0.228 &  2.33$^{*~}$ && 0.208 &  2.08$^{*~}$ && 0.231 &  2.35$^{*~}$ && 0.251 &  2.53$^{*~}$ && 0.260 &  2.66$^{**}$ && 0.261 &  3.14$^{**}$ && 0.242 &  3.79$^{**}$ && 0.244 &  4.25$^{**}$ && 0.249 &  4.17$^{**}$  \\
   6& 0.238 &  2.22$^{*~}$ && 0.194 &  1.95$^{~~}$ && 0.232 &  2.33$^{*~}$ && 0.249 &  2.47$^{*~}$ && 0.264 &  2.66$^{**}$ && 0.263 &  3.21$^{**}$ && 0.248 &  3.89$^{**}$ && 0.249 &  4.34$^{**}$ && 0.255 &  4.23$^{**}$  \\
   12& 0.227 &  2.11$^{*~}$ && 0.207 &  2.04$^{*~}$ && 0.241 &  2.40$^{*~}$ && 0.262 &  2.55$^{*~}$ && 0.272 &  2.80$^{**}$ && 0.274 &  3.35$^{**}$ && 0.261 &  4.05$^{**}$ && 0.259 &  4.47$^{**}$ && 0.267 &  4.32$^{**}$  \\
   18& 0.245 &  2.28$^{*~}$ && 0.220 &  2.14$^{*~}$ && 0.258 &  2.50$^{*~}$ && 0.282 &  2.69$^{**}$ && 0.291 &  2.95$^{**}$ && 0.291 &  3.50$^{**}$ && 0.277 &  4.18$^{**}$ && 0.277 &  4.60$^{**}$ && 0.285 &  4.45$^{**}$  \\
   24& 0.245 &  2.28$^{*~}$ && 0.235 &  2.28$^{*~}$ && 0.275 &  2.64$^{**}$ && 0.298 &  2.81$^{**}$ && 0.304 &  3.03$^{**}$ && 0.299 &  3.58$^{**}$ && 0.287 &  4.30$^{**}$ && 0.288 &  4.70$^{**}$ && 0.293 &  4.51$^{**}$  \\
   30& 0.271 &  2.47$^{*~}$ && 0.257 &  2.45$^{*~}$ && 0.289 &  2.77$^{**}$ && 0.308 &  2.89$^{**}$ && 0.313 &  3.11$^{**}$ && 0.307 &  3.71$^{**}$ && 0.293 &  4.49$^{**}$ && 0.295 &  4.85$^{**}$ && 0.305 &  4.57$^{**}$  \\
   36& 0.281 &  2.57$^{*~}$ && 0.273 &  2.56$^{*~}$ && 0.298 &  2.84$^{**}$ && 0.318 &  2.96$^{**}$ && 0.325 &  3.21$^{**}$ && 0.318 &  3.85$^{**}$ && 0.303 &  4.68$^{**}$ && 0.308 &  4.99$^{**}$ && 0.317 &  4.72$^{**}$  \\
   42& 0.285 &  2.58$^{*~}$ && 0.270 &  2.55$^{*~}$ && 0.300 &  2.86$^{**}$ && 0.323 &  3.01$^{**}$ && 0.325 &  3.27$^{**}$ && 0.322 &  3.93$^{**}$ && 0.315 &  4.78$^{**}$ && 0.318 &  5.12$^{**}$ && 0.322 &  4.81$^{**}$  \\
   48& 0.281 &  2.56$^{*~}$ && 0.279 &  2.64$^{**}$ && 0.308 &  2.92$^{**}$ && 0.325 &  3.05$^{**}$ && 0.328 &  3.26$^{**}$ && 0.324 &  3.90$^{**}$ && 0.319 &  4.81$^{**}$ && 0.319 &  5.20$^{**}$ && 0.324 &  4.85$^{**}$  \\
    \vspace{-3mm}\\
   \multicolumn{27}{l}{\textit{Panel B: Winner portfolio}}  \\
   1& 0.107 &  1.07$^{~~}$ && 0.172 &  1.88$^{~~}$ && 0.217 &  2.30$^{*~}$ && 0.236 &  2.39$^{*~}$ && 0.243 &  2.52$^{*~}$ && 0.239 &  2.94$^{**}$ && 0.227 &  3.47$^{**}$ && 0.227 &  3.94$^{**}$ && 0.229 &  3.95$^{**}$  \\
   6& 0.129 &  1.35$^{~~}$ && 0.186 &  1.99$^{*~}$ && 0.209 &  2.25$^{*~}$ && 0.230 &  2.34$^{*~}$ && 0.233 &  2.44$^{*~}$ && 0.235 &  2.77$^{**}$ && 0.226 &  3.21$^{**}$ && 0.222 &  3.63$^{**}$ && 0.220 &  3.77$^{**}$  \\
   12& 0.147 &  1.52$^{~~}$ && 0.168 &  1.86$^{~~}$ && 0.190 &  2.09$^{*~}$ && 0.210 &  2.18$^{*~}$ && 0.216 &  2.29$^{*~}$ && 0.219 &  2.60$^{*~}$ && 0.210 &  3.01$^{**}$ && 0.204 &  3.40$^{**}$ && 0.203 &  3.57$^{**}$  \\
   18& 0.123 &  1.28$^{~~}$ && 0.156 &  1.73$^{~~}$ && 0.181 &  1.99$^{*~}$ && 0.200 &  2.08$^{*~}$ && 0.208 &  2.20$^{*~}$ && 0.211 &  2.53$^{*~}$ && 0.197 &  2.87$^{**}$ && 0.188 &  3.30$^{**}$ && 0.188 &  3.41$^{**}$  \\
   24& 0.119 &  1.24$^{~~}$ && 0.155 &  1.70$^{~~}$ && 0.175 &  1.93$^{~~}$ && 0.191 &  2.04$^{*~}$ && 0.199 &  2.14$^{*~}$ && 0.202 &  2.44$^{*~}$ && 0.187 &  2.79$^{**}$ && 0.180 &  3.22$^{**}$ && 0.177 &  3.30$^{**}$  \\
   30& 0.120 &  1.24$^{~~}$ && 0.146 &  1.62$^{~~}$ && 0.170 &  1.89$^{~~}$ && 0.188 &  1.99$^{*~}$ && 0.198 &  2.09$^{*~}$ && 0.199 &  2.39$^{*~}$ && 0.182 &  2.73$^{**}$ && 0.174 &  3.16$^{**}$ && 0.169 &  3.22$^{**}$  \\
   36& 0.118 &  1.21$^{~~}$ && 0.145 &  1.60$^{~~}$ && 0.169 &  1.86$^{~~}$ && 0.185 &  1.96$^{~~}$ && 0.195 &  2.07$^{*~}$ && 0.196 &  2.35$^{*~}$ && 0.177 &  2.69$^{**}$ && 0.166 &  3.10$^{**}$ && 0.160 &  3.19$^{**}$  \\
   42& 0.116 &  1.19$^{~~}$ && 0.148 &  1.62$^{~~}$ && 0.169 &  1.86$^{~~}$ && 0.185 &  1.97$^{~~}$ && 0.199 &  2.10$^{*~}$ && 0.197 &  2.41$^{*~}$ && 0.174 &  2.76$^{**}$ && 0.169 &  3.19$^{**}$ && 0.164 &  3.27$^{**}$  \\
   48& 0.120 &  1.22$^{~~}$ && 0.139 &  1.53$^{~~}$ && 0.168 &  1.85$^{~~}$ && 0.190 &  2.02$^{*~}$ && 0.198 &  2.15$^{*~}$ && 0.195 &  2.49$^{*~}$ && 0.178 &  2.85$^{**}$ && 0.173 &  3.26$^{**}$ && 0.171 &  3.40$^{**}$  \\
   \vspace{-3mm}\\
   \multicolumn{27}{l}{\textit{Panel C: Contrarian portfolio}} \\
   1 & 0.120 &  3.58$^{**}$ && 0.037 &  2.17$^{*~}$ && 0.014 &  1.19$^{~~}$ && 0.015 &  1.60$^{~~}$ && 0.017 &  2.24$^{*~}$ && 0.022 &  2.43$^{*~}$ && 0.016 &  1.60$^{~~}$ && 0.017 &  1.84$^{~~}$ && 0.020 &  1.80$^{~~}$  \\
   6 & 0.110 &  2.82$^{**}$ && 0.008 &  0.26$^{~~}$ && 0.023 &  0.92$^{~~}$ && 0.018 &  1.01$^{~~}$ && 0.031 &  1.75$^{~~}$ && 0.028 &  1.47$^{~~}$ && 0.022 &  1.00$^{~~}$ && 0.027 &  1.29$^{~~}$ && 0.035 &  1.70$^{~~}$  \\
   12 & 0.080 &  1.95$^{~~}$ && 0.040 &  1.15$^{~~}$ && 0.051 &  1.75$^{~~}$ && 0.052 &  2.32$^{*~}$ && 0.056 &  2.83$^{**}$ && 0.055 &  2.62$^{*~}$ && 0.050 &  2.05$^{*~}$ && 0.055 &  2.20$^{*~}$ && 0.064 &  2.74$^{**}$  \\
   18 & 0.122 &  2.76$^{**}$ && 0.064 &  1.72$^{~~}$ && 0.076 &  2.40$^{*~}$ && 0.082 &  3.24$^{**}$ && 0.083 &  3.53$^{**}$ && 0.080 &  3.30$^{**}$ && 0.080 &  2.96$^{**}$ && 0.088 &  3.30$^{**}$ && 0.096 &  3.70$^{**}$  \\
   24 & 0.126 &  2.82$^{**}$ && 0.080 &  2.29$^{*~}$ && 0.099 &  3.05$^{**}$ && 0.106 &  3.99$^{**}$ && 0.106 &  4.43$^{**}$ && 0.097 &  4.09$^{**}$ && 0.100 &  3.67$^{**}$ && 0.108 &  3.94$^{**}$ && 0.116 &  4.08$^{**}$  \\
   30 & 0.151 &  3.27$^{**}$ && 0.111 &  2.97$^{**}$ && 0.118 &  3.65$^{**}$ && 0.120 &  4.72$^{**}$ && 0.115 &  4.83$^{**}$ && 0.108 &  4.26$^{**}$ && 0.111 &  3.87$^{**}$ && 0.121 &  4.10$^{**}$ && 0.136 &  4.47$^{**}$  \\
   36 & 0.164 &  3.44$^{**}$ && 0.128 &  3.43$^{**}$ && 0.129 &  4.13$^{**}$ && 0.133 &  5.14$^{**}$ && 0.130 &  5.41$^{**}$ && 0.122 &  4.77$^{**}$ && 0.126 &  4.49$^{**}$ && 0.142 &  4.82$^{**}$ && 0.156 &  5.08$^{**}$  \\
   42 & 0.169 &  3.31$^{**}$ && 0.122 &  3.27$^{**}$ && 0.130 &  4.08$^{**}$ && 0.137 &  5.46$^{**}$ && 0.126 &  5.54$^{**}$ && 0.125 &  5.31$^{**}$ && 0.141 &  5.38$^{**}$ && 0.148 &  5.18$^{**}$ && 0.159 &  5.17$^{**}$  \\
   48 & 0.161 &  3.24$^{**}$ && 0.141 &  3.81$^{**}$ && 0.140 &  4.62$^{**}$ && 0.135 &  5.49$^{**}$ && 0.129 &  6.06$^{**}$ && 0.129 &  5.72$^{**}$ && 0.141 &  5.83$^{**}$ && 0.146 &  5.43$^{**}$ && 0.153 &  4.87$^{**}$  \\
   \hline
   \end{tabular}
   \begin{tablenotes}
   \small
     \item This table reports the average annualized returns and the corresponding t-statistics adjusted for heteroscedasticity and autocorrelation of the loser, winner and contrarian portfolios, which are formed by ranking the stocks based on their $J$-month lagged returns, adopting the quintile grouping, and holding for $K$ months. The values of $J$ and $K$ for different strategies are indicated in the first collum and the first row respectively. The sample period is January 1997 to  December 2012. The superscripts * and ** denote the significance at 5\% and 1\% levels, respectively.
   \end{tablenotes}
\end{threeparttable}
\end{table}
\end{landscape}

\setlength\tabcolsep{2.0pt}
\begin{landscape}
\begin{table}[htb]
\centering
\begin{threeparttable}[b]
   \small
   \caption{The annualized returns of the loser, winner, and contrarian portfolios on the SZSE formed based on $J$-month lagged returns and held for $K$ months by adopting the quintile grouping for the whole sample period 1997-2012.}
   \label{TBS:Empirics:SZSE:Quintile}
   \begin{tabular}{ccccccccccccccccccccccccccc}
   \hline
   \cline{2-27}
    & \multicolumn{2}{c}{$K=1$} && \multicolumn{2}{c}{6} && \multicolumn{2}{c}{12} && \multicolumn{2}{c}{18} && \multicolumn{2}{c}{24} && \multicolumn{2}{c}{30} && \multicolumn{2}{c}{36} && \multicolumn{2}{c}{42} && \multicolumn{2}{c}{48} \\
   \cline{2-3} \cline{5-6} \cline{8-9} \cline{11-12} \cline{14-15} \cline{17-18} \cline{20-21} \cline{23-24} \cline{26-27}
    $J$ & Return & t-stat && Return & t-stat && Return & t-stat && Return & t-stat && Return & t-stat && Return & t-stat && Return & t-stat && Return & t-stat && Return & t-stat \\
   \hline
   \multicolumn{27}{l}{\textit{Panel A: Loser portfolio}} \\
   1& 0.203 &  2.13$^{*~}$ && 0.184 &  1.89$^{~~}$ && 0.213 &  2.18$^{*~}$ && 0.237 &  2.43$^{*~}$ && 0.250 &  2.59$^{*~}$ && 0.253 &  3.02$^{**}$ && 0.232 &  3.61$^{**}$ && 0.230 &  4.18$^{**}$ && 0.230 &  4.03$^{**}$  \\
   6& 0.206 &  2.00$^{*~}$ && 0.189 &  1.90$^{~~}$ && 0.224 &  2.27$^{*~}$ && 0.241 &  2.46$^{*~}$ && 0.255 &  2.61$^{*~}$ && 0.259 &  3.07$^{**}$ && 0.241 &  3.66$^{**}$ && 0.242 &  4.21$^{**}$ && 0.237 &  4.00$^{**}$  \\
   12& 0.242 &  2.16$^{*~}$ && 0.221 &  2.15$^{*~}$ && 0.250 &  2.52$^{*~}$ && 0.266 &  2.70$^{**}$ && 0.290 &  2.82$^{**}$ && 0.292 &  3.25$^{**}$ && 0.273 &  3.91$^{**}$ && 0.267 &  4.44$^{**}$ && 0.265 &  4.22$^{**}$  \\
   18& 0.260 &  2.36$^{*~}$ && 0.239 &  2.27$^{*~}$ && 0.266 &  2.59$^{*~}$ && 0.290 &  2.78$^{**}$ && 0.312 &  2.88$^{**}$ && 0.314 &  3.34$^{**}$ && 0.288 &  4.05$^{**}$ && 0.278 &  4.56$^{**}$ && 0.279 &  4.30$^{**}$  \\
   24& 0.259 &  2.36$^{*~}$ && 0.237 &  2.25$^{*~}$ && 0.271 &  2.59$^{*~}$ && 0.295 &  2.78$^{**}$ && 0.321 &  2.93$^{**}$ && 0.315 &  3.41$^{**}$ && 0.290 &  4.13$^{**}$ && 0.282 &  4.61$^{**}$ && 0.282 &  4.34$^{**}$  \\
   30& 0.265 &  2.44$^{*~}$ && 0.251 &  2.38$^{*~}$ && 0.280 &  2.70$^{**}$ && 0.311 &  2.90$^{**}$ && 0.330 &  3.07$^{**}$ && 0.322 &  3.54$^{**}$ && 0.295 &  4.23$^{**}$ && 0.288 &  4.71$^{**}$ && 0.284 &  4.37$^{**}$  \\
   36& 0.277 &  2.52$^{*~}$ && 0.266 &  2.49$^{*~}$ && 0.295 &  2.77$^{**}$ && 0.317 &  2.96$^{**}$ && 0.338 &  3.16$^{**}$ && 0.332 &  3.68$^{**}$ && 0.308 &  4.42$^{**}$ && 0.297 &  4.86$^{**}$ && 0.297 &  4.56$^{**}$  \\
   42& 0.302 &  2.66$^{**}$ && 0.276 &  2.56$^{*~}$ && 0.297 &  2.78$^{**}$ && 0.322 &  2.98$^{**}$ && 0.338 &  3.17$^{**}$ && 0.328 &  3.70$^{**}$ && 0.301 &  4.48$^{**}$ && 0.295 &  5.00$^{**}$ && 0.297 &  4.58$^{**}$  \\
   48& 0.292 &  2.61$^{*~}$ && 0.280 &  2.54$^{*~}$ && 0.295 &  2.75$^{**}$ && 0.324 &  2.97$^{**}$ && 0.346 &  3.25$^{**}$ && 0.330 &  3.84$^{**}$ && 0.308 &  4.69$^{**}$ && 0.304 &  5.11$^{**}$ && 0.308 &  4.68$^{**}$  \\
    \vspace{-3mm}\\
   \multicolumn{27}{l}{\textit{Panel B: Winner portfolio}}  \\
   1& 0.094 &  0.92$^{~~}$ && 0.160 &  1.73$^{~~}$ && 0.199 &  2.11$^{*~}$ && 0.224 &  2.22$^{*~}$ && 0.236 &  2.42$^{*~}$ && 0.234 &  2.87$^{**}$ && 0.217 &  3.47$^{**}$ && 0.213 &  3.90$^{**}$ && 0.216 &  3.80$^{**}$  \\
   6& 0.118 &  1.24$^{~~}$ && 0.170 &  1.79$^{~~}$ && 0.201 &  2.08$^{*~}$ && 0.228 &  2.18$^{*~}$ && 0.236 &  2.37$^{*~}$ && 0.230 &  2.73$^{**}$ && 0.217 &  3.28$^{**}$ && 0.219 &  3.62$^{**}$ && 0.224 &  3.59$^{**}$  \\
   12& 0.144 &  1.48$^{~~}$ && 0.160 &  1.69$^{~~}$ && 0.193 &  1.93$^{~~}$ && 0.217 &  2.07$^{*~}$ && 0.224 &  2.25$^{*~}$ && 0.222 &  2.61$^{*~}$ && 0.209 &  3.06$^{**}$ && 0.208 &  3.36$^{**}$ && 0.209 &  3.42$^{**}$  \\
   18& 0.107 &  1.10$^{~~}$ && 0.147 &  1.55$^{~~}$ && 0.180 &  1.83$^{~~}$ && 0.207 &  1.96$^{~~}$ && 0.222 &  2.13$^{*~}$ && 0.217 &  2.45$^{*~}$ && 0.197 &  2.85$^{**}$ && 0.195 &  3.21$^{**}$ && 0.198 &  3.30$^{**}$  \\
   24& 0.108 &  1.10$^{~~}$ && 0.149 &  1.55$^{~~}$ && 0.177 &  1.83$^{~~}$ && 0.208 &  1.92$^{~~}$ && 0.217 &  2.03$^{*~}$ && 0.209 &  2.35$^{*~}$ && 0.190 &  2.76$^{**}$ && 0.190 &  3.18$^{**}$ && 0.192 &  3.24$^{**}$  \\
   30& 0.121 &  1.21$^{~~}$ && 0.152 &  1.58$^{~~}$ && 0.187 &  1.86$^{~~}$ && 0.213 &  1.93$^{~~}$ && 0.218 &  2.01$^{*~}$ && 0.208 &  2.33$^{*~}$ && 0.191 &  2.78$^{**}$ && 0.190 &  3.15$^{**}$ && 0.193 &  3.18$^{**}$  \\
   36& 0.125 &  1.25$^{~~}$ && 0.161 &  1.65$^{~~}$ && 0.194 &  1.90$^{~~}$ && 0.213 &  1.95$^{~~}$ && 0.218 &  2.02$^{*~}$ && 0.211 &  2.36$^{*~}$ && 0.192 &  2.81$^{**}$ && 0.189 &  3.13$^{**}$ && 0.190 &  3.18$^{**}$  \\
   42& 0.134 &  1.32$^{~~}$ && 0.156 &  1.59$^{~~}$ && 0.189 &  1.83$^{~~}$ && 0.209 &  1.90$^{~~}$ && 0.218 &  2.00$^{*~}$ && 0.208 &  2.30$^{*~}$ && 0.186 &  2.66$^{**}$ && 0.180 &  2.99$^{**}$ && 0.183 &  3.13$^{**}$  \\
   48& 0.121 &  1.16$^{~~}$ && 0.150 &  1.53$^{~~}$ && 0.180 &  1.76$^{~~}$ && 0.204 &  1.85$^{~~}$ && 0.212 &  1.93$^{~~}$ && 0.198 &  2.19$^{*~}$ && 0.175 &  2.52$^{*~}$ && 0.174 &  2.94$^{**}$ && 0.175 &  3.09$^{**}$  \\
   \vspace{-3mm}\\
   \multicolumn{27}{l}{\textit{Panel C: Contrarian portfolio}} \\
   1 & 0.109 &  3.33$^{**}$ && 0.025 &  1.78$^{~~}$ && 0.014 &  1.18$^{~~}$ && 0.013 &  0.84$^{~~}$ && 0.014 &  1.24$^{~~}$ && 0.018 &  2.05$^{*~}$ && 0.015 &  1.77$^{~~}$ && 0.017 &  1.91$^{~~}$ && 0.014 &  1.29$^{~~}$  \\
   6 & 0.088 &  2.36$^{*~}$ && 0.020 &  0.61$^{~~}$ && 0.023 &  0.90$^{~~}$ && 0.013 &  0.51$^{~~}$ && 0.018 &  1.04$^{~~}$ && 0.029 &  1.84$^{~~}$ && 0.024 &  1.39$^{~~}$ && 0.022 &  1.01$^{~~}$ && 0.014 &  0.60$^{~~}$  \\
   12 & 0.098 &  2.07$^{*~}$ && 0.061 &  1.74$^{~~}$ && 0.057 &  1.77$^{~~}$ && 0.049 &  1.81$^{~~}$ && 0.067 &  3.23$^{**}$ && 0.070 &  3.27$^{**}$ && 0.065 &  2.41$^{*~}$ && 0.059 &  1.94$^{~~}$ && 0.056 &  1.87$^{~~}$  \\
   18 & 0.154 &  3.36$^{**}$ && 0.092 &  2.32$^{*~}$ && 0.086 &  2.58$^{*~}$ && 0.083 &  2.97$^{**}$ && 0.090 &  3.59$^{**}$ && 0.097 &  3.38$^{**}$ && 0.090 &  2.84$^{**}$ && 0.083 &  2.52$^{*~}$ && 0.082 &  2.69$^{**}$  \\
   24 & 0.151 &  2.86$^{**}$ && 0.088 &  2.12$^{*~}$ && 0.094 &  2.76$^{**}$ && 0.087 &  2.78$^{**}$ && 0.104 &  3.56$^{**}$ && 0.107 &  3.62$^{**}$ && 0.100 &  3.03$^{**}$ && 0.092 &  2.85$^{**}$ && 0.090 &  2.78$^{**}$  \\
   30 & 0.144 &  2.68$^{**}$ && 0.099 &  2.38$^{*~}$ && 0.093 &  2.54$^{*~}$ && 0.098 &  2.96$^{**}$ && 0.113 &  3.56$^{**}$ && 0.114 &  3.80$^{**}$ && 0.105 &  3.31$^{**}$ && 0.097 &  3.03$^{**}$ && 0.091 &  2.75$^{**}$  \\
   36 & 0.151 &  2.73$^{**}$ && 0.104 &  2.32$^{*~}$ && 0.101 &  2.50$^{*~}$ && 0.104 &  3.03$^{**}$ && 0.120 &  3.75$^{**}$ && 0.121 &  4.02$^{**}$ && 0.116 &  3.66$^{**}$ && 0.108 &  3.17$^{**}$ && 0.107 &  3.14$^{**}$  \\
   42 & 0.168 &  3.02$^{**}$ && 0.121 &  2.73$^{**}$ && 0.109 &  2.55$^{*~}$ && 0.112 &  3.28$^{**}$ && 0.120 &  3.90$^{**}$ && 0.120 &  4.04$^{**}$ && 0.115 &  3.52$^{**}$ && 0.115 &  3.44$^{**}$ && 0.114 &  3.62$^{**}$  \\
   48 & 0.172 &  2.98$^{**}$ && 0.130 &  2.76$^{**}$ && 0.114 &  2.70$^{**}$ && 0.120 &  3.56$^{**}$ && 0.134 &  4.21$^{**}$ && 0.132 &  4.10$^{**}$ && 0.133 &  4.00$^{**}$ && 0.130 &  3.95$^{**}$ && 0.133 &  4.60$^{**}$  \\
   \hline
   \end{tabular}
   \begin{tablenotes}
   \small
     \item This table reports the average annualized returns and the corresponding t-statistics adjusted for heteroscedasticity and autocorrelation of the loser, winner and contrarian portfolios, which are formed by ranking the stocks based on their $J$-month lagged returns, adopting the quintile grouping, and holding for $K$ months. The values of $J$ and $K$ for different strategies are indicated in the first collum and the first row respectively. The sample period is January 1997 to  December 2012. The superscripts * and ** denote the significance at 5\% and 1\% levels, respectively.
   \end{tablenotes}
\end{threeparttable}
\end{table}
\end{landscape}

\setlength\tabcolsep{2.0pt}
\begin{landscape}
\begin{table}[htb]
\centering
\begin{threeparttable}[b]
   \small
   \caption{The annualized returns of the loser, winner, and contrarian portfolios on the SHSE formed based on $J$-month lagged returns and held for $K$ months by adopting the tertile grouping for the whole sample period 1997-2012.}
   \label{TBS:Empirics:SHSE:tertile}
   \begin{tabular}{ccccccccccccccccccccccccccc}
   \hline
   \cline{2-27}
    & \multicolumn{2}{c}{$K=1$} && \multicolumn{2}{c}{6} && \multicolumn{2}{c}{12} && \multicolumn{2}{c}{18} && \multicolumn{2}{c}{24} && \multicolumn{2}{c}{30} && \multicolumn{2}{c}{36} && \multicolumn{2}{c}{42} && \multicolumn{2}{c}{48} \\
   \cline{2-3} \cline{5-6} \cline{8-9} \cline{11-12} \cline{14-15} \cline{17-18} \cline{20-21} \cline{23-24} \cline{26-27}
    $J$ & Return & t-stat && Return & t-stat && Return & t-stat && Return & t-stat && Return & t-stat && Return & t-stat && Return & t-stat && Return & t-stat && Return & t-stat \\
   \hline
   \multicolumn{27}{l}{\textit{Panel A: Loser portfolio}} \\
   1& 0.227 &  2.30$^{*~}$ && 0.210 &  2.11$^{*~}$ && 0.233 &  2.39$^{*~}$ && 0.254 &  2.55$^{*~}$ && 0.264 &  2.67$^{**}$ && 0.261 &  3.16$^{**}$ && 0.243 &  3.81$^{**}$ && 0.243 &  4.28$^{**}$ && 0.248 &  4.20$^{**}$  \\
   6& 0.234 &  2.18$^{*~}$ && 0.203 &  2.04$^{*~}$ && 0.237 &  2.40$^{*~}$ && 0.256 &  2.53$^{*~}$ && 0.264 &  2.70$^{**}$ && 0.264 &  3.25$^{**}$ && 0.247 &  3.95$^{**}$ && 0.247 &  4.40$^{**}$ && 0.253 &  4.29$^{**}$  \\
   12& 0.220 &  2.06$^{*~}$ && 0.214 &  2.11$^{*~}$ && 0.246 &  2.47$^{*~}$ && 0.269 &  2.63$^{**}$ && 0.278 &  2.85$^{**}$ && 0.276 &  3.40$^{**}$ && 0.259 &  4.10$^{**}$ && 0.260 &  4.55$^{**}$ && 0.267 &  4.38$^{**}$  \\
   18& 0.238 &  2.25$^{*~}$ && 0.224 &  2.21$^{*~}$ && 0.257 &  2.55$^{*~}$ && 0.281 &  2.73$^{**}$ && 0.288 &  2.94$^{**}$ && 0.288 &  3.48$^{**}$ && 0.271 &  4.19$^{**}$ && 0.271 &  4.61$^{**}$ && 0.278 &  4.45$^{**}$  \\
   24& 0.245 &  2.28$^{*~}$ && 0.236 &  2.29$^{*~}$ && 0.269 &  2.64$^{**}$ && 0.294 &  2.80$^{**}$ && 0.300 &  3.00$^{**}$ && 0.296 &  3.57$^{**}$ && 0.281 &  4.28$^{**}$ && 0.281 &  4.68$^{**}$ && 0.286 &  4.51$^{**}$  \\
   30& 0.253 &  2.36$^{*~}$ && 0.252 &  2.42$^{*~}$ && 0.282 &  2.75$^{**}$ && 0.301 &  2.88$^{**}$ && 0.305 &  3.07$^{**}$ && 0.301 &  3.65$^{**}$ && 0.284 &  4.42$^{**}$ && 0.285 &  4.84$^{**}$ && 0.291 &  4.64$^{**}$  \\
   36& 0.275 &  2.54$^{*~}$ && 0.263 &  2.52$^{*~}$ && 0.287 &  2.80$^{**}$ && 0.304 &  2.91$^{**}$ && 0.310 &  3.13$^{**}$ && 0.305 &  3.74$^{**}$ && 0.290 &  4.56$^{**}$ && 0.293 &  4.95$^{**}$ && 0.299 &  4.75$^{**}$  \\
   42& 0.264 &  2.45$^{*~}$ && 0.257 &  2.48$^{*~}$ && 0.284 &  2.79$^{**}$ && 0.305 &  2.94$^{**}$ && 0.310 &  3.17$^{**}$ && 0.307 &  3.82$^{**}$ && 0.296 &  4.64$^{**}$ && 0.297 &  5.05$^{**}$ && 0.298 &  4.83$^{**}$  \\
   48& 0.262 &  2.43$^{*~}$ && 0.264 &  2.53$^{*~}$ && 0.293 &  2.84$^{**}$ && 0.307 &  2.98$^{**}$ && 0.313 &  3.21$^{**}$ && 0.312 &  3.83$^{**}$ && 0.300 &  4.63$^{**}$ && 0.296 &  5.10$^{**}$ && 0.298 &  4.88$^{**}$  \\
    \vspace{-3mm}\\
   \multicolumn{27}{l}{\textit{Panel B: Winner portfolio}}  \\
   1& 0.126 &  1.24$^{~~}$ && 0.183 &  1.96$^{~~}$ && 0.224 &  2.37$^{*~}$ && 0.245 &  2.47$^{*~}$ && 0.252 &  2.62$^{*~}$ && 0.248 &  3.05$^{**}$ && 0.233 &  3.60$^{**}$ && 0.231 &  4.08$^{**}$ && 0.232 &  4.08$^{**}$  \\
   6& 0.140 &  1.45$^{~~}$ && 0.193 &  2.03$^{*~}$ && 0.217 &  2.33$^{*~}$ && 0.241 &  2.43$^{*~}$ && 0.247 &  2.55$^{*~}$ && 0.245 &  2.91$^{**}$ && 0.231 &  3.40$^{**}$ && 0.229 &  3.87$^{**}$ && 0.229 &  3.93$^{**}$  \\
   12& 0.148 &  1.52$^{~~}$ && 0.177 &  1.90$^{~~}$ && 0.204 &  2.20$^{*~}$ && 0.226 &  2.31$^{*~}$ && 0.233 &  2.41$^{*~}$ && 0.232 &  2.78$^{**}$ && 0.216 &  3.28$^{**}$ && 0.212 &  3.70$^{**}$ && 0.212 &  3.76$^{**}$  \\
   18& 0.138 &  1.42$^{~~}$ && 0.170 &  1.84$^{~~}$ && 0.201 &  2.17$^{*~}$ && 0.219 &  2.25$^{*~}$ && 0.228 &  2.37$^{*~}$ && 0.227 &  2.76$^{**}$ && 0.210 &  3.19$^{**}$ && 0.204 &  3.64$^{**}$ && 0.204 &  3.67$^{**}$  \\
   24& 0.131 &  1.35$^{~~}$ && 0.166 &  1.79$^{~~}$ && 0.193 &  2.09$^{*~}$ && 0.212 &  2.21$^{*~}$ && 0.222 &  2.34$^{*~}$ && 0.223 &  2.69$^{**}$ && 0.202 &  3.13$^{**}$ && 0.197 &  3.56$^{**}$ && 0.196 &  3.57$^{**}$  \\
   30& 0.123 &  1.27$^{~~}$ && 0.157 &  1.72$^{~~}$ && 0.185 &  2.03$^{*~}$ && 0.206 &  2.16$^{*~}$ && 0.218 &  2.29$^{*~}$ && 0.215 &  2.63$^{**}$ && 0.195 &  3.05$^{**}$ && 0.187 &  3.47$^{**}$ && 0.183 &  3.47$^{**}$  \\
   36& 0.124 &  1.27$^{~~}$ && 0.155 &  1.69$^{~~}$ && 0.184 &  2.01$^{*~}$ && 0.204 &  2.14$^{*~}$ && 0.214 &  2.26$^{*~}$ && 0.208 &  2.59$^{*~}$ && 0.186 &  2.99$^{**}$ && 0.177 &  3.40$^{**}$ && 0.175 &  3.43$^{**}$  \\
   42& 0.123 &  1.26$^{~~}$ && 0.159 &  1.72$^{~~}$ && 0.186 &  2.02$^{*~}$ && 0.208 &  2.16$^{*~}$ && 0.216 &  2.28$^{*~}$ && 0.210 &  2.61$^{*~}$ && 0.185 &  2.99$^{**}$ && 0.179 &  3.43$^{**}$ && 0.178 &  3.46$^{**}$  \\
   48& 0.136 &  1.37$^{~~}$ && 0.163 &  1.75$^{~~}$ && 0.192 &  2.05$^{*~}$ && 0.213 &  2.20$^{*~}$ && 0.218 &  2.33$^{*~}$ && 0.211 &  2.67$^{**}$ && 0.191 &  3.10$^{**}$ && 0.185 &  3.53$^{**}$ && 0.183 &  3.58$^{**}$  \\
   \vspace{-3mm}\\
   \multicolumn{27}{l}{\textit{Panel C: Contrarian portfolio}} \\
   1 & 0.101 &  3.62$^{**}$ && 0.028 &  2.14$^{*~}$ && 0.009 &  0.95$^{~~}$ && 0.009 &  1.29$^{~~}$ && 0.012 &  1.89$^{~~}$ && 0.013 &  1.69$^{~~}$ && 0.010 &  1.33$^{~~}$ && 0.012 &  1.60$^{~~}$ && 0.016 &  1.89$^{~~}$  \\
   6 & 0.094 &  3.07$^{**}$ && 0.010 &  0.41$^{~~}$ && 0.021 &  1.04$^{~~}$ && 0.015 &  1.02$^{~~}$ && 0.017 &  1.25$^{~~}$ && 0.019 &  1.25$^{~~}$ && 0.016 &  0.98$^{~~}$ && 0.018 &  1.12$^{~~}$ && 0.025 &  1.59$^{~~}$  \\
   12 & 0.073 &  2.29$^{*~}$ && 0.037 &  1.36$^{~~}$ && 0.042 &  1.82$^{~~}$ && 0.043 &  2.40$^{*~}$ && 0.045 &  2.82$^{**}$ && 0.045 &  2.60$^{*~}$ && 0.043 &  2.26$^{*~}$ && 0.048 &  2.46$^{*~}$ && 0.054 &  3.05$^{**}$  \\
   18 & 0.101 &  2.90$^{**}$ && 0.054 &  1.83$^{~~}$ && 0.056 &  2.21$^{*~}$ && 0.061 &  3.02$^{**}$ && 0.061 &  3.21$^{**}$ && 0.061 &  3.10$^{**}$ && 0.061 &  2.80$^{**}$ && 0.067 &  3.08$^{**}$ && 0.073 &  3.55$^{**}$  \\
   24 & 0.114 &  3.23$^{**}$ && 0.071 &  2.41$^{*~}$ && 0.077 &  2.93$^{**}$ && 0.082 &  3.83$^{**}$ && 0.078 &  4.01$^{**}$ && 0.073 &  3.67$^{**}$ && 0.079 &  3.58$^{**}$ && 0.084 &  3.70$^{**}$ && 0.091 &  3.88$^{**}$  \\
   30 & 0.129 &  3.61$^{**}$ && 0.095 &  3.17$^{**}$ && 0.097 &  3.69$^{**}$ && 0.095 &  4.56$^{**}$ && 0.087 &  4.60$^{**}$ && 0.086 &  4.28$^{**}$ && 0.089 &  4.11$^{**}$ && 0.097 &  4.26$^{**}$ && 0.108 &  4.51$^{**}$  \\
   36 & 0.151 &  4.18$^{**}$ && 0.108 &  3.61$^{**}$ && 0.103 &  4.03$^{**}$ && 0.100 &  4.96$^{**}$ && 0.096 &  5.38$^{**}$ && 0.097 &  5.14$^{**}$ && 0.104 &  4.82$^{**}$ && 0.116 &  5.05$^{**}$ && 0.124 &  5.09$^{**}$  \\
   42 & 0.141 &  3.77$^{**}$ && 0.097 &  3.34$^{**}$ && 0.099 &  3.90$^{**}$ && 0.097 &  5.06$^{**}$ && 0.093 &  5.32$^{**}$ && 0.097 &  5.26$^{**}$ && 0.112 &  5.35$^{**}$ && 0.118 &  5.54$^{**}$ && 0.120 &  5.32$^{**}$  \\
   48 & 0.127 &  3.36$^{**}$ && 0.101 &  3.34$^{**}$ && 0.102 &  3.92$^{**}$ && 0.094 &  4.77$^{**}$ && 0.096 &  5.39$^{**}$ && 0.101 &  5.43$^{**}$ && 0.110 &  6.02$^{**}$ && 0.111 &  5.84$^{**}$ && 0.114 &  5.35$^{**}$  \\
   \hline
   \end{tabular}
   \begin{tablenotes}
   \small
     \item This table reports the average annualized returns and the corresponding t-statistics adjusted for heteroscedasticity and autocorrelation of the loser, winner and contrarian portfolios, which are formed by ranking the stocks based on their $J$-month lagged returns, adopting the tertile grouping, and holding for $K$ months. The values of $J$ and $K$ for different strategies are indicated in the first collum and the first row respectively. The sample period is January 1997 to  December 2012. The superscripts * and ** denote the significance at 5\% and 1\% levels, respectively.
   \end{tablenotes}
\end{threeparttable}
\end{table}
\end{landscape}

\setlength\tabcolsep{2.0pt}
\begin{landscape}
\begin{table}[htb]
\centering
\begin{threeparttable}[b]
   \small
   \caption{The annualized returns of the loser, winner, and contrarian portfolios on the SZSE formed based on $J$-month lagged returns and held for $K$ months by adopting the tertile grouping for the whole sample period 1997-2012.}
   \label{TBS:Empirics:SZSE:tertile}
   \begin{tabular}{ccccccccccccccccccccccccccc}
   \hline
   \cline{2-27}
    & \multicolumn{2}{c}{$K=1$} && \multicolumn{2}{c}{6} && \multicolumn{2}{c}{12} && \multicolumn{2}{c}{18} && \multicolumn{2}{c}{24} && \multicolumn{2}{c}{30} && \multicolumn{2}{c}{36} && \multicolumn{2}{c}{42} && \multicolumn{2}{c}{48} \\
   \cline{2-3} \cline{5-6} \cline{8-9} \cline{11-12} \cline{14-15} \cline{17-18} \cline{20-21} \cline{23-24} \cline{26-27}
    $J$ & Return & t-stat && Return & t-stat && Return & t-stat && Return & t-stat && Return & t-stat && Return & t-stat && Return & t-stat && Return & t-stat && Return & t-stat \\
   \hline
   \multicolumn{27}{l}{\textit{Panel A: Loser portfolio}} \\
   1& 0.207 &  2.12$^{*~}$ && 0.193 &  1.95$^{~~}$ && 0.222 &  2.26$^{*~}$ && 0.246 &  2.47$^{*~}$ && 0.258 &  2.59$^{*~}$ && 0.255 &  3.04$^{**}$ && 0.234 &  3.70$^{**}$ && 0.230 &  4.24$^{**}$ && 0.232 &  4.06$^{**}$  \\
   6& 0.207 &  1.98$^{*~}$ && 0.198 &  1.96$^{~~}$ && 0.233 &  2.33$^{*~}$ && 0.252 &  2.50$^{*~}$ && 0.262 &  2.65$^{**}$ && 0.261 &  3.14$^{**}$ && 0.242 &  3.78$^{**}$ && 0.240 &  4.33$^{**}$ && 0.236 &  4.10$^{**}$  \\
   12& 0.232 &  2.08$^{*~}$ && 0.224 &  2.13$^{*~}$ && 0.250 &  2.48$^{*~}$ && 0.268 &  2.63$^{**}$ && 0.282 &  2.78$^{**}$ && 0.282 &  3.23$^{**}$ && 0.264 &  3.94$^{**}$ && 0.257 &  4.47$^{**}$ && 0.254 &  4.24$^{**}$  \\
   18& 0.245 &  2.24$^{*~}$ && 0.233 &  2.23$^{*~}$ && 0.256 &  2.52$^{*~}$ && 0.281 &  2.67$^{**}$ && 0.296 &  2.82$^{**}$ && 0.296 &  3.31$^{**}$ && 0.272 &  4.06$^{**}$ && 0.263 &  4.54$^{**}$ && 0.262 &  4.29$^{**}$  \\
   24& 0.250 &  2.29$^{*~}$ && 0.234 &  2.23$^{*~}$ && 0.261 &  2.53$^{*~}$ && 0.287 &  2.70$^{**}$ && 0.307 &  2.90$^{**}$ && 0.299 &  3.43$^{**}$ && 0.274 &  4.17$^{**}$ && 0.268 &  4.61$^{**}$ && 0.269 &  4.37$^{**}$  \\
   30& 0.254 &  2.32$^{*~}$ && 0.241 &  2.29$^{*~}$ && 0.269 &  2.60$^{*~}$ && 0.297 &  2.75$^{**}$ && 0.312 &  2.97$^{**}$ && 0.302 &  3.51$^{**}$ && 0.279 &  4.23$^{**}$ && 0.272 &  4.69$^{**}$ && 0.270 &  4.41$^{**}$  \\
   36& 0.255 &  2.35$^{*~}$ && 0.248 &  2.33$^{*~}$ && 0.277 &  2.63$^{**}$ && 0.305 &  2.82$^{**}$ && 0.317 &  3.05$^{**}$ && 0.311 &  3.59$^{**}$ && 0.286 &  4.34$^{**}$ && 0.276 &  4.78$^{**}$ && 0.276 &  4.51$^{**}$  \\
   42& 0.270 &  2.43$^{*~}$ && 0.259 &  2.41$^{*~}$ && 0.280 &  2.67$^{**}$ && 0.305 &  2.85$^{**}$ && 0.319 &  3.08$^{**}$ && 0.309 &  3.64$^{**}$ && 0.282 &  4.41$^{**}$ && 0.275 &  4.89$^{**}$ && 0.276 &  4.51$^{**}$  \\
   48& 0.264 &  2.38$^{*~}$ && 0.255 &  2.34$^{*~}$ && 0.277 &  2.61$^{*~}$ && 0.302 &  2.83$^{**}$ && 0.316 &  3.11$^{**}$ && 0.304 &  3.68$^{**}$ && 0.281 &  4.49$^{**}$ && 0.277 &  4.92$^{**}$ && 0.280 &  4.54$^{**}$  \\
    \vspace{-3mm}\\
   \multicolumn{27}{l}{\textit{Panel B: Winner portfolio}}  \\
   1& 0.110 &  1.06$^{~~}$ && 0.173 &  1.82$^{~~}$ && 0.214 &  2.22$^{*~}$ && 0.236 &  2.33$^{*~}$ && 0.247 &  2.52$^{*~}$ && 0.242 &  2.96$^{**}$ && 0.224 &  3.60$^{**}$ && 0.221 &  4.03$^{**}$ && 0.224 &  3.92$^{**}$  \\
   6& 0.123 &  1.27$^{~~}$ && 0.172 &  1.82$^{~~}$ && 0.206 &  2.16$^{*~}$ && 0.234 &  2.30$^{*~}$ && 0.245 &  2.47$^{*~}$ && 0.241 &  2.83$^{**}$ && 0.224 &  3.42$^{**}$ && 0.220 &  3.86$^{**}$ && 0.226 &  3.79$^{**}$  \\
   12& 0.139 &  1.41$^{~~}$ && 0.163 &  1.71$^{~~}$ && 0.202 &  2.05$^{*~}$ && 0.228 &  2.21$^{*~}$ && 0.239 &  2.36$^{*~}$ && 0.236 &  2.77$^{**}$ && 0.217 &  3.31$^{**}$ && 0.213 &  3.68$^{**}$ && 0.216 &  3.66$^{**}$  \\
   18& 0.124 &  1.26$^{~~}$ && 0.159 &  1.65$^{~~}$ && 0.197 &  2.01$^{*~}$ && 0.224 &  2.15$^{*~}$ && 0.236 &  2.31$^{*~}$ && 0.229 &  2.70$^{**}$ && 0.209 &  3.20$^{**}$ && 0.207 &  3.62$^{**}$ && 0.211 &  3.59$^{**}$  \\
   24& 0.120 &  1.23$^{~~}$ && 0.156 &  1.62$^{~~}$ && 0.191 &  1.96$^{~~}$ && 0.222 &  2.11$^{*~}$ && 0.230 &  2.23$^{*~}$ && 0.220 &  2.58$^{*~}$ && 0.202 &  3.09$^{**}$ && 0.202 &  3.53$^{**}$ && 0.206 &  3.52$^{**}$  \\
   30& 0.117 &  1.19$^{~~}$ && 0.153 &  1.60$^{~~}$ && 0.197 &  1.99$^{*~}$ && 0.221 &  2.11$^{*~}$ && 0.229 &  2.21$^{*~}$ && 0.221 &  2.57$^{*~}$ && 0.202 &  3.08$^{**}$ && 0.199 &  3.49$^{**}$ && 0.202 &  3.51$^{**}$  \\
   36& 0.127 &  1.27$^{~~}$ && 0.166 &  1.69$^{~~}$ && 0.198 &  2.00$^{*~}$ && 0.217 &  2.06$^{*~}$ && 0.227 &  2.19$^{*~}$ && 0.220 &  2.54$^{*~}$ && 0.200 &  3.02$^{**}$ && 0.197 &  3.41$^{**}$ && 0.198 &  3.46$^{**}$  \\
   42& 0.120 &  1.20$^{~~}$ && 0.159 &  1.62$^{~~}$ && 0.191 &  1.91$^{~~}$ && 0.214 &  2.02$^{*~}$ && 0.224 &  2.14$^{*~}$ && 0.216 &  2.47$^{*~}$ && 0.196 &  2.89$^{**}$ && 0.189 &  3.29$^{**}$ && 0.192 &  3.38$^{**}$  \\
   48& 0.133 &  1.31$^{~~}$ && 0.159 &  1.62$^{~~}$ && 0.190 &  1.89$^{~~}$ && 0.210 &  1.99$^{*~}$ && 0.220 &  2.08$^{*~}$ && 0.213 &  2.39$^{*~}$ && 0.190 &  2.78$^{**}$ && 0.184 &  3.22$^{**}$ && 0.186 &  3.34$^{**}$  \\
   \vspace{-3mm}\\
   \multicolumn{27}{l}{\textit{Panel C: Contrarian portfolio}} \\
   1 & 0.097 &  3.71$^{**}$ && 0.020 &  1.75$^{~~}$ && 0.008 &  0.92$^{~~}$ && 0.010 &  0.87$^{~~}$ && 0.012 &  1.36$^{~~}$ && 0.012 &  1.72$^{~~}$ && 0.010 &  1.51$^{~~}$ && 0.009 &  1.38$^{~~}$ && 0.008 &  0.91$^{~~}$  \\
   6 & 0.084 &  2.89$^{**}$ && 0.026 &  1.09$^{~~}$ && 0.027 &  1.38$^{~~}$ && 0.018 &  0.90$^{~~}$ && 0.018 &  1.19$^{~~}$ && 0.020 &  1.40$^{~~}$ && 0.019 &  1.25$^{~~}$ && 0.020 &  1.18$^{~~}$ && 0.011 &  0.57$^{~~}$  \\
   12 & 0.093 &  2.49$^{*~}$ && 0.061 &  2.16$^{*~}$ && 0.049 &  1.94$^{~~}$ && 0.040 &  2.05$^{*~}$ && 0.043 &  2.79$^{**}$ && 0.045 &  2.78$^{**}$ && 0.047 &  2.32$^{*~}$ && 0.044 &  1.92$^{~~}$ && 0.038 &  1.63$^{~~}$  \\
   18 & 0.121 &  3.32$^{**}$ && 0.074 &  2.37$^{*~}$ && 0.058 &  2.22$^{*~}$ && 0.057 &  2.72$^{**}$ && 0.060 &  3.18$^{**}$ && 0.067 &  3.08$^{**}$ && 0.064 &  2.62$^{*~}$ && 0.056 &  2.27$^{*~}$ && 0.051 &  2.11$^{*~}$  \\
   24 & 0.130 &  3.13$^{**}$ && 0.078 &  2.37$^{*~}$ && 0.071 &  2.65$^{**}$ && 0.066 &  2.80$^{**}$ && 0.077 &  3.32$^{**}$ && 0.079 &  3.24$^{**}$ && 0.072 &  2.81$^{**}$ && 0.066 &  2.51$^{*~}$ && 0.063 &  2.32$^{*~}$  \\
   30 & 0.137 &  3.23$^{**}$ && 0.089 &  2.73$^{**}$ && 0.072 &  2.50$^{*~}$ && 0.075 &  3.12$^{**}$ && 0.083 &  3.54$^{**}$ && 0.081 &  3.43$^{**}$ && 0.077 &  3.03$^{**}$ && 0.073 &  2.71$^{**}$ && 0.068 &  2.50$^{*~}$  \\
   36 & 0.128 &  3.02$^{**}$ && 0.083 &  2.47$^{*~}$ && 0.079 &  2.65$^{**}$ && 0.088 &  3.47$^{**}$ && 0.090 &  3.72$^{**}$ && 0.090 &  3.64$^{**}$ && 0.086 &  3.17$^{**}$ && 0.079 &  2.73$^{**}$ && 0.078 &  2.79$^{**}$  \\
   42 & 0.150 &  3.57$^{**}$ && 0.100 &  3.13$^{**}$ && 0.089 &  2.96$^{**}$ && 0.091 &  3.77$^{**}$ && 0.095 &  4.10$^{**}$ && 0.093 &  3.81$^{**}$ && 0.087 &  3.16$^{**}$ && 0.086 &  3.09$^{**}$ && 0.084 &  3.26$^{**}$  \\
   48 & 0.131 &  2.84$^{**}$ && 0.095 &  2.76$^{**}$ && 0.087 &  2.70$^{**}$ && 0.092 &  3.56$^{**}$ && 0.096 &  3.75$^{**}$ && 0.091 &  3.37$^{**}$ && 0.090 &  3.25$^{**}$ && 0.093 &  3.61$^{**}$ && 0.094 &  4.12$^{**}$  \\
   \hline
   \end{tabular}
   \begin{tablenotes}
   \small
     \item This table reports the average annualized returns and the corresponding t-statistics adjusted for heteroscedasticity and autocorrelation of the loser, winner and contrarian portfolios, which are formed by ranking the stocks based on their $J$-month lagged returns, adopting the tertile grouping, and holding for $K$ months. The values of $J$ and $K$ for different strategies are indicated in the first collum and the first row respectively. The sample period is January 1997 to  December 2012. The superscripts * and ** denote the significance at 5\% and 1\% levels, respectively.
   \end{tablenotes}
\end{threeparttable}
\end{table}
\end{landscape}


\begin{landscape}
\begin{table}[htb]
\centering
\begin{threeparttable}[b]
   \small
   \caption{The return difference of loser portfolios formed based on different grouping ways of the SHSE stocks.}
   \label{TBS:Empirics:Diff:Group:LOS:SHSE}
   \begin{tabular}{ccccccccccccccccccccccccccccc}
   \hline
          & \multicolumn{2}{c}{$K=1$} && \multicolumn{2}{c}{6} && \multicolumn{2}{c}{12} && \multicolumn{2}{c}{18} && \multicolumn{2}{c}{24} && \multicolumn{2}{c}{30} && \multicolumn{2}{c}{36} && \multicolumn{2}{c}{42} && \multicolumn{2}{c}{48} \\
   \cline{2-3} \cline{5-6} \cline{8-9} \cline{11-12} \cline{14-15} \cline{17-18} \cline{20-21} \cline{23-24} \cline{26-27}
     $J$  & $\Delta{R}$ & t-stat && $\Delta{R}$ & t-stat && $\Delta{R}$ & t-stat && $\Delta{R}$ & t-stat && $\Delta{R}$ & t-stat && $\Delta{R}$ & t-stat && $\Delta{R}$ & t-stat && $\Delta{R}$ & t-stat && $\Delta{R}$ & t-stat \\
   \hline
   \multicolumn{27}{l}{\textit{Panel A: $G_{5}-G_{3}$}} \\
 1& 0.001 &  0.14$~~$ && -0.002 & -0.80$~~$ && -0.003 & -1.00$~~$ && -0.003 & -1.27$~~$ && -0.004 & -1.67$^{~~}$ && -0.000 & -0.05$~~$ && -0.001 & -0.38$~~$ && 0.001 &  0.52$~~$ && 0.001 &  0.58$~~$  \\
 6& 0.005 &  0.70$~~$ && -0.009 & -3.14$^{**}$ && -0.005 & -2.17$^{*~}$ && -0.007 & -2.47$^{*~}$ && -0.000 & -0.10$~~$ && -0.000 & -0.08$~~$ && 0.001 &  0.36$~~$ && 0.002 &  1.02$~~$ && 0.001 &  0.51$~~$  \\
 12& 0.007 &  0.99$~~$ && -0.007 & -2.27$^{*~}$ && -0.005 & -1.96$^{~~}$ && -0.006 & -2.44$^{*~}$ && -0.005 & -2.14$^{*~}$ && -0.002 & -1.00$~~$ && 0.001 &  0.50$~~$ && -0.001 & -0.24$~~$ && -0.000 & -0.14$~~$  \\
 18& 0.007 &  0.93$~~$ && -0.004 & -1.22$~~$ && 0.001 &  0.33$~~$ && 0.001 &  0.49$~~$ && 0.002 &  0.86$~~$ && 0.003 &  1.13$~~$ && 0.006 &  2.36$^{*~}$ && 0.005 &  2.02$^{*~}$ && 0.007 &  2.57$^{*~}$  \\
 24& -0.000 & -0.01$~~$ && -0.001 & -0.35$~~$ && 0.005 &  1.82$^{~~}$ && 0.004 &  1.49$~~$ && 0.004 &  1.59$~~$ && 0.004 &  1.53$~~$ && 0.006 &  2.78$^{**}$ && 0.007 &  2.98$^{**}$ && 0.007 &  2.89$^{**}$  \\
 30& 0.018 &  2.31$^{*~}$ && 0.005 &  1.31$~~$ && 0.007 &  2.55$^{*~}$ && 0.007 &  2.82$^{**}$ && 0.008 &  2.75$^{**}$ && 0.006 &  1.99$^{*~}$ && 0.009 &  3.21$^{**}$ && 0.010 &  3.32$^{**}$ && 0.014 &  4.29$^{**}$  \\
 36& 0.006 &  0.67$~~$ && 0.010 &  2.80$^{**}$ && 0.011 &  4.06$^{**}$ && 0.014 &  4.84$^{**}$ && 0.015 &  5.55$^{**}$ && 0.013 &  4.26$^{**}$ && 0.013 &  4.71$^{**}$ && 0.015 &  5.07$^{**}$ && 0.017 &  5.37$^{**}$  \\
 42& 0.021 &  2.42$^{*~}$ && 0.013 &  3.39$^{**}$ && 0.016 &  4.79$^{**}$ && 0.017 &  5.76$^{**}$ && 0.015 &  5.09$^{**}$ && 0.015 &  5.26$^{**}$ && 0.019 &  6.54$^{**}$ && 0.021 &  6.38$^{**}$ && 0.024 &  6.53$^{**}$  \\
 48& 0.019 &  1.99$^{*~}$ && 0.016 &  4.04$^{**}$ && 0.015 &  4.67$^{**}$ && 0.017 &  5.14$^{**}$ && 0.014 &  4.30$^{**}$ && 0.012 &  3.58$^{**}$ && 0.018 &  5.35$^{**}$ && 0.023 &  6.20$^{**}$ && 0.026 &  6.99$^{**}$  \\
    \vspace{-3mm}\\
   \multicolumn{27}{l}{\textit{Panel B: $G_{10}-G_{5}$}} \\
 1& -0.014 & -1.60$~~$ && -0.004 & -1.02$~~$ && -0.006 & -1.74$^{~~}$ && -0.006 & -1.66$~~$ && -0.002 & -0.63$~~$ && -0.000 & -0.03$~~$ && 0.003 &  0.76$~~$ && 0.002 &  0.65$~~$ && -0.001 & -0.23$~~$  \\
 6& -0.002 & -0.16$~~$ && -0.007 & -1.56$~~$ && -0.012 & -3.20$^{**}$ && -0.007 & -1.85$^{~~}$ && -0.007 & -1.73$^{~~}$ && -0.007 & -1.84$^{~~}$ && -0.004 & -0.97$~~$ && -0.001 & -0.23$~~$ && -0.003 & -0.70$~~$  \\
 12& -0.006 & -0.58$~~$ && -0.013 & -2.65$^{**}$ && -0.016 & -4.29$^{**}$ && -0.009 & -2.53$^{*~}$ && -0.005 & -1.36$~~$ && 0.000 &  0.05$~~$ && 0.010 &  2.77$^{**}$ && 0.011 &  2.70$^{**}$ && 0.011 &  2.78$^{**}$  \\
 18& 0.002 &  0.19$~~$ && -0.014 & -2.45$^{*~}$ && -0.011 & -2.37$^{*~}$ && -0.010 & -2.01$^{*~}$ && -0.011 & -2.16$^{*~}$ && -0.005 & -1.10$~~$ && 0.003 &  0.64$~~$ && 0.008 &  1.86$^{~~}$ && 0.008 &  1.76$^{~~}$  \\
 24& 0.004 &  0.37$~~$ && -0.009 & -1.49$~~$ && -0.007 & -1.45$~~$ && -0.005 & -1.08$~~$ && -0.002 & -0.51$~~$ && 0.004 &  0.96$~~$ && 0.007 &  1.93$^{~~}$ && 0.011 &  2.37$^{*~}$ && 0.017 &  4.00$^{**}$  \\
 30& 0.007 &  0.56$~~$ && 0.006 &  1.08$~~$ && 0.002 &  0.53$~~$ && 0.006 &  1.47$~~$ && 0.010 &  2.72$^{**}$ && 0.021 &  5.43$^{**}$ && 0.022 &  5.80$^{**}$ && 0.023 &  6.03$^{**}$ && 0.026 &  6.04$^{**}$  \\
 36& 0.022 &  1.74$^{~~}$ && 0.005 &  0.64$~~$ && 0.005 &  1.19$~~$ && 0.008 &  1.92$^{~~}$ && 0.015 &  3.36$^{**}$ && 0.021 &  5.10$^{**}$ && 0.024 &  5.96$^{**}$ && 0.027 &  6.69$^{**}$ && 0.029 &  7.11$^{**}$  \\
 42& 0.016 &  1.13$~~$ && 0.013 &  1.81$^{~~}$ && 0.013 &  2.30$^{*~}$ && 0.019 &  3.68$^{**}$ && 0.026 &  5.28$^{**}$ && 0.031 &  5.48$^{**}$ && 0.028 &  4.73$^{**}$ && 0.032 &  5.41$^{**}$ && 0.034 &  6.56$^{**}$  \\
 48& 0.023 &  1.38$~~$ && 0.012 &  1.47$~~$ && 0.015 &  2.32$^{*~}$ && 0.020 &  3.43$^{**}$ && 0.032 &  4.89$^{**}$ && 0.037 &  5.79$^{**}$ && 0.034 &  4.77$^{**}$ && 0.034 &  4.78$^{**}$ && 0.037 &  5.83$^{**}$  \\
   \vspace{-3mm}\\
   \multicolumn{27}{l}{\textit{Panel C: $G_{10}-G_{3}$}} \\
 1 & -0.014 & -1.03$~~$ && -0.006 & -1.13$~~$ && -0.009 & -1.78$^{~~}$ && -0.009 & -1.90$^{~~}$ && -0.006 & -1.25$~~$ && -0.000 & -0.04$~~$ && 0.002 &  0.43$~~$ && 0.003 &  0.80$~~$ && 0.000 &  0.09$~~$  \\
 6 & 0.003 &  0.21$~~$ && -0.016 & -2.59$^{*~}$ && -0.018 & -3.67$^{**}$ && -0.014 & -2.73$^{**}$ && -0.007 & -1.42$~~$ && -0.007 & -1.48$~~$ && -0.003 & -0.67$~~$ && 0.001 &  0.28$~~$ && -0.002 & -0.35$~~$  \\
 12 & 0.001 &  0.05$~~$ && -0.020 & -3.12$^{**}$ && -0.021 & -4.06$^{**}$ && -0.015 & -3.01$^{**}$ && -0.011 & -2.00$^{*~}$ && -0.002 & -0.46$~~$ && 0.012 &  2.45$^{*~}$ && 0.010 &  1.91$^{~~}$ && 0.011 &  1.92$^{~~}$  \\
 18 & 0.009 &  0.54$~~$ && -0.018 & -2.23$^{*~}$ && -0.010 & -1.61$~~$ && -0.009 & -1.52$~~$ && -0.009 & -1.50$~~$ && -0.003 & -0.47$~~$ && 0.008 &  1.69$^{~~}$ && 0.013 &  2.41$^{*~}$ && 0.015 &  2.55$^{*~}$  \\
 24 & 0.004 &  0.26$~~$ && -0.010 & -1.31$~~$ && -0.002 & -0.32$~~$ && -0.001 & -0.21$~~$ && 0.002 &  0.28$~~$ && 0.007 &  1.45$~~$ && 0.014 &  2.66$^{**}$ && 0.018 &  3.11$^{**}$ && 0.024 &  4.21$^{**}$  \\
 30 & 0.025 &  1.49$~~$ && 0.011 &  1.49$~~$ && 0.009 &  1.66$~~$ && 0.013 &  2.75$^{**}$ && 0.018 &  3.62$^{**}$ && 0.027 &  4.97$^{**}$ && 0.032 &  5.77$^{**}$ && 0.033 &  6.02$^{**}$ && 0.040 &  6.29$^{**}$  \\
 36 & 0.028 &  1.56$~~$ && 0.014 &  1.59$~~$ && 0.017 &  3.11$^{**}$ && 0.022 &  4.31$^{**}$ && 0.030 &  5.08$^{**}$ && 0.034 &  5.52$^{**}$ && 0.037 &  6.42$^{**}$ && 0.042 &  7.05$^{**}$ && 0.046 &  7.03$^{**}$  \\
 42 & 0.037 &  1.97$^{~~}$ && 0.026 &  2.98$^{**}$ && 0.028 &  4.23$^{**}$ && 0.036 &  5.54$^{**}$ && 0.042 &  6.93$^{**}$ && 0.046 &  7.01$^{**}$ && 0.046 &  6.74$^{**}$ && 0.053 &  7.66$^{**}$ && 0.059 &  7.89$^{**}$  \\
 48 & 0.041 &  1.84$^{~~}$ && 0.028 &  2.79$^{**}$ && 0.029 &  3.64$^{**}$ && 0.037 &  5.41$^{**}$ && 0.046 &  5.64$^{**}$ && 0.049 &  5.97$^{**}$ && 0.052 &  6.10$^{**}$ && 0.057 &  6.72$^{**}$ && 0.063 &  7.03$^{**}$  \\
   \hline
   \end{tabular}
   \begin{tablenotes}
   \small
     \item This table reports the differences of the average annualized returns and the corresponding t-statistics adjusted for heteroscedasticity and autocorrelation of two loser strategies that are different only in the grouping methods for SHSE stocks. The three panels are for the loser, winner and contrarian portfolios, respectively. In the first row, $G_3$, $G_5$ and $G_{10}$ stand for tertile, quintile and decile groupings. The sample period is January 1997 to  December 2012. The superscripts * and ** denote the significance at 5\% and 1\% levels, respectively.
   \end{tablenotes}
\end{threeparttable}
\end{table}
\end{landscape}

\setlength\tabcolsep{2.0pt}
\begin{landscape}
\begin{table}[htb]
\centering
\begin{threeparttable}[b]
   \small
   \caption{The return difference of loser portfolios formed based on different grouping ways of the SZSE stocks.}
   \label{TBS:Empirics:Diff:Group:LOS:SZSE}
   \begin{tabular}{ccccccccccccccccccccccccccccc}
   \hline
          & \multicolumn{2}{c}{$K=1$} && \multicolumn{2}{c}{6} && \multicolumn{2}{c}{12} && \multicolumn{2}{c}{18} && \multicolumn{2}{c}{24} && \multicolumn{2}{c}{30} && \multicolumn{2}{c}{36} && \multicolumn{2}{c}{42} && \multicolumn{2}{c}{48} \\
   \cline{2-3} \cline{5-6} \cline{8-9} \cline{11-12} \cline{14-15} \cline{17-18} \cline{20-21} \cline{23-24} \cline{26-27}
     $J$  & $\Delta{R}$ & t-stat && $\Delta{R}$ & t-stat && $\Delta{R}$ & t-stat && $\Delta{R}$ & t-stat && $\Delta{R}$ & t-stat && $\Delta{R}$ & t-stat && $\Delta{R}$ & t-stat && $\Delta{R}$ & t-stat && $\Delta{R}$ & t-stat \\
   \hline
   \multicolumn{27}{l}{\textit{Panel A: $G_{5}-G_{3}$}} \\
 1& -0.004 & -0.54$~~$ && -0.008 & -2.44$^{*~}$ && -0.008 & -3.33$^{**}$ && -0.009 & -3.10$^{**}$ && -0.009 & -2.98$^{**}$ && -0.002 & -0.75$~~$ && -0.003 & -1.02$~~$ && -0.001 & -0.21$~~$ && -0.001 & -0.62$~~$  \\
 6& -0.001 & -0.06$~~$ && -0.009 & -2.08$^{*~}$ && -0.009 & -2.82$^{**}$ && -0.011 & -3.50$^{**}$ && -0.008 & -2.66$^{**}$ && -0.002 & -0.80$~~$ && -0.001 & -0.46$~~$ && 0.002 &  0.74$~~$ && 0.001 &  0.28$~~$  \\
 12& 0.011 &  1.35$~~$ && -0.003 & -0.73$~~$ && -0.001 & -0.19$~~$ && -0.001 & -0.32$~~$ && 0.009 &  2.23$^{*~}$ && 0.011 &  3.27$^{**}$ && 0.009 &  3.01$^{**}$ && 0.010 &  3.56$^{**}$ && 0.011 &  3.69$^{**}$  \\
 18& 0.016 &  1.47$~~$ && 0.006 &  1.62$~~$ && 0.011 &  3.28$^{**}$ && 0.010 &  2.70$^{**}$ && 0.016 &  4.29$^{**}$ && 0.018 &  4.65$^{**}$ && 0.015 &  4.63$^{**}$ && 0.015 &  4.76$^{**}$ && 0.017 &  4.91$^{**}$  \\
 24& 0.009 &  0.94$~~$ && 0.003 &  0.78$~~$ && 0.010 &  2.79$^{**}$ && 0.008 &  2.11$^{*~}$ && 0.014 &  2.92$^{**}$ && 0.016 &  3.33$^{**}$ && 0.016 &  3.75$^{**}$ && 0.014 &  3.80$^{**}$ && 0.013 &  3.59$^{**}$  \\
 30& 0.010 &  1.07$~~$ && 0.009 &  2.54$^{*~}$ && 0.011 &  3.73$^{**}$ && 0.014 &  4.00$^{**}$ && 0.018 &  4.68$^{**}$ && 0.020 &  4.89$^{**}$ && 0.017 &  5.02$^{**}$ && 0.016 &  4.99$^{**}$ && 0.015 &  4.07$^{**}$  \\
 36& 0.022 &  2.07$^{*~}$ && 0.017 &  3.58$^{**}$ && 0.018 &  4.50$^{**}$ && 0.012 &  2.94$^{**}$ && 0.021 &  5.15$^{**}$ && 0.022 &  5.43$^{**}$ && 0.021 &  5.97$^{**}$ && 0.021 &  6.80$^{**}$ && 0.021 &  6.03$^{**}$  \\
 42& 0.032 &  2.52$^{*~}$ && 0.017 &  3.19$^{**}$ && 0.017 &  4.02$^{**}$ && 0.017 &  4.30$^{**}$ && 0.019 &  4.55$^{**}$ && 0.018 &  4.27$^{**}$ && 0.018 &  5.29$^{**}$ && 0.020 &  6.06$^{**}$ && 0.021 &  5.75$^{**}$  \\
 48& 0.029 &  1.99$^{*~}$ && 0.025 &  4.10$^{**}$ && 0.018 &  4.35$^{**}$ && 0.022 &  5.24$^{**}$ && 0.030 &  6.67$^{**}$ && 0.026 &  6.40$^{**}$ && 0.028 &  7.10$^{**}$ && 0.027 &  6.13$^{**}$ && 0.028 &  7.62$^{**}$  \\
    \vspace{-3mm}\\
   \multicolumn{27}{l}{\textit{Panel B: $G_{10}-G_{5}$}} \\
 1& -0.009 & -0.84$~~$ && -0.002 & -0.29$~~$ && -0.011 & -2.80$^{**}$ && -0.009 & -1.94$^{~~}$ && -0.010 & -2.43$^{*~}$ && -0.008 & -1.63$~~$ && -0.002 & -0.54$~~$ && 0.000 &  0.07$~~$ && -0.002 & -0.54$~~$  \\
 6& -0.003 & -0.20$~~$ && -0.014 & -2.34$^{*~}$ && -0.006 & -1.40$~~$ && -0.006 & -1.46$~~$ && -0.002 & -0.47$~~$ && 0.008 &  1.98$^{~~}$ && 0.008 &  1.82$^{~~}$ && 0.005 &  1.33$~~$ && 0.007 &  1.84$^{~~}$  \\
 12& 0.022 &  1.44$~~$ && -0.010 & -1.90$^{~~}$ && -0.005 & -1.12$~~$ && -0.007 & -1.21$~~$ && 0.003 &  0.73$~~$ && 0.014 &  2.45$^{*~}$ && 0.013 &  2.46$^{*~}$ && 0.011 &  2.35$^{*~}$ && 0.016 &  3.43$^{**}$  \\
 18& 0.001 &  0.07$~~$ && -0.007 & -1.10$~~$ && -0.005 & -0.93$~~$ && -0.004 & -0.58$~~$ && 0.007 &  1.00$~~$ && 0.015 &  2.28$^{*~}$ && 0.020 &  3.14$^{**}$ && 0.019 &  3.45$^{**}$ && 0.019 &  3.53$^{**}$  \\
 24& 0.025 &  1.46$~~$ && 0.003 &  0.43$~~$ && -0.004 & -0.50$~~$ && 0.009 &  1.31$~~$ && 0.026 &  3.79$^{**}$ && 0.032 &  4.50$^{**}$ && 0.029 &  4.59$^{**}$ && 0.031 &  5.19$^{**}$ && 0.031 &  4.89$^{**}$  \\
 30& 0.016 &  0.98$~~$ && 0.014 &  2.11$^{*~}$ && 0.015 &  2.12$^{*~}$ && 0.016 &  2.03$^{*~}$ && 0.027 &  3.54$^{**}$ && 0.035 &  5.00$^{**}$ && 0.034 &  5.58$^{**}$ && 0.033 &  5.67$^{**}$ && 0.034 &  5.22$^{**}$  \\
 36& 0.024 &  1.39$~~$ && 0.023 &  2.80$^{**}$ && 0.022 &  2.95$^{**}$ && 0.026 &  3.43$^{**}$ && 0.028 &  3.49$^{**}$ && 0.033 &  4.32$^{**}$ && 0.030 &  4.28$^{**}$ && 0.033 &  4.83$^{**}$ && 0.038 &  5.37$^{**}$  \\
 42& 0.034 &  1.42$~~$ && 0.033 &  3.17$^{**}$ && 0.031 &  3.47$^{**}$ && 0.027 &  3.30$^{**}$ && 0.035 &  4.40$^{**}$ && 0.041 &  5.63$^{**}$ && 0.042 &  6.32$^{**}$ && 0.047 &  7.10$^{**}$ && 0.042 &  5.79$^{**}$  \\
 48& 0.033 &  1.12$~~$ && 0.027 &  2.48$^{*~}$ && 0.025 &  3.14$^{**}$ && 0.018 &  1.74$^{~~}$ && 0.023 &  2.38$^{*~}$ && 0.035 &  4.50$^{**}$ && 0.043 &  5.96$^{**}$ && 0.040 &  5.58$^{**}$ && 0.035 &  4.81$^{**}$  \\
   \vspace{-3mm}\\
   \multicolumn{27}{l}{\textit{Panel C: $G_{10}-G_{3}$}} \\
 1 & -0.013 & -0.85$~~$ && -0.010 & -1.39$~~$ && -0.020 & -3.77$^{**}$ && -0.018 & -2.69$^{**}$ && -0.019 & -3.34$^{**}$ && -0.010 & -2.00$^{*~}$ && -0.005 & -1.01$~~$ && -0.000 & -0.05$~~$ && -0.004 & -0.75$~~$  \\
 6 & -0.003 & -0.17$~~$ && -0.023 & -2.61$^{*~}$ && -0.016 & -2.58$^{*~}$ && -0.017 & -2.79$^{**}$ && -0.010 & -1.83$^{~~}$ && 0.006 &  1.05$~~$ && 0.006 &  1.19$~~$ && 0.008 &  1.32$~~$ && 0.008 &  1.46$~~$  \\
 12 & 0.032 &  1.62$~~$ && -0.013 & -1.68$^{~~}$ && -0.006 & -0.85$~~$ && -0.008 & -1.00$~~$ && 0.012 &  1.86$^{~~}$ && 0.025 &  3.42$^{**}$ && 0.022 &  3.14$^{**}$ && 0.021 &  3.26$^{**}$ && 0.026 &  4.01$^{**}$  \\
 18 & 0.017 &  0.81$~~$ && -0.001 & -0.08$~~$ && 0.005 &  0.67$~~$ && 0.006 &  0.74$~~$ && 0.023 &  2.64$^{**}$ && 0.033 &  3.82$^{**}$ && 0.035 &  4.19$^{**}$ && 0.034 &  4.48$^{**}$ && 0.037 &  4.72$^{**}$  \\
 24 & 0.034 &  1.51$~~$ && 0.006 &  0.63$~~$ && 0.006 &  0.80$~~$ && 0.017 &  1.93$^{~~}$ && 0.040 &  4.08$^{**}$ && 0.048 &  4.86$^{**}$ && 0.045 &  5.18$^{**}$ && 0.045 &  5.68$^{**}$ && 0.044 &  5.09$^{**}$  \\
 30 & 0.026 &  1.17$~~$ && 0.023 &  2.71$^{**}$ && 0.026 &  2.96$^{**}$ && 0.030 &  3.05$^{**}$ && 0.045 &  4.59$^{**}$ && 0.055 &  5.55$^{**}$ && 0.050 &  5.90$^{**}$ && 0.049 &  6.21$^{**}$ && 0.049 &  5.35$^{**}$  \\
 36 & 0.045 &  1.91$^{~~}$ && 0.041 &  3.47$^{**}$ && 0.040 &  3.88$^{**}$ && 0.038 &  3.67$^{**}$ && 0.048 &  4.61$^{**}$ && 0.054 &  5.34$^{**}$ && 0.051 &  5.84$^{**}$ && 0.055 &  6.26$^{**}$ && 0.059 &  6.25$^{**}$  \\
 42 & 0.066 &  1.96$^{~~}$ && 0.051 &  3.50$^{**}$ && 0.048 &  4.04$^{**}$ && 0.044 &  4.10$^{**}$ && 0.054 &  4.95$^{**}$ && 0.059 &  5.63$^{**}$ && 0.061 &  6.45$^{**}$ && 0.067 &  7.44$^{**}$ && 0.063 &  6.37$^{**}$  \\
 48 & 0.061 &  1.62$~~$ && 0.051 &  3.44$^{**}$ && 0.043 &  4.06$^{**}$ && 0.040 &  3.40$^{**}$ && 0.053 &  4.52$^{**}$ && 0.061 &  5.88$^{**}$ && 0.071 &  7.27$^{**}$ && 0.067 &  6.68$^{**}$ && 0.063 &  6.53$^{**}$  \\
   \hline
   \end{tabular}
   \begin{tablenotes}
   \small
     \item This table reports the differences of the average annualized returns and the corresponding t-statistics adjusted for heteroscedasticity and autocorrelation of two loser strategies that are different only in the grouping methods for SZSE stocks. The three panels are for the loser, winner and contrarian portfolios, respectively. In the first row, $G_3$, $G_5$ and $G_{10}$ stand for tertile, quintile and decile groupings. The sample period is January 1997 to  December 2012. The superscripts * and ** denote the significance at 5\% and 1\% levels, respectively.
   \end{tablenotes}
\end{threeparttable}
\end{table}
\end{landscape}

\setlength\tabcolsep{2.0pt}
\begin{landscape}
\begin{table}[htb]
\centering
\begin{threeparttable}[b]
   \small
   \caption{The return difference of winner portfolios formed based on different grouping ways of the SHSE stocks.}
   \label{TBS:Empirics:Diff:Group:WIN:SHSE}
   \begin{tabular}{ccccccccccccccccccccccccccccc}
   \hline
          & \multicolumn{2}{c}{$K=1$} && \multicolumn{2}{c}{6} && \multicolumn{2}{c}{12} && \multicolumn{2}{c}{18} && \multicolumn{2}{c}{24} && \multicolumn{2}{c}{30} && \multicolumn{2}{c}{36} && \multicolumn{2}{c}{42} && \multicolumn{2}{c}{48} \\
   \cline{2-3} \cline{5-6} \cline{8-9} \cline{11-12} \cline{14-15} \cline{17-18} \cline{20-21} \cline{23-24} \cline{26-27}
     $J$  & $\Delta{R}$ & t-stat && $\Delta{R}$ & t-stat && $\Delta{R}$ & t-stat && $\Delta{R}$ & t-stat && $\Delta{R}$ & t-stat && $\Delta{R}$ & t-stat && $\Delta{R}$ & t-stat && $\Delta{R}$ & t-stat && $\Delta{R}$ & t-stat \\
   \hline
   \multicolumn{27}{l}{\textit{Panel A: $G_{5}-G_{3}$}} \\
 1& -0.018 & -2.40$^{*~}$ && -0.011 & -3.37$^{**}$ && -0.008 & -2.85$^{**}$ && -0.009 & -3.84$^{**}$ && -0.009 & -3.88$^{**}$ && -0.009 & -3.71$^{**}$ && -0.006 & -2.48$^{*~}$ && -0.004 & -1.66$~~$ && -0.003 & -1.06$~~$  \\
 6& -0.011 & -1.33$~~$ && -0.007 & -2.07$^{*~}$ && -0.008 & -2.68$^{**}$ && -0.010 & -4.30$^{**}$ && -0.014 & -6.00$^{**}$ && -0.010 & -3.90$^{**}$ && -0.005 & -1.79$^{~~}$ && -0.007 & -2.81$^{**}$ && -0.009 & -3.98$^{**}$  \\
 12& -0.000 & -0.03$~~$ && -0.009 & -2.28$^{*~}$ && -0.014 & -4.43$^{**}$ && -0.016 & -5.21$^{**}$ && -0.017 & -6.27$^{**}$ && -0.013 & -4.61$^{**}$ && -0.006 & -1.81$^{~~}$ && -0.008 & -2.66$^{**}$ && -0.010 & -3.79$^{**}$  \\
 18& -0.015 & -1.69$^{~~}$ && -0.014 & -3.78$^{**}$ && -0.019 & -6.29$^{**}$ && -0.020 & -7.57$^{**}$ && -0.020 & -7.92$^{**}$ && -0.017 & -6.20$^{**}$ && -0.013 & -4.66$^{**}$ && -0.016 & -7.28$^{**}$ && -0.016 & -7.11$^{**}$  \\
 24& -0.012 & -1.38$~~$ && -0.011 & -2.71$^{**}$ && -0.017 & -5.50$^{**}$ && -0.021 & -8.39$^{**}$ && -0.024 & -8.80$^{**}$ && -0.021 & -8.79$^{**}$ && -0.015 & -5.76$^{**}$ && -0.018 & -7.86$^{**}$ && -0.018 & -8.96$^{**}$  \\
 30& -0.004 & -0.38$~~$ && -0.011 & -2.56$^{*~}$ && -0.015 & -5.09$^{**}$ && -0.018 & -6.84$^{**}$ && -0.020 & -7.22$^{**}$ && -0.016 & -5.68$^{**}$ && -0.013 & -4.26$^{**}$ && -0.014 & -6.07$^{**}$ && -0.014 & -6.59$^{**}$  \\
 36& -0.006 & -0.70$~~$ && -0.010 & -2.67$^{**}$ && -0.015 & -4.62$^{**}$ && -0.019 & -6.86$^{**}$ && -0.018 & -6.32$^{**}$ && -0.013 & -4.13$^{**}$ && -0.009 & -3.16$^{**}$ && -0.012 & -4.95$^{**}$ && -0.015 & -7.30$^{**}$  \\
 42& -0.007 & -0.73$~~$ && -0.012 & -2.92$^{**}$ && -0.016 & -5.21$^{**}$ && -0.022 & -7.95$^{**}$ && -0.017 & -6.75$^{**}$ && -0.013 & -4.72$^{**}$ && -0.010 & -4.30$^{**}$ && -0.009 & -3.26$^{**}$ && -0.014 & -4.99$^{**}$  \\
 48& -0.016 & -1.33$~~$ && -0.024 & -4.96$^{**}$ && -0.024 & -7.29$^{**}$ && -0.023 & -7.41$^{**}$ && -0.020 & -7.30$^{**}$ && -0.016 & -5.15$^{**}$ && -0.013 & -4.65$^{**}$ && -0.012 & -3.62$^{**}$ && -0.012 & -3.46$^{**}$  \\
    \vspace{-3mm}\\
   \multicolumn{27}{l}{\textit{Panel B: $G_{10}-G_{5}$}} \\
 1& -0.022 & -1.79$^{~~}$ && -0.007 & -1.64$~~$ && -0.006 & -1.44$~~$ && -0.005 & -1.00$~~$ && -0.006 & -1.67$^{~~}$ && -0.002 & -0.54$~~$ && -0.001 & -0.31$~~$ && 0.002 &  0.55$~~$ && 0.001 &  0.16$~~$  \\
 6& -0.015 & -1.30$~~$ && -0.010 & -2.07$^{*~}$ && -0.017 & -4.31$^{**}$ && -0.015 & -4.24$^{**}$ && -0.014 & -4.60$^{**}$ && -0.017 & -5.31$^{**}$ && -0.013 & -3.97$^{**}$ && -0.007 & -1.97$^{~~}$ && -0.008 & -2.23$^{*~}$  \\
 12& -0.007 & -0.66$~~$ && -0.003 & -0.55$~~$ && -0.005 & -1.27$~~$ && -0.009 & -2.60$^{*~}$ && -0.013 & -3.78$^{**}$ && -0.009 & -2.64$^{**}$ && -0.007 & -1.52$~~$ && -0.007 & -1.45$~~$ && -0.016 & -4.31$^{**}$  \\
 18& -0.005 & -0.38$~~$ && -0.009 & -1.50$~~$ && -0.012 & -2.77$^{**}$ && -0.015 & -3.98$^{**}$ && -0.015 & -4.08$^{**}$ && -0.015 & -3.64$^{**}$ && -0.009 & -2.09$^{*~}$ && -0.012 & -3.11$^{**}$ && -0.016 & -4.87$^{**}$  \\
 24& -0.004 & -0.35$~~$ && -0.013 & -2.21$^{*~}$ && -0.012 & -2.83$^{**}$ && -0.014 & -4.10$^{**}$ && -0.009 & -2.31$^{*~}$ && -0.012 & -3.05$^{**}$ && -0.010 & -2.92$^{**}$ && -0.010 & -2.87$^{**}$ && -0.010 & -3.43$^{**}$  \\
 30& -0.028 & -2.29$^{*~}$ && -0.015 & -3.09$^{**}$ && -0.015 & -3.63$^{**}$ && -0.013 & -3.38$^{**}$ && -0.010 & -2.81$^{**}$ && -0.009 & -2.49$^{*~}$ && -0.007 & -2.06$^{*~}$ && -0.008 & -2.30$^{*~}$ && -0.010 & -3.23$^{**}$  \\
 36& -0.014 & -1.02$~~$ && -0.006 & -1.00$~~$ && -0.006 & -1.35$~~$ && -0.006 & -1.53$~~$ && -0.005 & -1.17$~~$ && -0.004 & -0.90$~~$ && -0.005 & -1.63$~~$ && -0.008 & -2.70$^{**}$ && -0.013 & -4.62$^{**}$  \\
 42& -0.006 & -0.48$~~$ && -0.016 & -2.29$^{*~}$ && -0.011 & -2.41$^{*~}$ && -0.004 & -0.79$~~$ && -0.005 & -0.97$~~$ && -0.005 & -1.03$~~$ && -0.008 & -1.96$^{~~}$ && -0.017 & -4.51$^{**}$ && -0.021 & -5.63$^{**}$  \\
 48& -0.025 & -1.61$~~$ && -0.021 & -3.08$^{**}$ && -0.010 & -2.09$^{*~}$ && -0.006 & -1.19$~~$ && -0.008 & -1.55$~~$ && -0.009 & -1.79$^{~~}$ && -0.008 & -1.70$^{~~}$ && -0.019 & -4.27$^{**}$ && -0.025 & -4.75$^{**}$  \\
   \vspace{-3mm}\\
   \multicolumn{27}{l}{\textit{Panel C: $G_{10}-G_{3}$}} \\
 1 & -0.040 & -2.45$^{*~}$ && -0.019 & -3.08$^{**}$ && -0.014 & -2.54$^{*~}$ && -0.014 & -2.53$^{*~}$ && -0.015 & -3.32$^{**}$ && -0.011 & -2.07$^{*~}$ && -0.007 & -1.42$~~$ && -0.002 & -0.55$~~$ && -0.002 & -0.49$~~$  \\
 6 & -0.026 & -1.51$~~$ && -0.018 & -2.50$^{*~}$ && -0.025 & -4.45$^{**}$ && -0.025 & -5.25$^{**}$ && -0.028 & -6.69$^{**}$ && -0.027 & -6.22$^{**}$ && -0.017 & -3.83$^{**}$ && -0.014 & -2.73$^{**}$ && -0.017 & -3.50$^{**}$  \\
 12 & -0.007 & -0.44$~~$ && -0.012 & -1.51$~~$ && -0.019 & -3.15$^{**}$ && -0.025 & -4.72$^{**}$ && -0.029 & -6.25$^{**}$ && -0.022 & -4.25$^{**}$ && -0.013 & -1.87$^{~~}$ && -0.015 & -2.15$^{*~}$ && -0.026 & -5.19$^{**}$  \\
 18 & -0.019 & -1.08$~~$ && -0.023 & -2.79$^{**}$ && -0.031 & -4.92$^{**}$ && -0.035 & -6.74$^{**}$ && -0.035 & -7.10$^{**}$ && -0.031 & -5.46$^{**}$ && -0.022 & -3.42$^{**}$ && -0.028 & -5.49$^{**}$ && -0.032 & -7.42$^{**}$  \\
 24 & -0.016 & -0.90$~~$ && -0.023 & -2.84$^{**}$ && -0.029 & -4.76$^{**}$ && -0.035 & -7.20$^{**}$ && -0.032 & -6.87$^{**}$ && -0.032 & -7.09$^{**}$ && -0.025 & -4.98$^{**}$ && -0.027 & -5.81$^{**}$ && -0.029 & -6.87$^{**}$  \\
 30 & -0.031 & -1.64$~~$ && -0.026 & -3.33$^{**}$ && -0.029 & -4.87$^{**}$ && -0.032 & -5.98$^{**}$ && -0.030 & -5.86$^{**}$ && -0.025 & -4.71$^{**}$ && -0.019 & -3.53$^{**}$ && -0.021 & -4.63$^{**}$ && -0.024 & -5.66$^{**}$  \\
 36 & -0.021 & -1.03$~~$ && -0.016 & -1.86$^{~~}$ && -0.021 & -3.07$^{**}$ && -0.025 & -4.46$^{**}$ && -0.023 & -3.85$^{**}$ && -0.016 & -2.49$^{*~}$ && -0.014 & -2.65$^{**}$ && -0.019 & -4.79$^{**}$ && -0.028 & -6.90$^{**}$  \\
 42 & -0.013 & -0.69$~~$ && -0.027 & -2.85$^{**}$ && -0.028 & -4.34$^{**}$ && -0.026 & -4.57$^{**}$ && -0.022 & -3.45$^{**}$ && -0.018 & -2.87$^{**}$ && -0.018 & -3.56$^{**}$ && -0.026 & -6.00$^{**}$ && -0.035 & -8.04$^{**}$  \\
 48 & -0.041 & -2.10$^{*~}$ && -0.045 & -4.40$^{**}$ && -0.034 & -5.14$^{**}$ && -0.029 & -4.99$^{**}$ && -0.028 & -4.63$^{**}$ && -0.025 & -4.63$^{**}$ && -0.021 & -3.58$^{**}$ && -0.031 & -6.31$^{**}$ && -0.037 & -8.01$^{**}$  \\
   \hline
   \end{tabular}
   \begin{tablenotes}
   \small
     \item This table reports the differences of the average annualized returns and the corresponding t-statistics adjusted for heteroscedasticity and autocorrelation of two winner strategies that are different only in the grouping methods for SHSE stocks. The three panels are for the loser, winner and contrarian portfolios, respectively. In the first row, $G_3$, $G_5$ and $G_{10}$ stand for tertile, quintile and decile groupings. The sample period is January 1997 to  December 2012. The superscripts * and ** denote the significance at 5\% and 1\% levels, respectively.
   \end{tablenotes}
\end{threeparttable}
\end{table}
\end{landscape}

\setlength\tabcolsep{2.0pt}
\begin{landscape}
\begin{table}[htb]
\centering
\begin{threeparttable}[b]
   \small
   \caption{The return difference of winner portfolios formed based on different grouping ways of the SZSE stocks.}
   \label{TBS:Empirics:Diff:Group:WIN:SZSE}
   \begin{tabular}{ccccccccccccccccccccccccccccc}
   \hline
          & \multicolumn{2}{c}{$K=1$} && \multicolumn{2}{c}{6} && \multicolumn{2}{c}{12} && \multicolumn{2}{c}{18} && \multicolumn{2}{c}{24} && \multicolumn{2}{c}{30} && \multicolumn{2}{c}{36} && \multicolumn{2}{c}{42} && \multicolumn{2}{c}{48} \\
   \cline{2-3} \cline{5-6} \cline{8-9} \cline{11-12} \cline{14-15} \cline{17-18} \cline{20-21} \cline{23-24} \cline{26-27}
     $J$  & $\Delta{R}$ & t-stat && $\Delta{R}$ & t-stat && $\Delta{R}$ & t-stat && $\Delta{R}$ & t-stat && $\Delta{R}$ & t-stat && $\Delta{R}$ & t-stat && $\Delta{R}$ & t-stat && $\Delta{R}$ & t-stat && $\Delta{R}$ & t-stat \\
   \hline
   \multicolumn{27}{l}{\textit{Panel A: $G_{5}-G_{3}$}} \\
 1& -0.016 & -1.76$^{~~}$ && -0.014 & -3.78$^{**}$ && -0.014 & -4.85$^{**}$ && -0.012 & -4.21$^{**}$ && -0.011 & -3.69$^{**}$ && -0.008 & -3.18$^{**}$ && -0.008 & -3.46$^{**}$ && -0.008 & -3.66$^{**}$ && -0.008 & -3.14$^{**}$  \\
 6& -0.005 & -0.50$~~$ && -0.003 & -0.65$~~$ && -0.006 & -1.70$^{~~}$ && -0.006 & -1.61$~~$ && -0.008 & -2.43$^{*~}$ && -0.011 & -3.24$^{**}$ && -0.006 & -2.14$^{*~}$ && -0.000 & -0.12$~~$ && -0.003 & -0.87$~~$  \\
 12& 0.006 &  0.54$~~$ && -0.003 & -0.86$~~$ && -0.009 & -2.66$^{**}$ && -0.010 & -2.29$^{*~}$ && -0.015 & -4.19$^{**}$ && -0.014 & -4.55$^{**}$ && -0.008 & -2.39$^{*~}$ && -0.005 & -1.43$~~$ && -0.007 & -2.31$^{*~}$  \\
 18& -0.017 & -1.58$~~$ && -0.012 & -2.73$^{**}$ && -0.017 & -5.44$^{**}$ && -0.017 & -4.74$^{**}$ && -0.014 & -3.76$^{**}$ && -0.012 & -3.38$^{**}$ && -0.012 & -3.20$^{**}$ && -0.012 & -3.32$^{**}$ && -0.013 & -4.95$^{**}$  \\
 24& -0.012 & -1.14$~~$ && -0.008 & -1.58$~~$ && -0.013 & -3.86$^{**}$ && -0.013 & -2.86$^{**}$ && -0.013 & -2.72$^{**}$ && -0.011 & -2.83$^{**}$ && -0.012 & -3.12$^{**}$ && -0.012 & -3.88$^{**}$ && -0.014 & -4.67$^{**}$  \\
 30& 0.004 &  0.31$~~$ && -0.001 & -0.22$~~$ && -0.011 & -2.54$^{*~}$ && -0.008 & -1.64$~~$ && -0.011 & -2.13$^{*~}$ && -0.013 & -3.25$^{**}$ && -0.011 & -3.08$^{**}$ && -0.008 & -2.68$^{**}$ && -0.009 & -2.79$^{**}$  \\
 36& -0.001 & -0.09$~~$ && -0.004 & -0.79$~~$ && -0.005 & -1.01$~~$ && -0.004 & -0.99$~~$ && -0.009 & -2.13$^{*~}$ && -0.009 & -2.64$^{**}$ && -0.008 & -3.08$^{**}$ && -0.008 & -3.16$^{**}$ && -0.009 & -3.51$^{**}$  \\
 42& 0.014 &  1.18$~~$ && -0.004 & -0.64$~~$ && -0.002 & -0.41$~~$ && -0.004 & -1.00$~~$ && -0.006 & -1.27$~~$ && -0.008 & -2.55$^{*~}$ && -0.010 & -3.01$^{**}$ && -0.009 & -2.86$^{**}$ && -0.009 & -3.33$^{**}$  \\
 48& -0.013 & -0.99$~~$ && -0.010 & -1.52$~~$ && -0.009 & -2.03$^{*~}$ && -0.006 & -1.26$~~$ && -0.008 & -1.93$^{~~}$ && -0.015 & -4.12$^{**}$ && -0.015 & -4.94$^{**}$ && -0.010 & -3.17$^{**}$ && -0.011 & -4.41$^{**}$  \\
    \vspace{-3mm}\\
   \multicolumn{27}{l}{\textit{Panel B: $G_{10}-G_{5}$}} \\
 1& -0.003 & -0.25$~~$ && -0.003 & -0.62$~~$ && -0.004 & -0.89$~~$ && -0.001 & -0.15$~~$ && -0.004 & -0.74$~~$ && -0.004 & -0.76$~~$ && -0.003 & -0.71$~~$ && -0.001 & -0.34$~~$ && -0.001 & -0.22$~~$  \\
 6& -0.014 & -1.02$~~$ && -0.011 & -1.76$^{~~}$ && -0.011 & -2.11$^{*~}$ && -0.007 & -1.33$~~$ && -0.011 & -2.43$^{*~}$ && -0.016 & -3.52$^{**}$ && -0.017 & -4.48$^{**}$ && -0.010 & -2.24$^{*~}$ && -0.011 & -2.56$^{*~}$  \\
 12& -0.003 & -0.23$~~$ && -0.002 & -0.31$~~$ && -0.005 & -1.09$~~$ && -0.009 & -2.28$^{*~}$ && -0.014 & -2.67$^{**}$ && -0.011 & -2.20$^{*~}$ && -0.007 & -1.51$~~$ && -0.007 & -1.78$^{~~}$ && -0.008 & -2.27$^{*~}$  \\
 18& 0.002 &  0.14$~~$ && -0.001 & -0.19$~~$ && -0.004 & -0.81$~~$ && -0.008 & -1.80$^{~~}$ && -0.009 & -1.88$^{~~}$ && -0.006 & -1.66$^{~~}$ && -0.003 & -0.62$~~$ && -0.004 & -0.98$~~$ && -0.009 & -2.39$^{*~}$  \\
 24& 0.012 &  0.65$~~$ && 0.001 &  0.21$~~$ && -0.004 & -0.76$~~$ && -0.006 & -1.12$~~$ && -0.000 & -0.03$~~$ && -0.004 & -0.84$~~$ && -0.007 & -1.63$~~$ && -0.010 & -2.13$^{*~}$ && -0.011 & -2.13$^{*~}$  \\
 30& -0.005 & -0.33$~~$ && -0.007 & -0.93$~~$ && -0.001 & -0.24$~~$ && 0.004 &  0.73$~~$ && 0.002 &  0.31$~~$ && -0.011 & -2.29$^{*~}$ && -0.017 & -4.13$^{**}$ && -0.013 & -3.01$^{**}$ && -0.011 & -2.41$^{*~}$  \\
 36& 0.024 &  1.59$~~$ && 0.006 &  0.79$~~$ && 0.007 &  1.03$~~$ && 0.005 &  0.67$~~$ && -0.003 & -0.49$~~$ && -0.014 & -3.06$^{**}$ && -0.015 & -3.74$^{**}$ && -0.014 & -3.43$^{**}$ && -0.016 & -4.37$^{**}$  \\
 42& -0.004 & -0.27$~~$ && 0.017 &  2.10$^{*~}$ && 0.012 &  1.33$~~$ && 0.003 &  0.37$~~$ && -0.009 & -1.64$~~$ && -0.017 & -3.56$^{**}$ && -0.016 & -3.57$^{**}$ && -0.013 & -2.77$^{**}$ && -0.018 & -3.77$^{**}$  \\
 48& -0.006 & -0.36$~~$ && 0.008 &  0.85$~~$ && 0.004 &  0.40$~~$ && -0.007 & -1.02$~~$ && -0.016 & -3.03$^{**}$ && -0.015 & -3.24$^{**}$ && -0.013 & -3.61$^{**}$ && -0.015 & -3.65$^{**}$ && -0.018 & -3.73$^{**}$  \\
   \vspace{-3mm}\\
   \multicolumn{27}{l}{\textit{Panel C: $G_{10}-G_{3}$}} \\
 1 & -0.019 & -1.00$~~$ && -0.017 & -2.33$^{*~}$ && -0.018 & -3.00$^{**}$ && -0.013 & -1.93$^{~~}$ && -0.015 & -2.20$^{*~}$ && -0.012 & -2.07$^{*~}$ && -0.011 & -2.02$^{*~}$ && -0.010 & -2.05$^{*~}$ && -0.008 & -1.69$^{~~}$  \\
 6 & -0.019 & -0.92$~~$ && -0.014 & -1.56$~~$ && -0.017 & -2.44$^{*~}$ && -0.013 & -1.80$^{~~}$ && -0.020 & -2.94$^{**}$ && -0.027 & -4.22$^{**}$ && -0.023 & -4.53$^{**}$ && -0.010 & -1.60$~~$ && -0.014 & -2.20$^{*~}$  \\
 12 & 0.003 &  0.13$~~$ && -0.005 & -0.59$~~$ && -0.015 & -2.06$^{*~}$ && -0.020 & -2.90$^{**}$ && -0.029 & -4.61$^{**}$ && -0.025 & -3.90$^{**}$ && -0.015 & -2.46$^{*~}$ && -0.012 & -1.98$^{~~}$ && -0.015 & -2.82$^{**}$  \\
 18 & -0.015 & -0.67$~~$ && -0.013 & -1.37$~~$ && -0.022 & -3.07$^{**}$ && -0.025 & -3.86$^{**}$ && -0.023 & -3.23$^{**}$ && -0.018 & -3.02$^{**}$ && -0.014 & -2.19$^{*~}$ && -0.016 & -2.40$^{*~}$ && -0.022 & -4.19$^{**}$  \\
 24 & -0.001 & -0.02$~~$ && -0.006 & -0.63$~~$ && -0.017 & -2.36$^{*~}$ && -0.019 & -2.37$^{*~}$ && -0.013 & -1.47$~~$ && -0.015 & -2.17$^{*~}$ && -0.019 & -2.80$^{**}$ && -0.022 & -3.36$^{**}$ && -0.025 & -3.73$^{**}$  \\
 30 & -0.002 & -0.07$~~$ && -0.008 & -0.71$~~$ && -0.012 & -1.38$~~$ && -0.004 & -0.43$~~$ && -0.009 & -0.85$~~$ && -0.024 & -3.16$^{**}$ && -0.028 & -4.37$^{**}$ && -0.021 & -3.50$^{**}$ && -0.020 & -3.37$^{**}$  \\
 36 & 0.023 &  0.94$~~$ && 0.001 &  0.10$~~$ && 0.002 &  0.21$~~$ && 0.001 &  0.06$~~$ && -0.012 & -1.27$~~$ && -0.023 & -3.65$^{**}$ && -0.023 & -4.44$^{**}$ && -0.022 & -4.15$^{**}$ && -0.025 & -5.14$^{**}$  \\
 42 & 0.010 &  0.43$~~$ && 0.014 &  1.15$~~$ && 0.010 &  0.87$~~$ && -0.002 & -0.20$~~$ && -0.015 & -1.73$^{~~}$ && -0.026 & -4.17$^{**}$ && -0.025 & -4.56$^{**}$ && -0.021 & -3.60$^{**}$ && -0.027 & -4.98$^{**}$  \\
 48 & -0.018 & -0.76$~~$ && -0.002 & -0.15$~~$ && -0.006 & -0.46$~~$ && -0.013 & -1.31$~~$ && -0.024 & -3.21$^{**}$ && -0.030 & -4.86$^{**}$ && -0.028 & -5.68$^{**}$ && -0.025 & -4.95$^{**}$ && -0.030 & -5.36$^{**}$  \\
   \hline
   \end{tabular}
   \begin{tablenotes}
   \small
     \item This table reports the differences of the average annualized returns and the corresponding t-statistics adjusted for heteroscedasticity and autocorrelation of two winner strategies that are different only in the grouping methods for SZSE stocks. The three panels are for the loser, winner and contrarian portfolios, respectively. In the first row, $G_3$, $G_5$ and $G_{10}$ stand for tertile, quintile and decile groupings. The sample period is January 1997 to  December 2012. The superscripts * and ** denote the significance at 5\% and 1\% levels, respectively.
   \end{tablenotes}
\end{threeparttable}
\end{table}
\end{landscape}

\setlength\tabcolsep{2.0pt}
\begin{landscape}
\begin{table}[htb]
\centering
\begin{threeparttable}[b]
   \small
   \caption{The return difference of contrarian portfolios formed based on different grouping ways of the SHSE stocks.}
   \label{TBS:Empirics:Diff:Group:CON:SHSE}
   \begin{tabular}{ccccccccccccccccccccccccccccc}
   \hline
          & \multicolumn{2}{c}{$K=1$} && \multicolumn{2}{c}{6} && \multicolumn{2}{c}{12} && \multicolumn{2}{c}{18} && \multicolumn{2}{c}{24} && \multicolumn{2}{c}{30} && \multicolumn{2}{c}{36} && \multicolumn{2}{c}{42} && \multicolumn{2}{c}{48} \\
   \cline{2-3} \cline{5-6} \cline{8-9} \cline{11-12} \cline{14-15} \cline{17-18} \cline{20-21} \cline{23-24} \cline{26-27}
     $J$  & $\Delta{R}$ & t-stat && $\Delta{R}$ & t-stat && $\Delta{R}$ & t-stat && $\Delta{R}$ & t-stat && $\Delta{R}$ & t-stat && $\Delta{R}$ & t-stat && $\Delta{R}$ & t-stat && $\Delta{R}$ & t-stat && $\Delta{R}$ & t-stat \\
   \hline
   \multicolumn{27}{l}{\textit{Panel A: $G_{5}-G_{3}$}} \\
 1& 0.019 &  1.95$^{~~}$ && 0.009 &  2.06$^{*~}$ && 0.005 &  1.23$~~$ && 0.006 &  1.71$^{~~}$ && 0.005 &  1.24$~~$ && 0.009 &  2.24$^{*~}$ && 0.005 &  1.47$~~$ && 0.005 &  1.54$~~$ && 0.004 &  1.12$~~$  \\
 6& 0.016 &  1.37$~~$ && -0.002 & -0.38$~~$ && 0.002 &  0.57$~~$ && 0.004 &  0.98$~~$ && 0.014 &  4.00$^{**}$ && 0.009 &  2.74$^{**}$ && 0.005 &  1.61$~~$ && 0.009 &  2.72$^{**}$ && 0.010 &  2.92$^{**}$  \\
 12& 0.007 &  0.55$~~$ && 0.002 &  0.44$~~$ && 0.009 &  2.05$^{*~}$ && 0.009 &  2.52$^{*~}$ && 0.011 &  3.26$^{**}$ && 0.011 &  2.70$^{**}$ && 0.007 &  1.61$~~$ && 0.007 &  1.80$^{~~}$ && 0.009 &  2.22$^{*~}$  \\
 18& 0.021 &  1.82$^{~~}$ && 0.010 &  1.82$^{~~}$ && 0.020 &  4.65$^{**}$ && 0.021 &  5.41$^{**}$ && 0.022 &  5.58$^{**}$ && 0.019 &  5.29$^{**}$ && 0.019 &  5.26$^{**}$ && 0.021 &  5.94$^{**}$ && 0.023 &  6.00$^{**}$  \\
 24& 0.012 &  0.96$~~$ && 0.009 &  1.85$^{~~}$ && 0.022 &  4.86$^{**}$ && 0.025 &  6.86$^{**}$ && 0.028 &  7.36$^{**}$ && 0.024 &  7.73$^{**}$ && 0.021 &  5.86$^{**}$ && 0.024 &  7.49$^{**}$ && 0.025 &  7.52$^{**}$  \\
 30& 0.022 &  1.61$~~$ && 0.016 &  2.88$^{**}$ && 0.022 &  5.15$^{**}$ && 0.025 &  6.71$^{**}$ && 0.028 &  6.68$^{**}$ && 0.022 &  5.19$^{**}$ && 0.022 &  4.99$^{**}$ && 0.024 &  5.72$^{**}$ && 0.028 &  6.66$^{**}$  \\
 36& 0.012 &  0.88$~~$ && 0.020 &  3.66$^{**}$ && 0.026 &  5.93$^{**}$ && 0.033 &  8.12$^{**}$ && 0.033 &  7.60$^{**}$ && 0.025 &  5.78$^{**}$ && 0.022 &  5.32$^{**}$ && 0.026 &  6.37$^{**}$ && 0.032 &  7.63$^{**}$  \\
 42& 0.028 &  1.84$^{~~}$ && 0.025 &  3.93$^{**}$ && 0.032 &  5.94$^{**}$ && 0.040 &  8.44$^{**}$ && 0.033 &  7.52$^{**}$ && 0.028 &  7.22$^{**}$ && 0.029 &  7.45$^{**}$ && 0.030 &  6.39$^{**}$ && 0.039 &  7.41$^{**}$  \\
 48& 0.034 &  2.16$^{*~}$ && 0.040 &  6.39$^{**}$ && 0.038 &  7.72$^{**}$ && 0.041 &  8.13$^{**}$ && 0.034 &  7.76$^{**}$ && 0.028 &  6.08$^{**}$ && 0.031 &  6.36$^{**}$ && 0.035 &  6.60$^{**}$ && 0.038 &  6.51$^{**}$  \\
    \vspace{-3mm}\\
   \multicolumn{27}{l}{\textit{Panel B: $G_{10}-G_{5}$}} \\
 1& 0.007 &  0.51$~~$ && 0.004 &  0.57$~~$ && 0.000 &  0.08$~~$ && -0.002 & -0.24$~~$ && 0.004 &  0.74$~~$ && 0.002 &  0.34$~~$ && 0.004 &  0.74$~~$ && 0.000 &  0.08$~~$ && -0.001 & -0.28$~~$  \\
 6& 0.014 &  0.85$~~$ && 0.003 &  0.44$~~$ && 0.005 &  0.97$~~$ && 0.008 &  1.56$~~$ && 0.007 &  1.41$~~$ && 0.011 &  2.14$^{*~}$ && 0.009 &  1.69$^{~~}$ && 0.006 &  1.16$~~$ && 0.005 &  0.86$~~$  \\
 12& 0.001 &  0.09$~~$ && -0.010 & -1.31$~~$ && -0.011 & -1.93$^{~~}$ && 0.000 &  0.06$~~$ && 0.007 &  1.46$~~$ && 0.009 &  1.73$^{~~}$ && 0.018 &  3.17$^{**}$ && 0.018 &  2.90$^{**}$ && 0.028 &  4.98$^{**}$  \\
 18& 0.007 &  0.38$~~$ && -0.005 & -0.61$~~$ && 0.001 &  0.15$~~$ && 0.005 &  0.67$~~$ && 0.004 &  0.70$~~$ && 0.009 &  1.35$~~$ && 0.012 &  1.83$^{~~}$ && 0.020 &  3.38$^{**}$ && 0.025 &  4.61$^{**}$  \\
 24& 0.009 &  0.51$~~$ && 0.004 &  0.46$~~$ && 0.005 &  0.75$~~$ && 0.009 &  1.36$~~$ && 0.006 &  0.86$~~$ && 0.015 &  2.44$^{*~}$ && 0.017 &  3.03$^{**}$ && 0.020 &  3.18$^{**}$ && 0.027 &  5.06$^{**}$  \\
 30& 0.035 &  2.00$^{*~}$ && 0.022 &  2.91$^{**}$ && 0.017 &  3.01$^{**}$ && 0.020 &  3.20$^{**}$ && 0.021 &  3.73$^{**}$ && 0.030 &  5.83$^{**}$ && 0.029 &  5.84$^{**}$ && 0.031 &  6.10$^{**}$ && 0.036 &  7.16$^{**}$  \\
 36& 0.036 &  1.91$^{~~}$ && 0.011 &  1.27$~~$ && 0.012 &  1.87$^{~~}$ && 0.014 &  2.36$^{*~}$ && 0.020 &  3.16$^{**}$ && 0.025 &  4.62$^{**}$ && 0.030 &  5.95$^{**}$ && 0.035 &  7.35$^{**}$ && 0.042 &  9.17$^{**}$  \\
 42& 0.022 &  1.26$~~$ && 0.028 &  3.49$^{**}$ && 0.024 &  3.36$^{**}$ && 0.022 &  3.24$^{**}$ && 0.031 &  4.62$^{**}$ && 0.036 &  6.14$^{**}$ && 0.035 &  5.82$^{**}$ && 0.049 &  7.66$^{**}$ && 0.055 & 10.06$^{**}$  \\
 48& 0.048 &  2.26$^{*~}$ && 0.033 &  3.07$^{**}$ && 0.025 &  2.84$^{**}$ && 0.026 &  3.43$^{**}$ && 0.041 &  5.56$^{**}$ && 0.046 &  7.38$^{**}$ && 0.042 &  5.37$^{**}$ && 0.053 &  7.72$^{**}$ && 0.062 &  9.34$^{**}$  \\
   \vspace{-3mm}\\
   \multicolumn{27}{l}{\textit{Panel C: $G_{10}-G_{3}$}} \\
 1 & 0.027 &  1.31$~~$ && 0.013 &  1.51$~~$ && 0.006 &  0.78$~~$ && 0.004 &  0.59$~~$ && 0.009 &  1.17$~~$ && 0.011 &  1.28$~~$ && 0.009 &  1.25$~~$ && 0.005 &  0.91$~~$ && 0.003 &  0.39$~~$  \\
 6 & 0.029 &  1.23$~~$ && 0.002 &  0.15$~~$ && 0.007 &  1.00$~~$ && 0.011 &  1.73$^{~~}$ && 0.021 &  3.15$^{**}$ && 0.020 &  3.11$^{**}$ && 0.014 &  2.05$^{*~}$ && 0.015 &  2.15$^{*~}$ && 0.015 &  1.97$^{~~}$  \\
 12 & 0.008 &  0.34$~~$ && -0.008 & -0.73$~~$ && -0.002 & -0.21$~~$ && 0.010 &  1.37$~~$ && 0.019 &  2.81$^{**}$ && 0.020 &  2.71$^{**}$ && 0.025 &  3.02$^{**}$ && 0.025 &  2.94$^{**}$ && 0.037 &  4.94$^{**}$  \\
 18 & 0.029 &  1.05$~~$ && 0.004 &  0.34$~~$ && 0.021 &  2.18$^{*~}$ && 0.026 &  3.10$^{**}$ && 0.026 &  3.65$^{**}$ && 0.029 &  3.36$^{**}$ && 0.031 &  3.66$^{**}$ && 0.041 &  5.48$^{**}$ && 0.048 &  6.67$^{**}$  \\
 24 & 0.021 &  0.81$~~$ && 0.013 &  1.17$~~$ && 0.027 &  3.20$^{**}$ && 0.033 &  4.44$^{**}$ && 0.034 &  4.27$^{**}$ && 0.039 &  5.45$^{**}$ && 0.039 &  5.02$^{**}$ && 0.044 &  5.69$^{**}$ && 0.052 &  7.43$^{**}$  \\
 30 & 0.057 &  2.09$^{*~}$ && 0.037 &  3.61$^{**}$ && 0.039 &  4.97$^{**}$ && 0.044 &  6.37$^{**}$ && 0.048 &  6.51$^{**}$ && 0.052 &  7.02$^{**}$ && 0.051 &  6.37$^{**}$ && 0.054 &  7.35$^{**}$ && 0.064 &  8.76$^{**}$  \\
 36 & 0.049 &  1.75$^{~~}$ && 0.031 &  2.62$^{*~}$ && 0.038 &  4.45$^{**}$ && 0.047 &  6.32$^{**}$ && 0.053 &  6.08$^{**}$ && 0.050 &  5.89$^{**}$ && 0.052 &  6.75$^{**}$ && 0.062 &  8.74$^{**}$ && 0.074 &  9.69$^{**}$  \\
 42 & 0.050 &  1.88$^{~~}$ && 0.053 &  4.39$^{**}$ && 0.056 &  5.72$^{**}$ && 0.062 &  7.14$^{**}$ && 0.064 &  7.33$^{**}$ && 0.064 &  7.96$^{**}$ && 0.065 &  7.82$^{**}$ && 0.079 &  9.46$^{**}$ && 0.094 & 10.63$^{**}$  \\
 48 & 0.082 &  2.74$^{**}$ && 0.073 &  4.96$^{**}$ && 0.063 &  5.42$^{**}$ && 0.067 &  7.23$^{**}$ && 0.074 &  7.79$^{**}$ && 0.074 &  7.76$^{**}$ && 0.073 &  6.54$^{**}$ && 0.088 &  8.51$^{**}$ && 0.100 &  9.22$^{**}$  \\
   \hline
   \end{tabular}
   \begin{tablenotes}
   \small
     \item This table reports the differences of the average annualized returns and the corresponding t-statistics adjusted for heteroscedasticity and autocorrelation of two contrarian strategies that are different only in the grouping methods for SHSE stocks. The three panels are for the loser, winner and contrarian portfolios, respectively. In the first row, $G_3$, $G_5$ and $G_{10}$ stand for tertile, quintile and decile groupings. The sample period is January 1997 to  December 2012. The superscripts * and ** denote the significance at 5\% and 1\% levels, respectively.
   \end{tablenotes}
\end{threeparttable}
\end{table}
\end{landscape}

\setlength\tabcolsep{2.0pt}
\begin{landscape}
\begin{table}[htb]
\centering
\begin{threeparttable}[b]
   \small
   \caption{The return difference of contrarian portfolios formed based on different grouping ways of the SZSE stocks.}
   \label{TBS:Empirics:Diff:Group:CON:SZSE}
   \begin{tabular}{ccccccccccccccccccccccccccccc}
   \hline
          & \multicolumn{2}{c}{$K=1$} && \multicolumn{2}{c}{6} && \multicolumn{2}{c}{12} && \multicolumn{2}{c}{18} && \multicolumn{2}{c}{24} && \multicolumn{2}{c}{30} && \multicolumn{2}{c}{36} && \multicolumn{2}{c}{42} && \multicolumn{2}{c}{48} \\
   \cline{2-3} \cline{5-6} \cline{8-9} \cline{11-12} \cline{14-15} \cline{17-18} \cline{20-21} \cline{23-24} \cline{26-27}
     $J$  & $\Delta{R}$ & t-stat && $\Delta{R}$ & t-stat && $\Delta{R}$ & t-stat && $\Delta{R}$ & t-stat && $\Delta{R}$ & t-stat && $\Delta{R}$ & t-stat && $\Delta{R}$ & t-stat && $\Delta{R}$ & t-stat && $\Delta{R}$ & t-stat \\
   \hline
   \multicolumn{27}{l}{\textit{Panel A: $G_{5}-G_{3}$}} \\
 1& 0.011 &  1.04$~~$ && 0.005 &  1.09$~~$ && 0.006 &  1.40$~~$ && 0.003 &  0.71$~~$ && 0.002 &  0.48$~~$ && 0.006 &  1.77$^{~~}$ && 0.005 &  1.80$^{~~}$ && 0.008 &  2.29$^{*~}$ && 0.006 &  1.74$^{~~}$  \\
 6& 0.004 &  0.34$~~$ && -0.006 & -0.98$~~$ && -0.004 & -0.77$~~$ && -0.004 & -0.80$~~$ && 0.001 &  0.13$~~$ && 0.009 &  1.75$^{~~}$ && 0.005 &  1.39$~~$ && 0.003 &  0.60$~~$ && 0.003 &  0.82$~~$  \\
 12& 0.005 &  0.38$~~$ && 0.000 &  0.06$~~$ && 0.009 &  1.56$~~$ && 0.009 &  1.34$~~$ && 0.024 &  3.98$^{**}$ && 0.025 &  5.32$^{**}$ && 0.017 &  3.89$^{**}$ && 0.015 &  3.22$^{**}$ && 0.018 &  3.90$^{**}$  \\
 18& 0.033 &  2.08$^{*~}$ && 0.018 &  2.79$^{**}$ && 0.028 &  5.62$^{**}$ && 0.026 &  5.01$^{**}$ && 0.030 &  6.53$^{**}$ && 0.030 &  6.57$^{**}$ && 0.027 &  5.87$^{**}$ && 0.027 &  5.49$^{**}$ && 0.030 &  6.61$^{**}$  \\
 24& 0.021 &  1.42$~~$ && 0.010 &  1.69$^{~~}$ && 0.023 &  4.30$^{**}$ && 0.021 &  3.34$^{**}$ && 0.027 &  4.38$^{**}$ && 0.028 &  4.77$^{**}$ && 0.027 &  4.85$^{**}$ && 0.026 &  5.64$^{**}$ && 0.027 &  5.97$^{**}$  \\
 30& 0.007 &  0.44$~~$ && 0.011 &  1.44$~~$ && 0.021 &  3.66$^{**}$ && 0.022 &  3.39$^{**}$ && 0.030 &  4.41$^{**}$ && 0.033 &  6.27$^{**}$ && 0.028 &  6.02$^{**}$ && 0.024 &  6.15$^{**}$ && 0.024 &  5.14$^{**}$  \\
 36& 0.023 &  1.34$~~$ && 0.022 &  2.65$^{**}$ && 0.023 &  3.02$^{**}$ && 0.016 &  2.52$^{*~}$ && 0.029 &  5.52$^{**}$ && 0.031 &  6.63$^{**}$ && 0.029 &  7.18$^{**}$ && 0.029 &  7.92$^{**}$ && 0.029 &  7.11$^{**}$  \\
 42& 0.017 &  0.98$~~$ && 0.021 &  2.52$^{*~}$ && 0.019 &  2.40$^{*~}$ && 0.021 &  3.19$^{**}$ && 0.025 &  4.24$^{**}$ && 0.027 &  5.16$^{**}$ && 0.028 &  5.96$^{**}$ && 0.028 &  6.84$^{**}$ && 0.030 &  6.84$^{**}$  \\
 48& 0.041 &  2.05$^{*~}$ && 0.034 &  3.84$^{**}$ && 0.028 &  4.00$^{**}$ && 0.028 &  5.08$^{**}$ && 0.038 &  8.22$^{**}$ && 0.041 &  9.38$^{**}$ && 0.043 &  9.24$^{**}$ && 0.037 &  6.78$^{**}$ && 0.039 &  8.73$^{**}$  \\
    \vspace{-3mm}\\
   \multicolumn{27}{l}{\textit{Panel B: $G_{10}-G_{5}$}} \\
 1& -0.006 & -0.40$~~$ && 0.002 &  0.24$~~$ && -0.007 & -1.22$~~$ && -0.008 & -1.15$~~$ && -0.006 & -0.89$~~$ && -0.004 & -0.56$~~$ && 0.001 &  0.15$~~$ && 0.002 &  0.29$~~$ && -0.001 & -0.21$~~$  \\
 6& 0.012 &  0.60$~~$ && -0.003 & -0.35$~~$ && 0.005 &  0.70$~~$ && 0.001 &  0.12$~~$ && 0.009 &  1.38$~~$ && 0.024 &  3.92$^{**}$ && 0.024 &  4.75$^{**}$ && 0.016 &  2.93$^{**}$ && 0.019 &  3.33$^{**}$  \\
 12& 0.025 &  1.23$~~$ && -0.008 & -0.91$~~$ && 0.000 &  0.04$~~$ && 0.003 &  0.32$~~$ && 0.017 &  2.27$^{*~}$ && 0.025 &  3.51$^{**}$ && 0.020 &  2.93$^{**}$ && 0.018 &  3.00$^{**}$ && 0.024 &  4.01$^{**}$  \\
 18& -0.001 & -0.06$~~$ && -0.006 & -0.68$~~$ && -0.001 & -0.15$~~$ && 0.004 &  0.53$~~$ && 0.016 &  1.85$^{~~}$ && 0.021 &  2.82$^{**}$ && 0.022 &  3.09$^{**}$ && 0.023 &  3.45$^{**}$ && 0.029 &  4.36$^{**}$  \\
 24& 0.014 &  0.57$~~$ && 0.002 &  0.20$~~$ && 0.001 &  0.06$~~$ && 0.014 &  1.56$~~$ && 0.026 &  2.92$^{**}$ && 0.036 &  4.47$^{**}$ && 0.037 &  5.24$^{**}$ && 0.041 &  6.26$^{**}$ && 0.042 &  6.74$^{**}$  \\
 30& 0.021 &  0.95$~~$ && 0.020 &  2.02$^{*~}$ && 0.017 &  1.63$~~$ && 0.012 &  1.01$~~$ && 0.025 &  2.49$^{*~}$ && 0.046 &  6.12$^{**}$ && 0.050 &  7.67$^{**}$ && 0.045 &  7.05$^{**}$ && 0.045 &  6.26$^{**}$  \\
 36& -0.001 & -0.03$~~$ && 0.018 &  1.51$~~$ && 0.015 &  1.28$~~$ && 0.021 &  1.82$^{~~}$ && 0.031 &  2.96$^{**}$ && 0.046 &  5.19$^{**}$ && 0.044 &  5.87$^{**}$ && 0.048 &  6.78$^{**}$ && 0.054 &  7.83$^{**}$  \\
 42& 0.038 &  1.28$~~$ && 0.016 &  1.08$~~$ && 0.019 &  1.39$~~$ && 0.025 &  2.05$^{*~}$ && 0.044 &  4.40$^{**}$ && 0.058 &  6.67$^{**}$ && 0.058 &  8.32$^{**}$ && 0.059 &  8.66$^{**}$ && 0.060 &  8.47$^{**}$  \\
 48& 0.039 &  1.13$~~$ && 0.019 &  1.29$~~$ && 0.021 &  1.55$~~$ && 0.025 &  1.83$^{~~}$ && 0.039 &  3.31$^{**}$ && 0.050 &  5.36$^{**}$ && 0.056 &  8.22$^{**}$ && 0.055 &  9.11$^{**}$ && 0.054 &  7.85$^{**}$  \\
   \vspace{-3mm}\\
   \multicolumn{27}{l}{\textit{Panel C: $G_{10}-G_{3}$}} \\
 1 & 0.005 &  0.24$~~$ && 0.007 &  0.68$~~$ && -0.001 & -0.18$~~$ && -0.005 & -0.55$~~$ && -0.004 & -0.48$~~$ && 0.002 &  0.29$~~$ && 0.006 &  0.84$~~$ && 0.009 &  1.36$~~$ && 0.005 &  0.64$~~$  \\
 6 & 0.016 &  0.57$~~$ && -0.009 & -0.72$~~$ && 0.001 &  0.11$~~$ && -0.003 & -0.31$~~$ && 0.010 &  1.05$~~$ && 0.033 &  3.51$^{**}$ && 0.029 &  4.51$^{**}$ && 0.018 &  2.36$^{*~}$ && 0.022 &  2.73$^{**}$  \\
 12 & 0.029 &  1.01$~~$ && -0.008 & -0.63$~~$ && 0.009 &  0.80$~~$ && 0.012 &  0.93$~~$ && 0.041 &  4.15$^{**}$ && 0.050 &  5.40$^{**}$ && 0.037 &  4.26$^{**}$ && 0.033 &  3.77$^{**}$ && 0.042 &  4.74$^{**}$  \\
 18 & 0.032 &  1.01$~~$ && 0.012 &  0.95$~~$ && 0.027 &  2.52$^{*~}$ && 0.031 &  2.76$^{**}$ && 0.046 &  4.29$^{**}$ && 0.051 &  5.34$^{**}$ && 0.049 &  5.12$^{**}$ && 0.050 &  5.06$^{**}$ && 0.059 &  6.38$^{**}$  \\
 24 & 0.035 &  1.04$~~$ && 0.012 &  0.86$~~$ && 0.024 &  2.00$^{*~}$ && 0.036 &  2.67$^{**}$ && 0.053 &  4.42$^{**}$ && 0.064 &  6.07$^{**}$ && 0.064 &  6.29$^{**}$ && 0.067 &  7.41$^{**}$ && 0.069 &  7.87$^{**}$  \\
 30 & 0.028 &  0.82$~~$ && 0.031 &  1.99$^{*~}$ && 0.038 &  2.63$^{**}$ && 0.034 &  2.06$^{*~}$ && 0.055 &  3.71$^{**}$ && 0.079 &  7.18$^{**}$ && 0.078 &  7.91$^{**}$ && 0.069 &  7.94$^{**}$ && 0.068 &  7.21$^{**}$  \\
 36 & 0.022 &  0.65$~~$ && 0.040 &  2.23$^{*~}$ && 0.038 &  2.18$^{*~}$ && 0.037 &  2.23$^{*~}$ && 0.060 &  4.42$^{**}$ && 0.077 &  6.81$^{**}$ && 0.074 &  7.90$^{**}$ && 0.077 &  9.10$^{**}$ && 0.084 &  9.23$^{**}$  \\
 42 & 0.055 &  1.31$~~$ && 0.037 &  1.80$^{~~}$ && 0.038 &  2.05$^{*~}$ && 0.046 &  2.81$^{**}$ && 0.069 &  4.91$^{**}$ && 0.085 &  7.12$^{**}$ && 0.086 &  8.95$^{**}$ && 0.088 &  9.81$^{**}$ && 0.090 & 10.00$^{**}$  \\
 48 & 0.080 &  1.73$^{~~}$ && 0.053 &  2.50$^{*~}$ && 0.048 &  2.70$^{**}$ && 0.053 &  3.10$^{**}$ && 0.077 &  5.34$^{**}$ && 0.091 &  8.06$^{**}$ && 0.099 & 10.76$^{**}$ && 0.092 & 10.63$^{**}$ && 0.093 & 10.87$^{**}$  \\
   \hline
   \end{tabular}
   \begin{tablenotes}
   \small
     \item This table reports the differences of the average annualized returns and the corresponding t-statistics adjusted for heteroscedasticity and autocorrelation of two contrarian strategies that are different only in the grouping methods for SZSE stocks. The three panels are for the loser, winner and contrarian portfolios, respectively. In the first row, $G_3$, $G_5$ and $G_{10}$ stand for tertile, quintile and decile groupings. The sample period is January 1997 to  December 2012. The superscripts * and ** denote the significance at 5\% and 1\% levels, respectively.
   \end{tablenotes}
\end{threeparttable}
\end{table}
\end{landscape}

\end{document}